\documentclass[a4paper]{article}
\usepackage[german, english]{babel}
\usepackage{a4wide}
\usepackage{amsmath}
\usepackage{amssymb}
\usepackage{ifthen}
\usepackage{epsfig}
\usepackage[hang,nooneline]{subfigure}
\newcounter{fig}   

\newcommand{\vphi}{\varphi}

\begin{document}

\title{\bf Boson Shells Harbouring Charged Black Holes}
\vspace{1.5truecm}
\author{
{\bf Burkhard Kleihaus$^1$, Jutta Kunz$^1$,
Claus L\"ammerzahl$^{2,1}$, Meike List$^{2,1}$}\\[10pt]
$^1$ Institut f\"ur  Physik, Universit\"at Oldenburg, Postfach 2503\\
D-26111 Oldenburg, Germany\\
$^2$ ZARM, Universit\"at Bremen, Am Fallturm\\
D-28359 Bremen, Germany
}

\vspace{1.5truecm}

\date{\today}

\maketitle
\vspace{1.0truecm}

\begin{abstract}
We consider boson shells in scalar electrodynamics
coupled to Einstein gravity.
The interior of the shells can be empty space,
or harbour a black hole or a naked singularity.
We analyze the properties of these types of solutions
and determine their domains of existence.
We investigate the energy conditions
and present mass formulae for the composite
black hole--boson shell systems.
We demonstrate that
these types of solutions violate black hole uniqueness.
\end{abstract}

\section{Introduction}

A complex scalar field theory with a suitable self-interaction
can lead to stationary localized solutions
called $Q$-balls \cite{Friedberg:1976me,Coleman:1985ki},
because the global phase invariance of the scalar field theory
is associated with a conserved charge $Q$.
When the theory is coupled to electromagnetism,
this charge represents the electromagnetic charge of the $Q$-balls
\cite{Friedberg:1976me}.

Recently, a {special} type of scalar potential was considered
\cite{Arodz:2008jk,Arodz:2008nm,Lis:2009au}.
Leading to the signum-Gordon equation for the scalar field,
this potential gives rise to spatially compact $Q$-balls 
\cite{Arodz:2008jk}.
The scalar field of these spherically symmetric configurations
is finite inside a ball of radius $r_{\rm o}$,
but vanishes identically outside this radius.
In this respect the compact $Q$-balls resemble stars.

Interestingly,
when coupled to electromagnetism, the balance of forces
allows for shell-like configurations
\cite{Arodz:2008nm}.
In these $Q$-shells the scalar field vanishes identically both inside
a certain radius $r_{\rm i}$ and outside a certain radius $r_{\rm o}$,
thus forming a shell of charged matter
of size $r_{\rm i} < r < r_{\rm o}$.
With increasing charge the shell radii increase,
allowing for arbitrarily large $Q$-shells. 
{
Note, that the shells considered here are thick shells,
in contrast to the often considered thin shells 
(see e.g.~Ref.~\cite{Gao:2008jy} and references therein).}

When the scalar field is coupled to gravity, branches of
globally regular self-gravitating solutions emerge
from the $Q$-ball solutions, corresponding to boson stars
\cite{Lee:1991ax,Jetzer:1991jr,Mielke:2000mh,Schunck:2003kk}.
The compact $Q$-balls then give rise to compact boson stars.
Likewise, in the presence of gravity
the compact $Q$-shells turn into compact shells of
self-gravitating charged matter,
gravitating boson shells \cite{Kleihaus:2009kr}.
Recently, we have analyzed the properties and
the domain of existence of these compact boson stars and
gravitating boson shells \cite{Kleihaus:2009kr}.

The gravitating boson shells surround a flat Minkowski-like interior region,
$r < r_{\rm i}$,
while their exterior region $r>r_{\rm o}$ is described
by the part $r>r_{\rm o}$ of a Reissner-Nordstr\"om solution,
which possesses the same charge and mass as the shell.
Unlike the $Q$-shells in flat space, however,
the gravitating boson shells cannot carry arbitrarily large charge.
As the charge increases the mass increases as well, 
until a limiting solution is reached.
Here the mass within the shell radius $r_{\rm o}$ becomes too big
for a regular shell space-time to persist,
and a throat is formed at $r_{\rm o}$.
The exterior space-time $r>r_{\rm o}$ then corresponds precisely
to the exterior of an extremal Reissner-Nordstr\"om solution.

However, the shells need not be empty in their interior $r<r_{\rm i}$.
Instead of flat Minkowski space the gravitating boson shells
can harbour a Schwarz\-schild like black hole there
\cite{Kleihaus:2009kr}.
This black hole has its event horizon $r_{\rm H}$ in the interior region
$0< r_{\rm H} < r_{\rm i}$, 
where the scalar field vanishes and the gauge potential is constant.
Since the black hole is surrounded by a shell of charged boson matter,
the presence of the scalar field outside the event horizon
may be interpreted at scalar hair.
Thus the black hole theorems forbidding scalar hair
under a great variety of circumstances
\cite{Bekenstein:1971hc,Bekenstein:1995un,Mayo:1996mv}
can be eluded in this model.

Here we show, that
the shells may also harbour charge in their interior $r<r_{\rm i}$.
The flat space shells may contain point charges at their center,
whereas the gravitating boson shells may either harbour
charged black holes or naked singularities.
These interior solutions are then described by
Reissner-Nordstr\"om like solutions.
Thus a space-time arises, that consists of a Reissner-Nordstr\"om like
interior $r<r_{\rm i}$, a boson shell $r_{\rm i} < r < r_{\rm o}$, and a
Reissner-Nordstr\"om like exterior $r_{\rm o} < r$.
Depending on the values of the mass and charge,
the interior solution will be a subextremal black hole,
an extremal black hole or a naked singularity.

We analyze the properties of these boson shell solution
carrying charge also in their interior $r<r_{\rm i}$, 
and we determine their domain of existence.
We show that as in the case of the uncharged interior solutions,
a throat can develop at the outer radius $r_{\rm o}$.
This renders the exterior space-time
an exterior extremal Reissner-Nordstr\"om space-time.
At the same time the temperature of the black hole event horizon $r_{\rm H}$ 
in the interior tends to zero.
When the interior black hole solution becomes extremal,
a further increase of the charge in the interior
results in a naked singularity surrounded only by the shell.
Moreover, space-times exist which
possess both an extremal interior black hole solution
as well as an outer throat.

The paper is organized as follows.
In section 2 we recall the action,
the equations of motion and the ansatz for the fields.
We also specify the boundary conditions,
discuss the global charges,
and present mass theorems for the solutions.
In section 3 we discuss boson shells
with a Minkowski like interior and with a Schwarzschild like interior.
We then turn in section 4 to the boson shells
carrying charge in their interior,
which may be point like or in the form of
a charged black hole.
We analyze the energy conditions for all of these solutions
in section 5,
and end with our conclusions in section 6.
A discussion of the special spiral like behaviour of the solutions,
which occurs when the event horizon approaches the inner shell radius,
is presented in the Appendix.

\section{Action, Equations, Boundary Conditions}

\subsection{Action and equations of motion}

We consider the action of a self-interacting complex scalar field
$\Phi$ coupled to a U(1) gauge field and to Einstein gravity
\begin{equation}
S=\int \left[ \frac{R}{16\pi G}
   - \frac{1}{4} F^{\mu\nu} F_{\mu\nu}
   -  \left( D_\mu \Phi \right)^* \left( D^\mu \Phi \right)
 - U( \left| \Phi \right|) 
 \right] \sqrt{-g} d^4x
 , \label{action}
\end{equation}
with field strength tensor
\begin{equation}
F_{\mu\nu} = \partial_\mu A_\nu - \partial_\nu A_\mu 
 , \label{action2}
\end{equation}
covariant derivative
\begin{equation}
D_\mu \Phi = \partial_\mu \Phi + i e A_\mu \Phi
 , \label{action3}
\end{equation}
curvature scalar $R$, Newton's constant $G$, 
gauge coupling constant $e$,
and the asterisk denotes complex conjugation.
The scalar potential $U$ is chosen as
\cite{Arodz:2008jk,Arodz:2008nm,Lis:2009au}
\begin{equation}
U(|\Phi|) =  \lambda  |\Phi| 
 . \label{U} \end{equation} 

Variation of the action with respect to the metric and the matter fields
leads, respectively, to the Einstein equations
\begin{equation}
G_{\mu\nu}= R_{\mu\nu}-\frac{1}{2}g_{\mu\nu}R = 8\pi G T_{\mu\nu}
\  \label{ee} \end{equation}
with stress-energy tensor
\begin{eqnarray}
T_{\mu\nu} &=& g_{\mu\nu}{L}_M
-2 \frac{\partial {L}_M}{\partial g^{\mu\nu}}
=
    ( F_{\mu\alpha} F_{\nu\beta} g^{\alpha\beta}
   -\frac{1}{4} g_{\mu\nu} F_{\alpha\beta} F^{\alpha\beta})
\nonumber\\
&-& 
   \frac{1}{2} g_{\mu\nu} \left(     (D_\alpha \Phi)^* (D_\beta \Phi)
  + (D_\beta \Phi)^* (D_\alpha \Phi)    \right) g^{\alpha\beta}
  + (D_\mu \Phi)^* (D_\nu \Phi) + (D_\nu \Phi)^* (D_\mu \Phi)
\nonumber\\
&-& 
    {\lambda} g_{\mu\nu}  |\Phi| 
 , \label{tmunu}
\end{eqnarray}
and the matter field equations,
\begin{eqnarray}
& & \partial_\mu \left ( \sqrt{-g} F^{\mu\nu} \right) =
   \sqrt{-g} e \Phi^* D^\nu \Phi 
\label{feqA} \end{eqnarray}
\begin{eqnarray}
& &D_\mu\left(\sqrt{-g}  D^\mu \Phi \right) =
    -\sqrt{-g} \frac{\lambda}{2} \frac{\Phi}{|\Phi|}
 . \label{feqH} \end{eqnarray}

\subsection{Ansatz}

To construct spherically symmetric solutions
we employ Schwarz\-schild like coordinates and adopt
the metric
\begin{equation}
ds^2=g_{\mu\nu}dx^\mu dx^\nu=
  -A^2N dt^2 + N^{-1} dr^2 + r^2 (d\theta^2 + \sin^2\theta d\phi^2) 
 . \end{equation}
For solutions with vanishing magnetic field
the Ansatz for the matter fields has the form
\begin{equation}
 \Phi = \phi(r) e^{i \omega t}
 , \label{phi} \end{equation}
\begin{equation}
 A_\mu d x^\mu = A_0(r) dt
 . \label{A_0} \end{equation}
 
For notational simplicity, we 
introduce new coupling constants \cite{Arodz:2008nm}
\begin{equation}
\alpha^2 = a = 4\pi G \frac{\beta^{1/3}}{e^2},
\ \ \
\beta = \frac{\lambda e}{\sqrt{2}} 
 , \label{constants} \end{equation}
and redefine the matter field functions,
\begin{equation}
h(r) = \sqrt{2}e \phi(r) ,
\ \ \
b(r) = \omega + e A_0(r)
 . \label{functions} \end{equation}
The latter corresponds to
performing a gauge transformation to make the scalar field real
and absorbing the frequency $\omega$ of the scalar field into the gauge
transformed vector potential.
Note, that the parameter $\beta$ can be removed by rescaling 
and will therefore be set to one
\cite{Arodz:2008nm}.
Thus the only parameter left is the gravitational coupling $\alpha$.
{In the following we will consider $h(r)$ as non-negative.}

With the above ansatz the Einstein equations 
$G_t^t = 2\alpha^2 T_t^t$, $G_r^r = 2\alpha^2 T_r^r$, 
$G_\theta^\theta = 2\alpha^2 T_\theta^\theta$ reduce to
\begin{eqnarray}
\frac{-1}{r^2}\left[r\left(1-N\right)\right]' 
& = & 
-\frac{\alpha^2}{A^2 N}\left(A^2 N^2 h'^2 + N b'^2 +2 A^2 N h+ b^2 h^2\right) \ ,
\label{E_00}\\
\frac{2 r A' N -A\left[r\left(1-N\right)\right]'}{A r^2}
& = & 
\frac{\alpha^2}{A^2 N}
\left(A^2 N^2 h'^2-N b'^2 -2 A^2 N h+ b^2 h^2\right) \ ,
\label{E_rr}\\
\frac{2r\left[rA'N\right]' + A\left[r^2 N'\right]'}{2 A r^2}
& = & 
\frac{\alpha^2}{A^2 N}
\left(-A^2 N^2 h'^2 + N b'^2 -2 A^2 N h+ b^2 h^2\right) \ ,
\label{E_tt}
\end{eqnarray}
respectively, {where the prime denotes differentiation with respect to $r$.}
Solving eqs.~(\ref{E_00}) and (\ref{E_rr}) for 
$N'$ and $A'$ yields
\begin{eqnarray}
N' & = & \frac{1-N}{r} 
-\frac{\alpha^2 r}{A^2 N}
\left(A^2 N^2 h'^2 + N b'^2 +2 A^2 N h+ b^2 h^2\right) \ ,
\label{eq_N}\\
A' & = & 
\frac{\alpha^2 r}{A N^2}\left(A^2 N^2 h'^2 + b^2 h^2\right) \ ,
\label{eq_A}\\
\end{eqnarray}
The field equations 
$\left[\frac{\partial L_M}{\partial h'}\right]' = \frac{\partial L_M}{\partial h}$ 
and 
$\left[\frac{\partial L_M}{\partial b'}\right]' = \frac{\partial L_M}{\partial b}$ 
read
\begin{eqnarray}
\left[A N r^2 h'\right]' 
&  = & 
\frac{r^2}{AN}\left(A^2N {\rm sign}(h) -b^2 h\right)\   ,
\label{eqq_H}\\
\left[\frac{r^2 b'}{A}\right]' 
&  = & 
\frac{b h^2 r^2}{AN} \   .
\label{eqq_b}
\end{eqnarray}
where ${\rm sign}(h)=1$ for $h>0$, but ${\rm sign}(0)=0$.
After elimination of $A'$ and $N'$ we obtain
\begin{eqnarray}
h'' & = & 
\frac{\alpha^2}{A^2N} r h' \left(2A^2 h +b'^2\right)
-\frac{h'\left(N+1\right)}{rN}
+\frac{A^2N {\rm sign}(h)-b^2 h}{A^2 N^2} \ ,
\label{eq_H}\\
b'' & = & 
\frac{\alpha^2}{A^2 N^2} rb'\left(A^2 N^2 h'^2 + b^2 h^2\right)
-\frac{2 b'}{r} + \frac{b h^2}{N}
\label{eq_b}
\end{eqnarray}

In order to solve the ODEs (\ref{eq_N}), (\ref{eq_A}), (\ref{eq_H}), (\ref{eq_b})
numerically we introduce a new coordinate $x$ via 
\begin{equation}
r= r_{\rm i} +x (r_{\rm o}-r_{\rm i})\ , \ \ \ \ 0\leq x \leq 1 \ .
\label{coord_x}
\end{equation}
Thus the inner and outer boundaries of the shell are always at $x=0$, respectively $x=1$,
while their radii $r_{\rm i}$ and $r_{\rm o}$ become free parameters.

\subsection{Boundary conditions}

Let us now specify the boundary conditions for the 
metric and matter functions.
For the metric function $A$ we adopt
\begin{equation}
A(r_{\rm o})=1
\ , \end{equation}
where $r_{\rm o}$ is the outer radius,
thus fixing the time coordinate. 
For the metric function $N(r)$ we require for globally regular ball-like
boson star solutions
\begin{equation}
N(0)=1 
, \end{equation}
for globally regular shell-like solutions 
\begin{equation}
N(r_{\rm i})=1 
, \end{equation}
where $r_{\rm i}$ is the inner radius of the shell.
For globally regular boson star solutions we require
at the origin and at the outer radius $r_{\rm o}$
the conditions
\begin{equation}
b'(0)=0 , \ \ \ h'(0)=0 , \ \ \ h(r_{\rm o})=0 , \ \ \ h'(r_{\rm o})=0 \ .
 \end{equation}
In order to choose also the value of $b(0)$ as a boundary condition,
we make the outer radius $r_{\rm o}$ an auxiliary (constant) variable,
and thus add the differential equation $r_{\rm o}\,'=0$,
without imposing a boundary condition.

For globally regular shell solutions 
we require at the inner radius $r_{\rm i}$ and at the outer radius $r_{\rm o}$
the conditions
\begin{equation}
b'(r_{\rm i})=0 , \ \ \ h(r_{\rm i})=0 , \ \ \ h'(r_{\rm i})=0 , \ \ \ h(r_{\rm o})=0 , \ \ \ h'(r_{\rm o})=0
 . \end{equation}
In order to choose also $b(r_{\rm i})=b_i$ as a boundary condition,
we now also make the ratio of inner and outer radius $r_{\rm i}/r_{\rm o}$
an auxiliary (constant) variable.
Alternatively to demanding a certain value for $b(r_{\rm i})$,
we may also specify the value of the electric charge $Q$
{i.e., $b'(r_{\rm o})=-Q/r_{\rm o}^2$.}

For electrically charged black hole solutions in the interior
of the shell, $r < r_{\rm i}$, we consider the exact solution
\begin{eqnarray}
N(r)& = & \left(1-\frac{r_{\rm H}}{r}\right)\left(1-\frac{r_{\rm C}}{r}\right) \ ,
\nonumber\\
b(r)& = & \beta_{\rm i} - \frac{Q_{\rm H}}{r} A_{\rm i} \ ,
\nonumber\\
A(r) & = & A_{\rm i} \ , 
\nonumber\\
h(r) & = & 0\ . 
\nonumber
\end{eqnarray}
where $A_{\rm i}$ and $\beta_{\rm i}$ are constants. The event horizon radius $r_{\rm H}$ 
and Cauchy horizon radius $r_{\rm C}\leq r_{\rm H}$ are related to the 
horizon charge $Q_{\rm H}$ by 
\begin{equation}
\alpha^2 Q_{\rm H}^2 =  r_{\rm H} r_{\rm C} \ .
\label{A_H}
\end{equation}

Evaluation at the inner radius of the shell yields the boundary conditions
\begin{eqnarray}
N(r_{\rm i})& = & \left(1-\frac{r_{\rm H}}{r_{\rm i}}\right)\left(1-\frac{r_{\rm C}}{r_{\rm i}}\right) \ ,
\nonumber\\
b(r_{\rm i}) & = & b_{\rm i} \ , 
\nonumber\\
b'(r_{\rm i})& = & \mp \frac{A_{\rm i}}{\alpha}\frac{\sqrt{r_{\rm H} r_{\rm C}}}{r_{\rm i}^2} \ ,
\nonumber\\
h(r_{\rm i}) & = & 0\ ,
\nonumber\\
h'(r_{\rm i}) & = & 0\ . 
\nonumber
\end{eqnarray}
At the outer radius of the shell the boundary conditions are the same
as for the ball-like boson star solutions and the shell-like solutions.

Thus the solutions are specified by the parameters
$\alpha$, $r_{\rm H}$, $r_{\rm C}$, and $b_{\rm i}$ (or $Q$).

\subsection{Charge and mass}

Let us define the electric charge $Q_{\rm S}$
localized within a given 2-sphere $S$ 
by\footnote{{ Note that our convention for the charge \cite{Arodz:2008nm} 
differs in sign from the usual one.}}
\begin{equation}
Q_{\rm S} = \frac{1}{4 \pi} \int_{S}
 {^*{F}_{\theta\varphi}}   d\theta d\varphi
\ . \label{def_CalQ} \end{equation}
The global charge of the solutions is then obtained
by taking the surface $S$ to infinity, yielding
\begin{equation}
Q= \frac{1}{4 \pi} \int_{\infty}
 {^*{F}_{\theta\varphi}}   d\theta d\varphi \ .
\end{equation}
When a charged black hole is located inside the shell,
the horizon charge $Q_{\rm H}$ of the black hole
is obtained, by locating the surface $S$ at the event horizon.
Furthermore, a point charge $Q_{\rm i}$
sitting at the origin may be considered.

The bosons contribute to the global charge $Q$ via the 
conserved current
\begin{eqnarray}
j^{\mu} & = &  - i \left( \Phi^* D^{\mu} \Phi
 - \Phi D^{\mu}\Phi ^* \right) \ , \ \ \
j^{\mu} _{\ ; \, \mu}  =  0 \ .
\end{eqnarray}
For the above ansatz the expression for the
time component of the electromagnetic current is
\begin{equation}
j^0 = -\frac{h^2 b}{A^2 N} \ .
\label{eq_j0}
\end{equation}
The charge contribution from
the boson shells is then obtained as the spatial integral
\begin{equation}
Q_{\rm sh} =  -\frac{1}{4\pi}\int_{r_{\rm i}}^{r_{\rm o}} j^0 \sqrt{-g} dr d\theta d\varphi
. \end{equation}
The global charge therefore
consists of the charge carried by the bosons forming the shells
and the charge localized in the interior,
$Q_{\rm i}$ or $Q_{\rm H}$.

The mass $M$ of the stationary asymptotically flat space-times
is obtained from the corresponding Komar expression.
For globally regular space-times like boson stars
and shells of boson matter the mass is given by
\begin{equation}
M = 
 \frac{1}{{4\pi G}} \int_{\Sigma}
 R_{\mu\nu}n^\mu\xi^\nu dV
 , \label{komarM1}
\end{equation}
where $\Sigma$ denotes an asymptotically flat spacelike hypersurface,
$n^\mu$ is normal to $\Sigma$ with $n_\mu n^\mu = -1$,
$dV$ is the natural volume element on $\Sigma$,
and $\xi$ denotes an asymptotically timelike Killing vector field
\cite{wald}.
Replacing the Ricci tensor via the Einstein equations by the
stress-energy tensor yields
\begin{equation}
M = 
  \, 2 \int_{\Sigma} \left(  T_{\mu \nu}
-\frac{1}{2} \, g_{\mu\nu} \, T_{\gamma}^{\ \gamma}
 \right) n^{\mu }\xi^{\nu} dV
 . \label{komarM2}
\end{equation}

For black hole space-times the corresponding Komar expression is given by
\begin{equation}
M = M_{\rm H} +
  \, 2 \int_{\Sigma} \left(  T_{\mu \nu}
-\frac{1}{2} \, g_{\mu\nu} \, T_{\gamma}^{\ \gamma}
 \right) n^{\mu }\xi^{\nu} dV 
 , \label{komarM3}
\end{equation}
where $M_{\rm H}$ is the horizon mass of the black hole.
The mass of all gravitating solutions
can be directly obtained from the asymptotic form of their metric.
In the units employed, we find
\begin{equation}
M= \frac{1}{\alpha^2}\, \lim_{r \rightarrow \infty} m(r)
 , \label{mass} \end{equation} 
where the mass function $m(r)$ is related to the metric function $N(r)$
by $N(r)= 1-2 m(r)/r$.

Finally we recall that, for fixed gravitational coupling constant $\alpha$,
regular solutions satisfy \cite{Arodz:2008nm,Kleihaus:2009kr}
\begin{equation}
d M = b(\infty) d Q 
 , \label{Mreg2} \end{equation}
{where $b(\infty)$ represents the electrostatic potential at infinity.} 
Thus integration yields the mass relation
\begin{equation}
M_2 =  M_1 + M_Q =
 M_1 + \int_{Q_1}^{Q_2} b(\infty) d Q
 , \label{Mreg} \end{equation}
where the mass $M_2$ of a regular solution with charge $Q_2$
is obtained
by integrating from any regular solution $M_1$ with charge $Q_1$
along the curve of intermediate solutions of the set.

The respective mass relation 
for black holes space-times within boson shells
follows in the isolated horizon framework
\cite{Ashtekar:2004cn}.
The latter states that the mass $M$
of a black hole space-time with horizon radius $r_{\rm H}$
and the mass $M_{\rm reg}$
of the corresponding globally regular space-time
obtained in the limit $r_{\rm H} \rightarrow 0$ are related via
\cite{Corichi:1999nw,Ashtekar:2000nx,Ashtekar:2004cn}
\begin{equation}
M = M_{\rm reg} + M_\Delta  ,
\label{IHmu} \end{equation}
where the mass contribution $M_\Delta$ is defined by
\begin{equation}
M_\Delta = \frac{1}{\alpha^2} \, \int_0^{r_{\rm H}} \kappa(r'_{\rm H})r'_{\rm H} d r'_{\rm H}  .
\label{IHmuD}
\end{equation}
Here $\kappa(r_{\rm H})$ represents the surface gravity
of the black hole with horizon radius $r_{\rm H}$,
$\kappa = 2 \pi T$.
Accordingly, the mass $M$
of a space-time with a black hole
with horizon radius $r_{\rm H}$ within a boson shell with global charge $Q$
is obtained as the sum of the globally regular gravitating boson shell
with charge $Q$ and the integral $M_\Delta$ along the set of black hole
space-times, obtained by increasing the horizon radius for fixed charge
from zero to $r_{\rm H}$ \cite{Kleihaus:2009kr}.

When the charge is allowed to vary, too, the above
relation generalizes. In accordance with (\ref{Mreg2})
and the first law (in the units employed)\footnote{{ Note that the 
signs of the last two terms follow from our definition of the 
electric charge.}}, i.e.,
\begin{equation}
dM = \frac{\kappa}{8 \pi \alpha^2} d{\cal A} + b(\infty) dQ
     -b(r_{\rm H}) dQ_{\rm H}
 , \label{firstlaw} \end{equation}
where ${\cal A} = 4 \pi r_{\rm H}^2$ denotes the area of the horizon,
the generalized relation reads \cite{Kleihaus:2009kr}
\begin{equation}
M = M_{\rm reg} + M_\Delta + M_Q =
 M_{\rm reg} + M_\Delta 
  +  \int_{Q_{\rm reg}}^{Q} b(\infty) d Q' 
  -  \int_{0}^{Q_{\rm H}} b(r_{\rm H}) d Q'_{\rm H} .
\label{IHmuDQ2}
\end{equation}

\section{Boson shells with neutral interior}

To set the stage for the discussion of the 
boson shell solutions with neutral black holes in their interior,
we begin by briefly recalling
the basic properties and the domain of existence
of the boson shells with Minkowski like interior
\cite{Kleihaus:2009kr}.

\subsection{Empty boson shells}

\begin{figure}[h]
\begin{center}
\mbox{\hspace{-1.5cm}
\subfigure[][]{
\includegraphics[height=.27\textheight, angle =0]{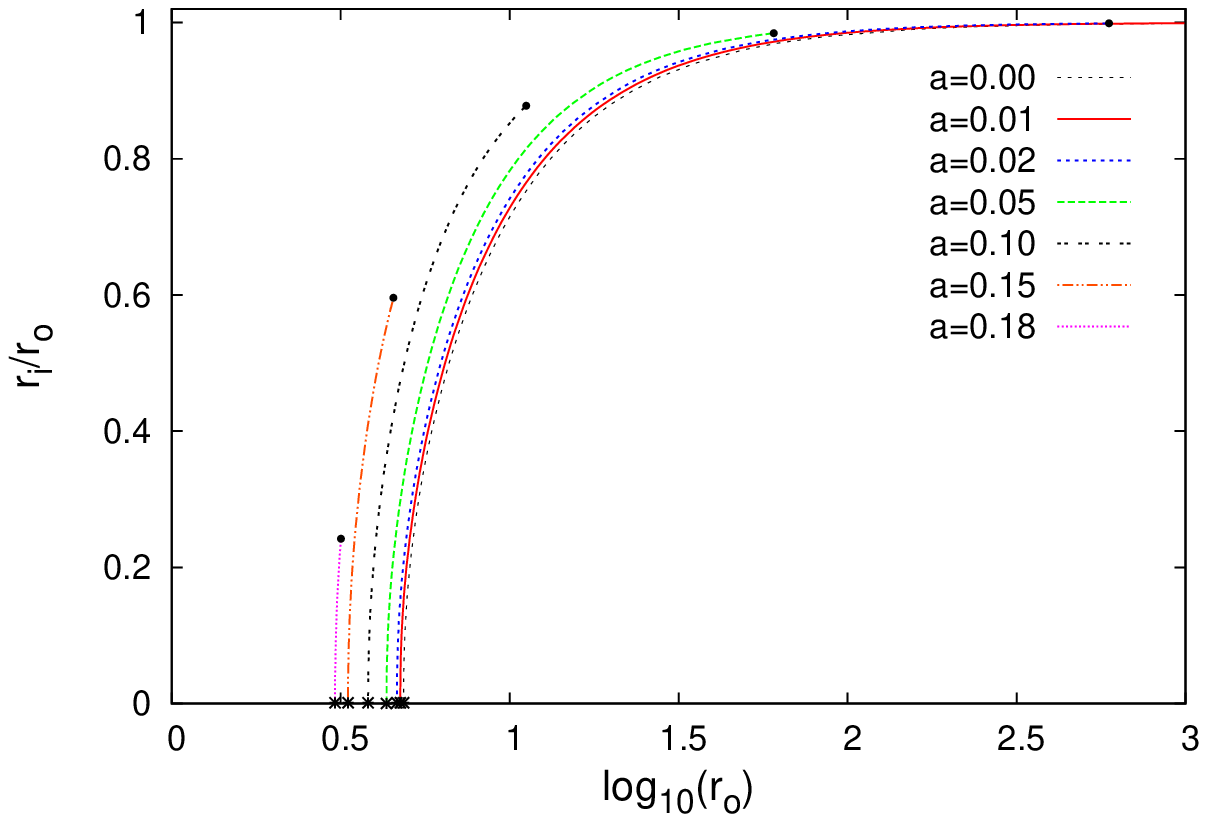}
\label{Sha}
}
\subfigure[][]{\hspace{-0.5cm}
\includegraphics[height=.27\textheight, angle =0]{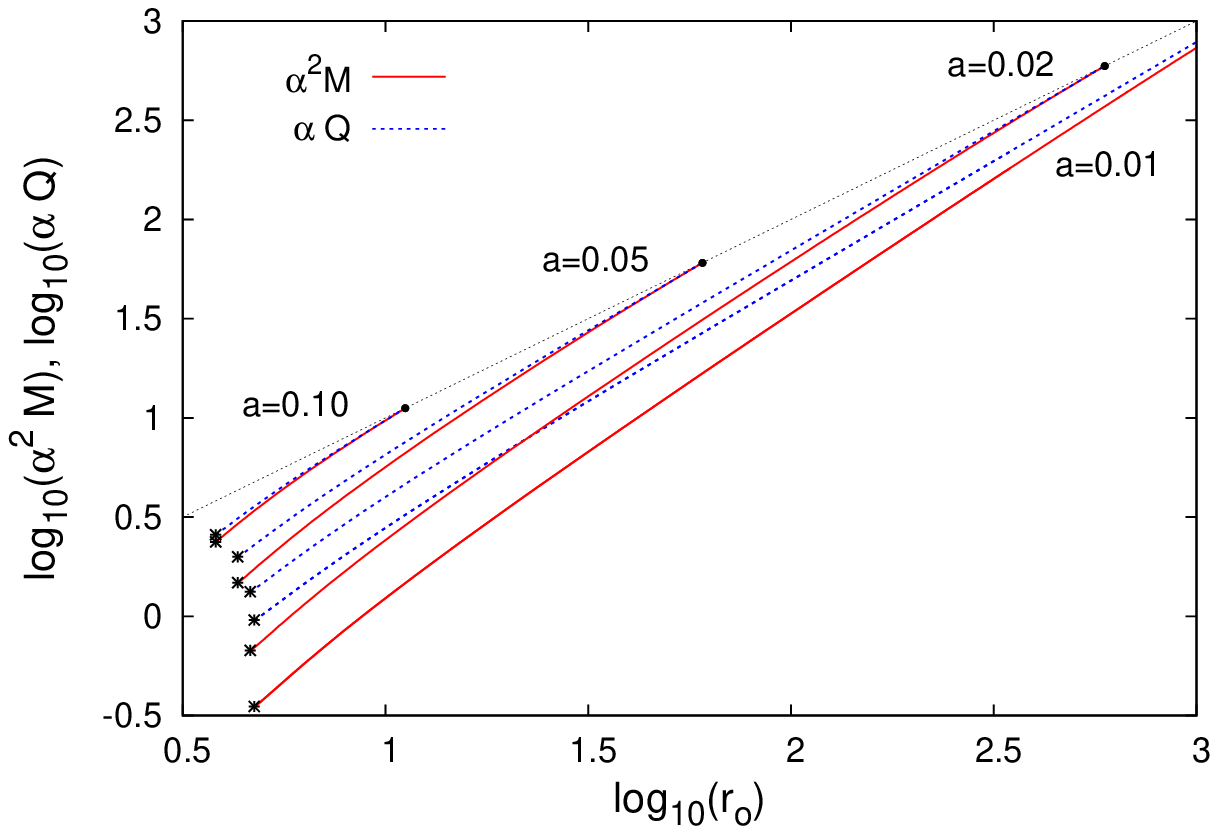}
\label{Shb}
}
}
\end{center}
\vspace{-0.5cm}
\caption{Properties of gravitating boson shells shown
versus the outer radius $r_{\rm o}$ of the shells
for several values of the gravitational coupling constant $a=\alpha^2$:
(a) the ratio of the inner to outer radius $r_{\rm i}/r_{\rm o}$ of the shell;
(b) the scaled mass $\alpha^2 M$ and the scaled charge $\alpha Q$
of the shell,
together with the condition for throat formation
$r_0=\alpha^2 M = \alpha Q$.
The large asterisks mark the bifuraction
points from boson stars ($Q$-balls) to boson shells,
the small asterisks mark the solutions with throats.
\label{sh1}
}
\end{figure}

In boson shells we need to distinguish 3 regions of the space-time.
In the inner region $0  \le r < r_{\rm i}$
the gauge potential is constant and the scalar field vanishes.
Consequently, it is Minkowski-like,
with $N(r)=1$ and $A(r)={\rm const}<1$.
The middle region $r_{\rm i} < r < r_{\rm o}$ represents the shell of
charged boson matter. The outer region $r_{\rm o} < r < \infty$, finally,
equals to the outer part of a Reissner-Nordstr\"om space-time
(for a naked singularity since the charge of the boson shell
is larger than the mass).
In this outer region the gauge field exhibits the standard Coulomb fall-off,
while the scalar field vanishes identically.
(A typical boson shell solution is seen in Fig.~\ref{S0a},
where it represents the beginning of a branch of
solutions with Schwarzschild like interior.)

The domain of existence of these gravitating boson shells
depends on the strength of the gravitational coupling 
$a = \alpha^2$.
For a given finite value of the gravitational coupling,
boson shells emerge from the boson star solutions,
when the scalar field vanishes at the origin, $h(0)=0$.
The value of the outer radius $r_{\rm o}$ at the bifurcation point
depends on the strength of the gravitational coupling.
{It decreases monotonically
from $r_{\rm o}\approx 4.9$ for vanishing coupling $\alpha=0$ to
$r_{\rm o} \approx 3.0$ for the maximal coupling $\alpha_{\rm cr}$,
for which boson shells exist.}

When the value of the inner shell radius $r_{\rm i}$ is increased from zero
while the gravitational coupling constant is kept fixed,
the corresponding branch of boson shells is obtained. 
With increasing inner shell radius $r_{\rm i}$ also the outer radius $r_{\rm o}$
increases. This is seen in Fig.~\ref{Sha}, where
the ratio $r_{\rm i}/r_{\rm o}$ of both shell radii is shown versus the outer radius $r_{\rm o}$.
Along with the shell radii also the mass $M$ and the charge $Q$
of the shells increase, as seen in Fig.~\ref{Shb}.

Since the mass increases faster than the outer shell radius
(and also faster than the charge),
one can expect, that at a given point there will be too much mass
within a region of radius $r_{\rm o}$ to still allow for globally regular solutions. 
Indeed, the branches of boson shells end,
when a throat is formed at the outer radius $r_{\rm o}$.
As this happens, the value of the gauge field function $b(r)$
reaches zero at the inner radius $r_{\rm i}$
(or equivalently $b(0)\rightarrow 0$,
since $b(r)$ is constant in the interior, $0 \le r \le r_{\rm i}$).
The outer space-time $r > r_{\rm o}$ then corresponds to the exterior of an
extremal Reissner-Nordstr\"om space-time.
Such an extremal Reissner-Nordstr\"om space-time requires a certain relation
between the horizon radius $r_{\rm H}$, the mass $M$ and the charge $Q$.
In our units and with the horizon at the outer shell radius $r_{\rm o}$ 
this relation becomes
\begin{equation}
r_{\rm o}= \alpha^2 M = \alpha Q \ .
\end{equation}
As seen in Fig.~\ref{Shb}, this relation indeed holds
at the endpoints of the boson shell branches.

We finally note, that
for large gravitational coupling, the boson shell branches 
end long before the inner and outer shell radii become
comparable in size.
In contrast, for small gravitational coupling, the boson shell branches 
extend much further and end only when the two shell radii are almost
of the same size.
However, as long as the gravitational coupling is finite,
the growth of gravitating boson shells is limited by gravity.
Only in flat space boson shells can have arbitrarily large 
mass, charge and size
\cite{Arodz:2008nm}.

\subsection{Boson shells with Schwarzschild like black holes: $Q_{\rm H} =0$}

\begin{figure}[h]
\begin{center}
\vspace{-0.5cm}
\mbox{\hspace{-1.5cm}
\subfigure[][]{
\includegraphics[height=.27\textheight, angle =0]{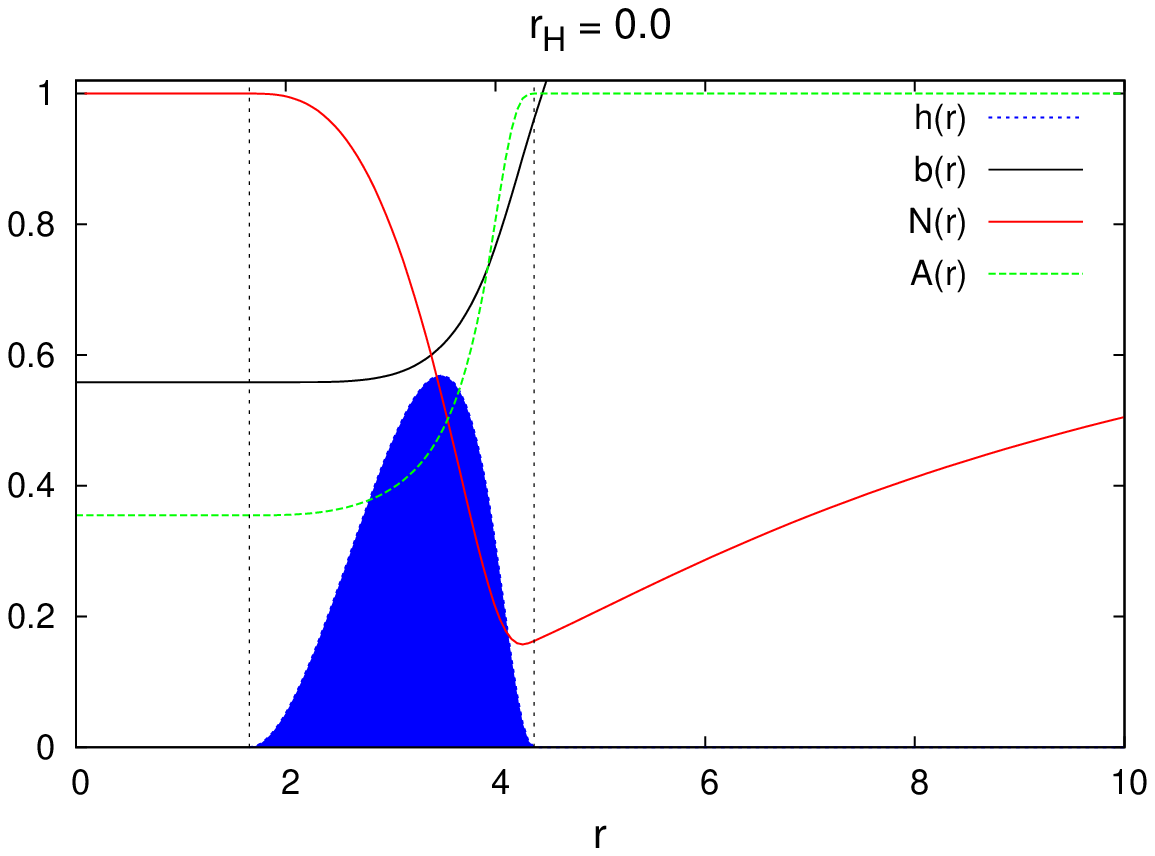}
\label{S0a}
}
\subfigure[][]{\hspace{-0.5cm}
\includegraphics[height=.27\textheight, angle =0]{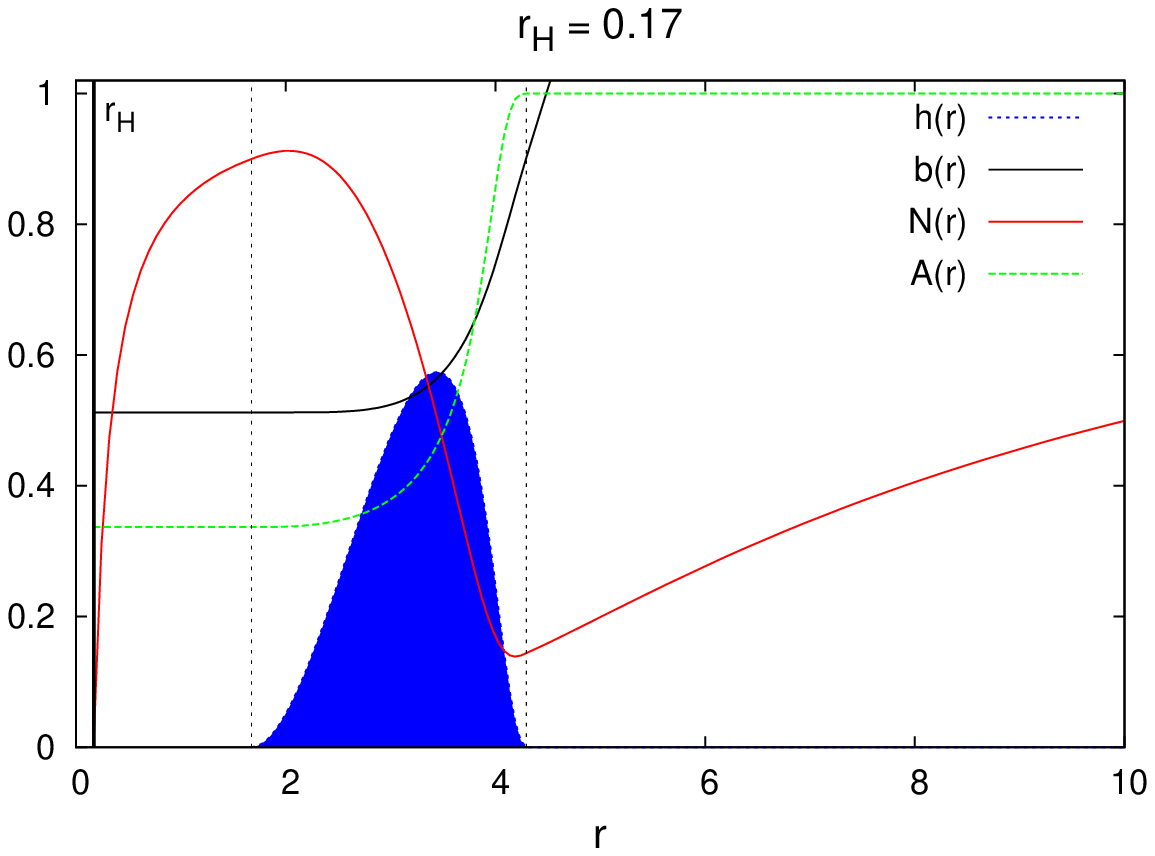}
\label{S0b}
}
}
\vspace{-0.5cm}
\mbox{\hspace{-1.5cm}
\subfigure[][]{
\includegraphics[height=.27\textheight, angle =0]{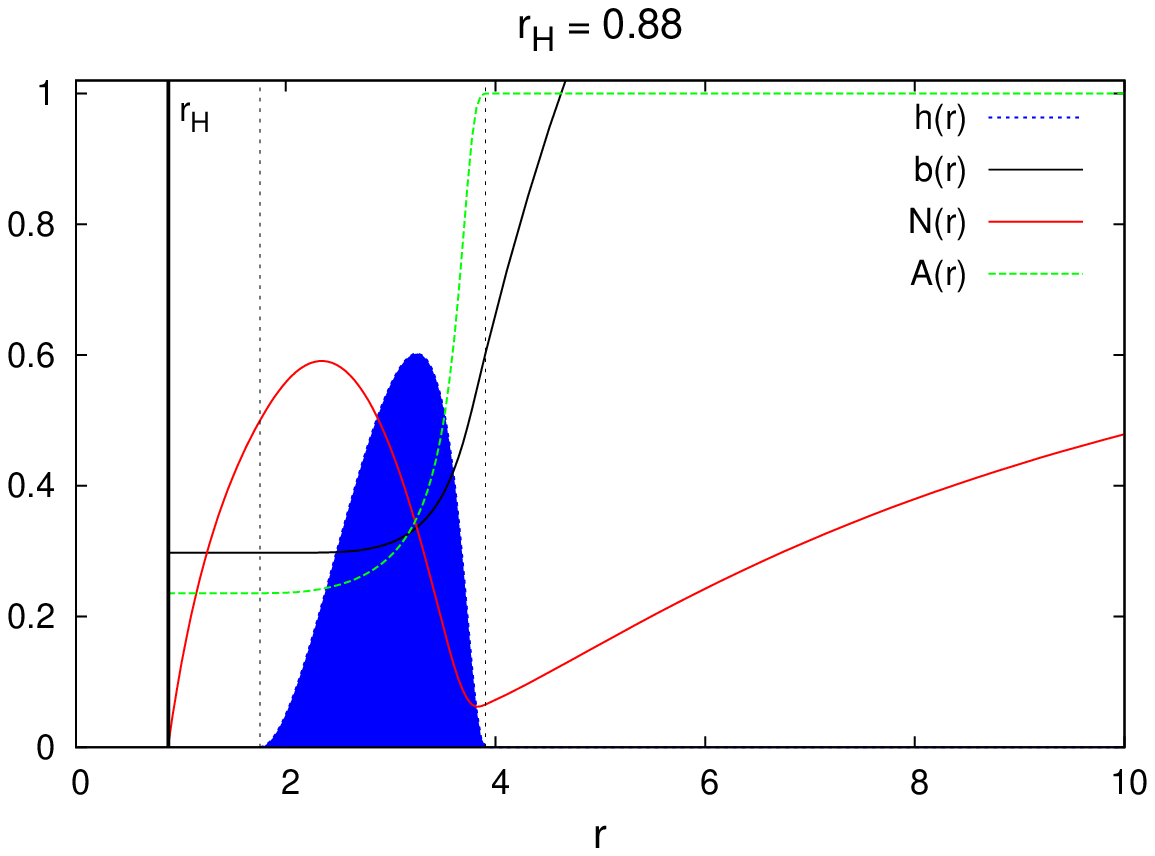}
\label{S0c}
}
\subfigure[][]{\hspace{-0.5cm}
\includegraphics[height=.27\textheight, angle =0]{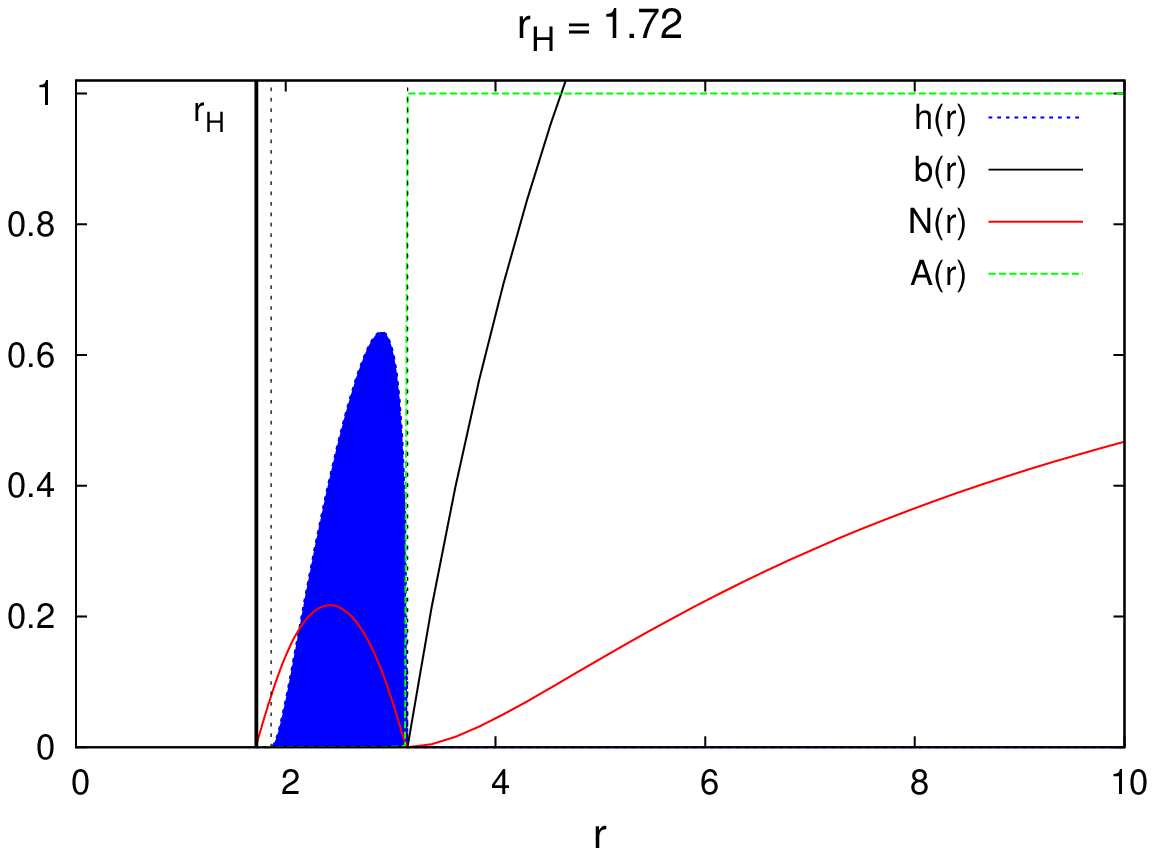}
\label{S0d}
}
}

\end{center}
\caption{Functions $h(r)$, $b(r)$, $N(r)$ and $A(r)$ versus
the radial coordinate $r$
with $a=\alpha^2=0.1$, $Q=10$
for boson shells with Minkowski and Schwarzschild like interior:
(a) $r_{\rm H}=0$; 
(b) $r_{\rm H}=0.17$; 
(c) $r_{\rm H}=0.88$; 
(d) $r_{\rm H}=1.72$.
\label{bhS0}
}
\end{figure}

\begin{figure}[p]
\begin{center}
\vspace{-1.5cm}
\mbox{\hspace{-1.5cm}
\subfigure[][]{
\includegraphics[height=.27\textheight, angle =0]{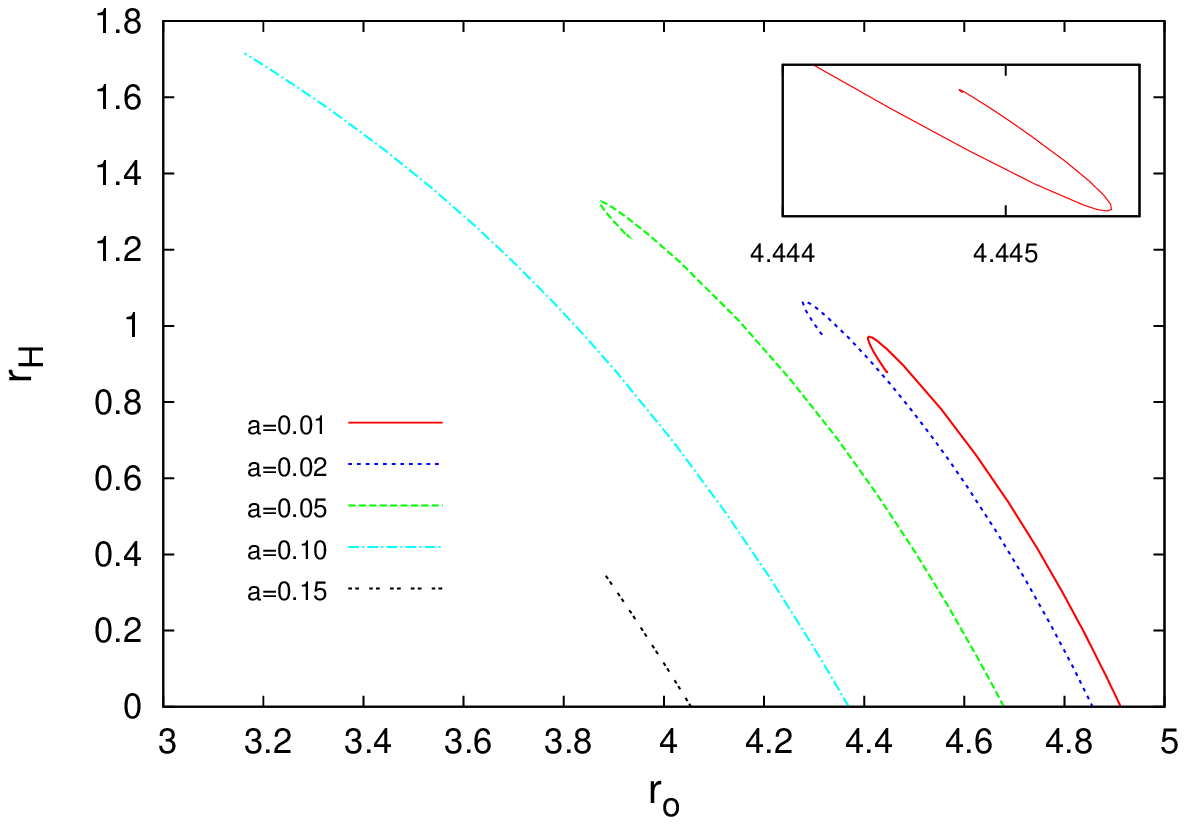}
\label{S1a}
}
\subfigure[][]{\hspace{-0.5cm}
\includegraphics[height=.27\textheight, angle =0]{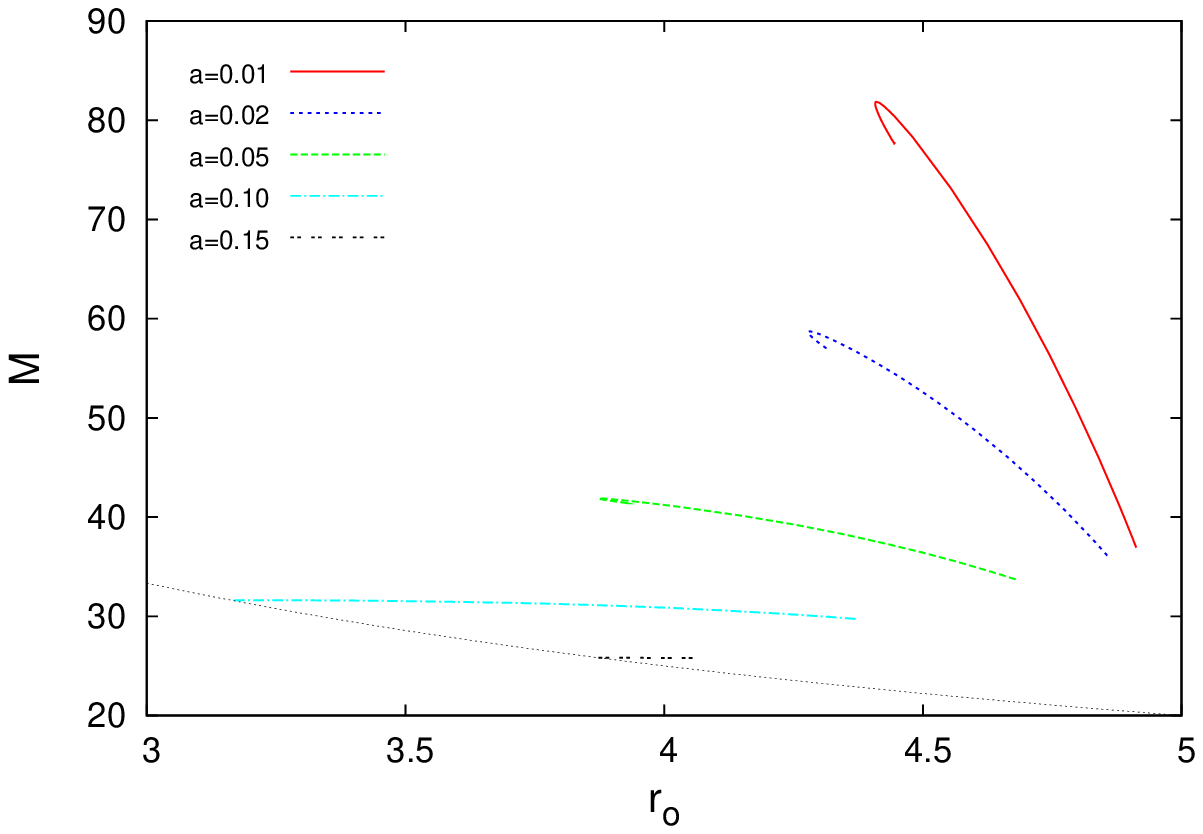}
\label{S1b}
}
}
\vspace{-0.5cm}
\mbox{\hspace{-1.5cm}
\subfigure[][]{
\includegraphics[height=.27\textheight, angle =0]{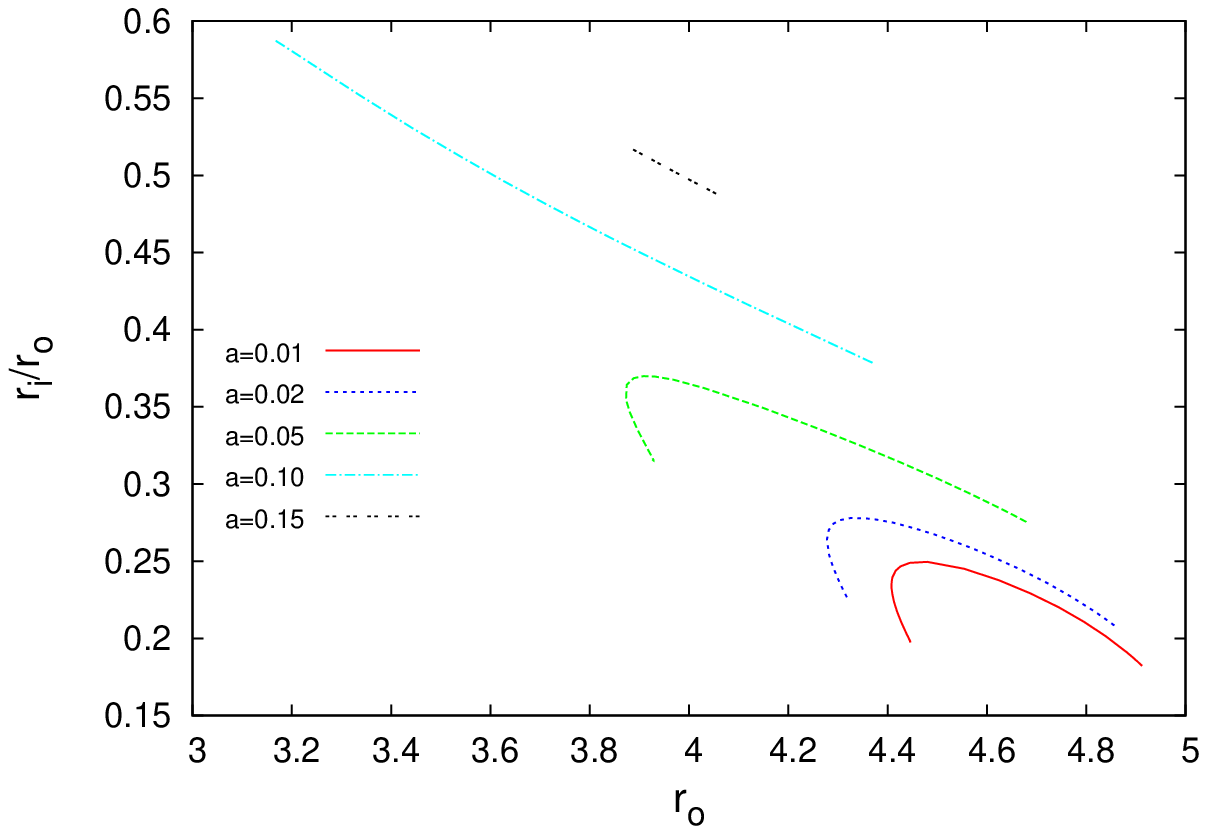}
\label{S1c}
}
\subfigure[][]{\hspace{-0.5cm}
\includegraphics[height=.27\textheight, angle =0]{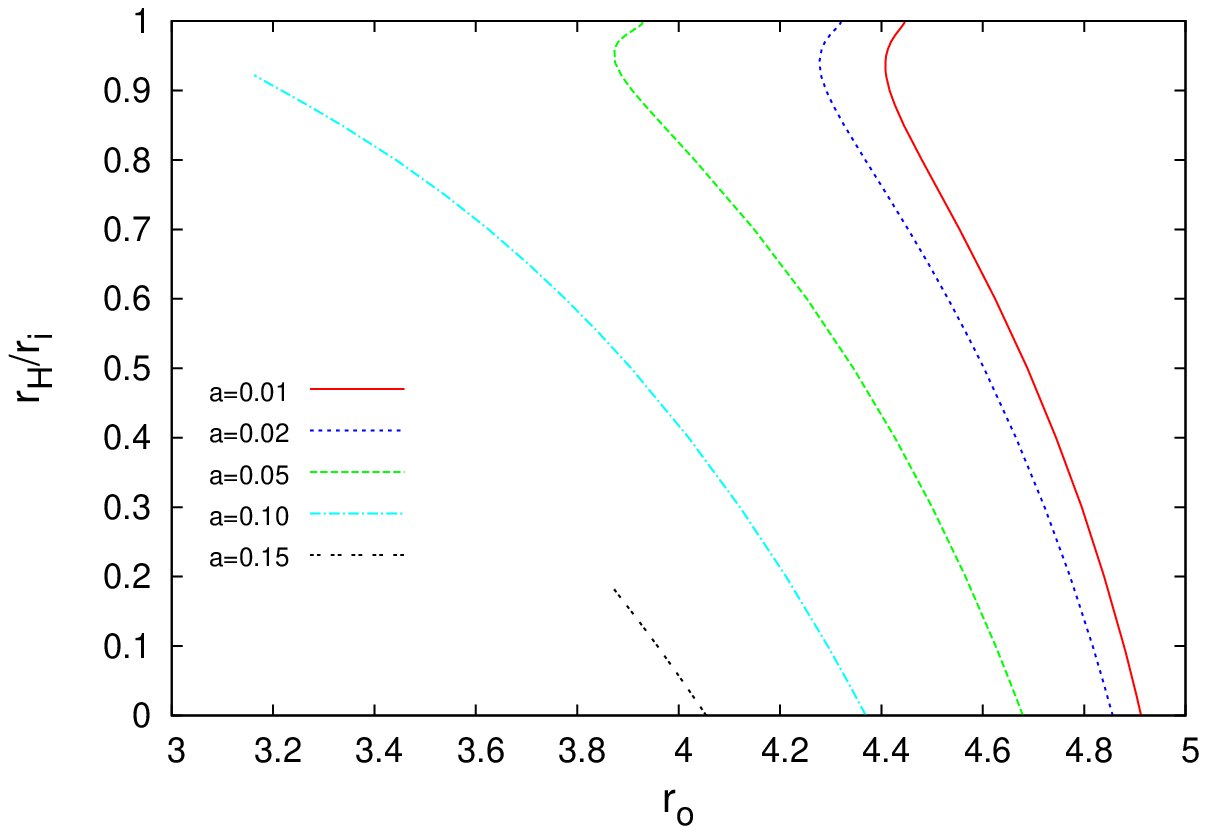}
\label{S1d}
}
}

\vspace{0.5cm}
\mbox{\hspace{-1.5cm}
\subfigure[][]{
\includegraphics[height=.27\textheight, angle =0]{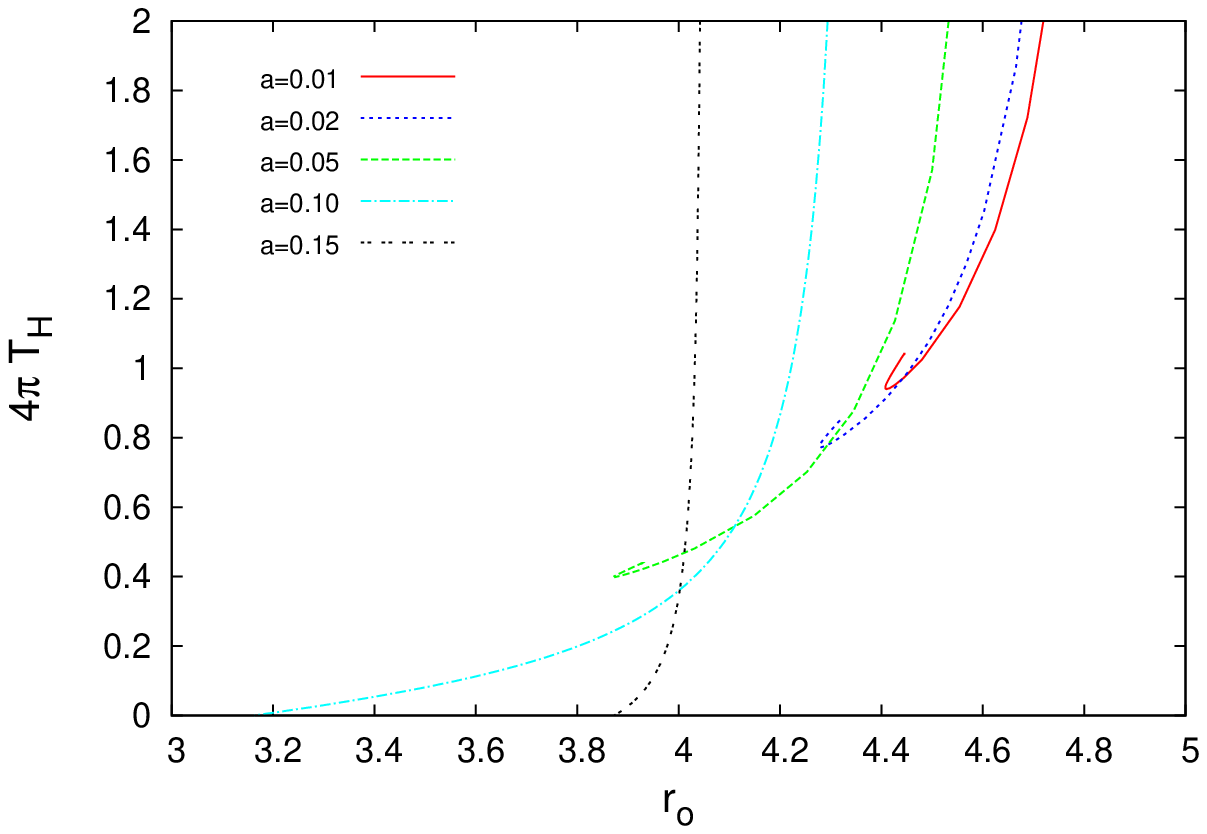}
\label{S1e}
}
\subfigure[][]{\hspace{-0.5cm}
\includegraphics[height=.27\textheight, angle =0]{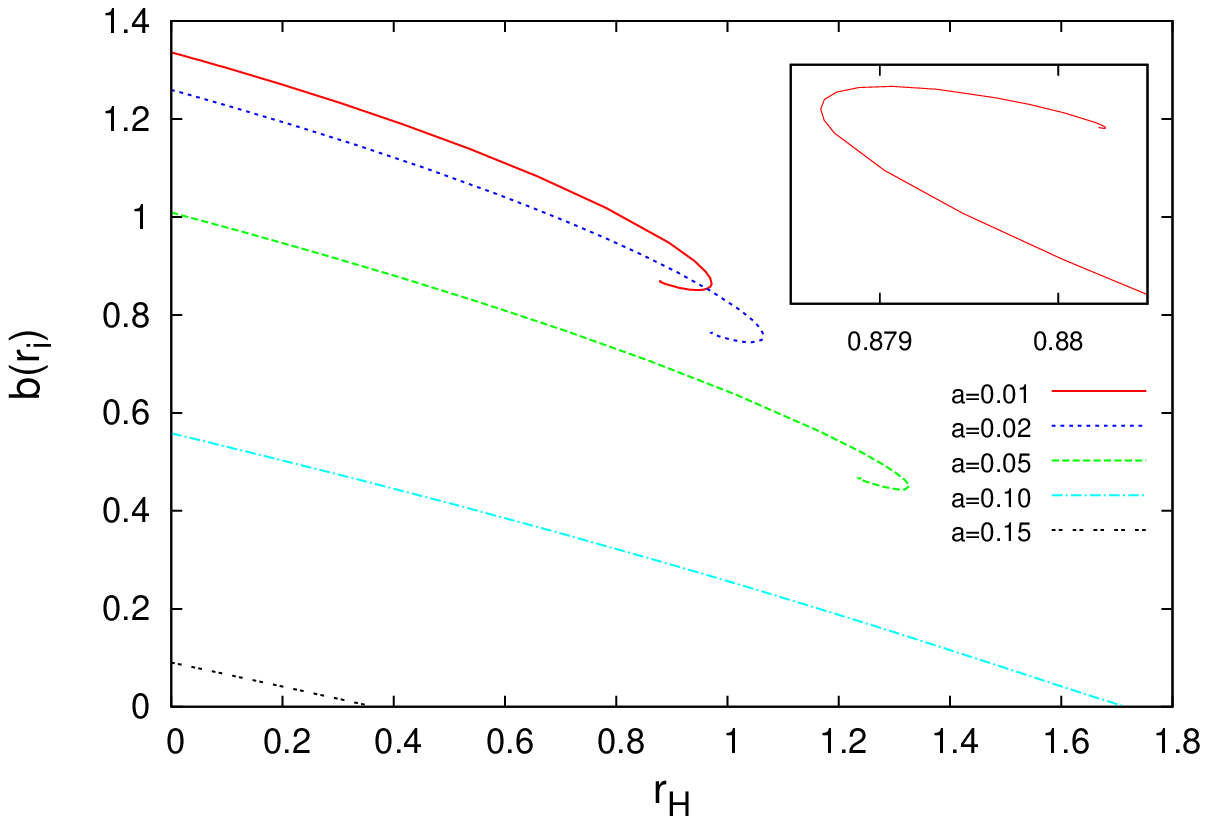}
\label{S1f}
}
}

\end{center}
\caption{Properties of boson shells with charge $Q=10$
and Schwarzschild like black holes in their interior:
(a) the event horizon radius $r_{\rm H}$ 
versus the outer shell radius $r_{\rm o}$;
(b) the mass $M$ vs.~$r_{\rm o}$,
the thin dotted curve corresponds to the
extremal limit where a throat is formed;
(c) the ratio $r_{\rm i}/r_{\rm o}$ vs.~$r_{\rm o}$;
(d) the ratio $r_{\rm H}/r_{\rm i}$ vs.~$r_{\rm o}$;
(e) the temperature $T_{\rm H}$ vs.~$r_{\rm o}$;
(f) the value $b(r_{\rm i})$ of the gauge field function $b(r)$ at the 
inner shell radius $r_{\rm i}$ vs.~$r_{\rm H}$.
The gravitational coupling $a=\alpha^2$ assumes the values
0.01, 0.02, 0.05, 0.10 and 0.15.
\label{bhS1}
}
\end{figure}

Let us now address the case, when a Schwarzschild like black hole
is immersed in the interior region $r<r_{\rm i}$ of the boson shell,
first discussed in ref.~\cite{Kleihaus:2009kr}.
Thus the Minkowski like inner region $r < r_{\rm i}$ of the space-time
of gravitating boson shell solutions
is replaced by the inner part of a curved Schwarzschild like space-time,
where the event horizon resides at $r_{\rm H} < r_{\rm i}$.
The metric in the interior region $r < r_{\rm i}$
is then determined by the Schwarzschild metric function
$N(r)= 1 - (r_{\rm H}/r)$
and a constant metric function $A(r)=A(r_{\rm i})$.
A set of such solutions with increasing event horizon $r_{\rm H}$
is exhibited in Fig.~\ref{bhS0}.

The presence of the boson shell outside the event horizon
of the black hole influences the metric in the interior region
of the shell.
{
Since $A(r)$ is monotonically increasing for $r_{\rm i}<r<r_{\rm o}$
and constant for $r\leq r_{\rm i}$ and $r \geq r_{\rm o}$, 
this implies $A(r_{\rm i})< A(r_{\rm o})$.
Consequently, for an asymptotically flat space-time, 
i.e., $A(\infty)=A(r_{\rm o})=1$,
the metric function $A(r)$ assumes some value $A(r)=A(r_{\rm i}) \leq 1$ 
in the interior region $r<r_{\rm i}$ of the space-time,
instead of the (usual) Schwarzschild value $A(r)=1$.
}

The boson shells with Schwarzschild like black holes 
in their interior are obtained
from the empty boson shells, by imposing 
the presence of an event horizon $r_{\rm H}$
and increasing its value from zero.
In the following we discuss the main features of these boson shell space-times
with electrically neutral black holes,
by considering some generic cases \cite{Kleihaus:2009kr}. 
In particular,
we consider solutions with small charge and with large charge.
Clearly, solutions with large charge are possible only for small 
values of the gravitational coupling,
whereas solutions with small charge are also possible for larger values of the
gravitational coupling. 
{
Indeed, for a given value of the charge the
gravitational coupling cannot exceed a critical value, which
decreases with increasing charge.
}

\subsubsection{Small $Q$}

We begin with Schwarzschild like black holes within boson shells
which carry only a small charge, and choose $Q=10$ for definiteness.
Varying the gravitational coupling between $a=0.01$ and 0.15,
we illustrate in Fig.~\ref{bhS1} the main properties of these solutions.
Their domain of existence is seen in Fig.~\ref{S1a},
where the radius of the event horizon $r_{\rm H}$ is exhibited 
versus the outer shell radius $r_{\rm o}$.
When the event horizon is increased from the limiting value of zero
(of the empty shells)
the outer shell radius $r_{\rm o}$ decreases.
This decrease may be interpreted as due to the gravitational
attraction from the black hole in the interior of the shell.

Obviously, the event horizon is limited in size, since the shell is
limited in size.
However, the physical reasons that limit the growth
of the event horizon, and thus the growth of the inner black hole,
differ for small and large values of the gravitational coupling.
For small values of the gravitational coupling constant, 
the branches of boson shells with Schwarzschild like black hole solutions 
in their interior exhibit
a spiral like behaviour, when approaching their endpoints.
Here the event horizon $r_{\rm H}$ and the outer shell radius $r_{\rm o}$
exhibit an oscillatory behaviour, as they converge towards limiting values.
(The first few branches of the spirals are apparent in Fig.~\ref{S1a}
and enlarged in the inlet for a representative value
of the gravitational coupling constant, $\alpha^2=0.01$,
while the higher branches are too small to be resolved there.)
{
In contrast, for larger values of the gravitational coupling constant
the branches of boson shells with Schwarzschild like black hole solutions 
in their interior exhibit a monotonic behaviour.
Here the radii $r_{\rm H}$ and $r_{\rm o}$ approach
their maximal values at the endpoints of the branches monotonically.
}

The mass $M$ of the solutions is exhibited in Fig.~\ref{S1b}
and follows this pattern. For small values of the gravitational coupling
the mass exhibits a spiral like behaviour.
Thus for given values of the charge and the mass,
there are several different solutions. 
These solutions form a countable set, 
where the number of solutions increases 
as the values of the mass and the charge
are chosen closer to the values of the mass and the charge
of the limiting solution of the spiral.
Thus for those limiting values at the endpoint of the
spiral a maximal (possibly infinite) number of discrete
solutions should be present, all with the same mass and the same charge.
Consequently these black hole space-times surrounded by boson shells
violate uniqueness \cite{Kleihaus:2009kr}.
These black holes carry scalar hair in the form of 
compact boson shells.

Let us now address the endpoints of the branches
of boson shells with Schwarzschild like black holes more closely, in order 
to understand the physical reasons causing the branches to end.
{
For that purpose we consider the behaviour of the
inner shell radius $r_{\rm i}$ and the horizon radius $r_{\rm H}$
as we move into the spiral. We note that while both oscillate,
their ratio $r_{\rm H}/r_{\rm i}$ increases monotonically and 
converges to the limiting value of one, 
as we approach the endpoint of the spiral.
This is seen in Fig.~\ref{S1c} and Fig.~\ref{S1d}.
Since the horizon cannot become bigger than the inner shell radius,
these branches of black holes terminate when the event horizon
coincides with the inner boundary of the shell.
}
A further discussion of the spirals and their limiting solution
is given in the Appendix.


As we turn to the larger values of the gravitational constant
{we note that} the branches end before the ratio $r_{\rm H}/r_{\rm i}$ tends to one.
{
Here the physical mechanism causing the branches to terminate
consists in the formation of a throat 
at the outer shell radius $r_{\rm o}$.
}
To understand the reason let us compare the mass $M$ and the
charge $Q$ of the solutions along these branches.
In terms of Reissner-Nordstr\"om solutions,
which describe the exterior solution $r \ge r_{\rm o}$,
these values of the mass and the charge
would correspond to Reissner-Nordstr\"om solutions
with naked singularities at their center.
Along the branches we keep $Q$ fixed and increase the horizon radius $r_{\rm H}$ and
with it the mass $M$. Thus along the branches the mass is getting 
steadily closer to the extremal value, 
where a degenerate horizon would be present
in the Reissner-Nordstr\"om solution. 
When the conditions for extremal Reissner-Nordstr\"om solutions,
$r_{\rm o}= \alpha^2 M = \alpha Q$, finally become satisfied,
a throat is formed at the outer shell radius $r_{\rm o}$.
{
Note that this analysis implies that uniqueness is not violated for 
these solutions.}

As the throat forms,
the temperature $T_{\rm H}$ at the event horizon $r_{\rm H} < r_{\rm i}$
of the Schwarzschild like black hole in the interior of the shell tends to zero.
This is seen in Fig.~\ref{S1e}.
At first this vanishing of the temperature appears unexpected,
since the Schwarzschild like black hole is not charged.
However, the reason for the vanishing of the temperature $T_{\rm H}$
can be understood from the behaviour of the metric function $A(r)$
in $g_{tt}$. 
{
The temperature $T_{\rm H}$ of the black hole within the boson shell 
is given by
\begin{equation}
T_{\rm H}= {A(r_{\rm H})}{T_{\rm S}}(r_{\rm H}) \ .
\end{equation}
where $T_{\rm S}(r_{\rm H})$ denotes the temperature
of a Schwarzschild black hole, 
$T_{\rm S}(r_{\rm H})= (4 \pi r_{\rm H})^{-1}$.
}
Indeed, we observe, that the function
$A(r)$ tends to zero in the interior,
when the throat is formed
as seen in Fig.~\ref{S0d}, 
{thus leading to a vanishing temperature $T_{\rm H}$.}
%
On the other hand, the divergence of the temperature $T_{\rm H}$ with vanishing 
event horizon $r_{\rm H} \to 0$
just reflects the usual Schwarzschild behaviour.

\subsubsection{Larger $Q$}

\begin{figure}[p]
\begin{center}
\vspace{-1.5cm}
\mbox{\hspace{-1.5cm}
\subfigure[][]{
\includegraphics[height=.27\textheight, angle =0]{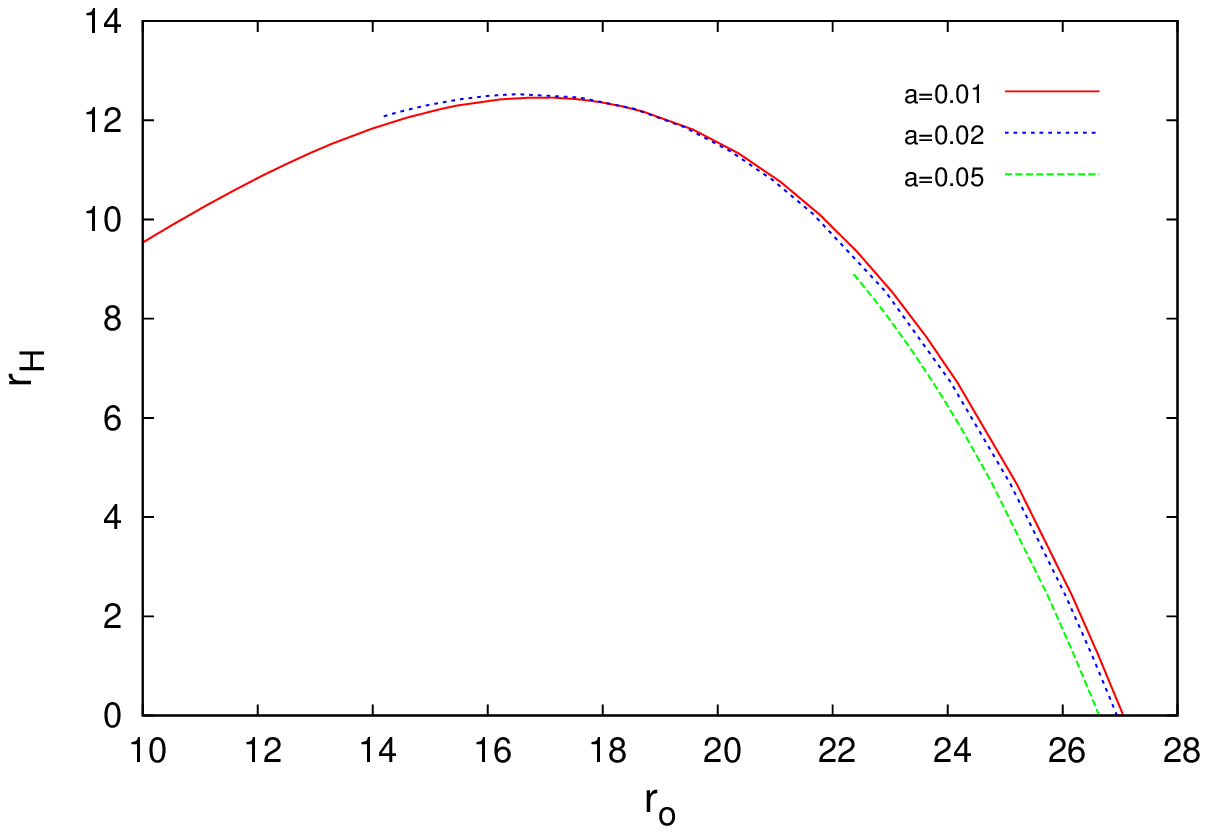}
\label{S2a}
}
\subfigure[][]{\hspace{-0.5cm}
\includegraphics[height=.27\textheight, angle =0]{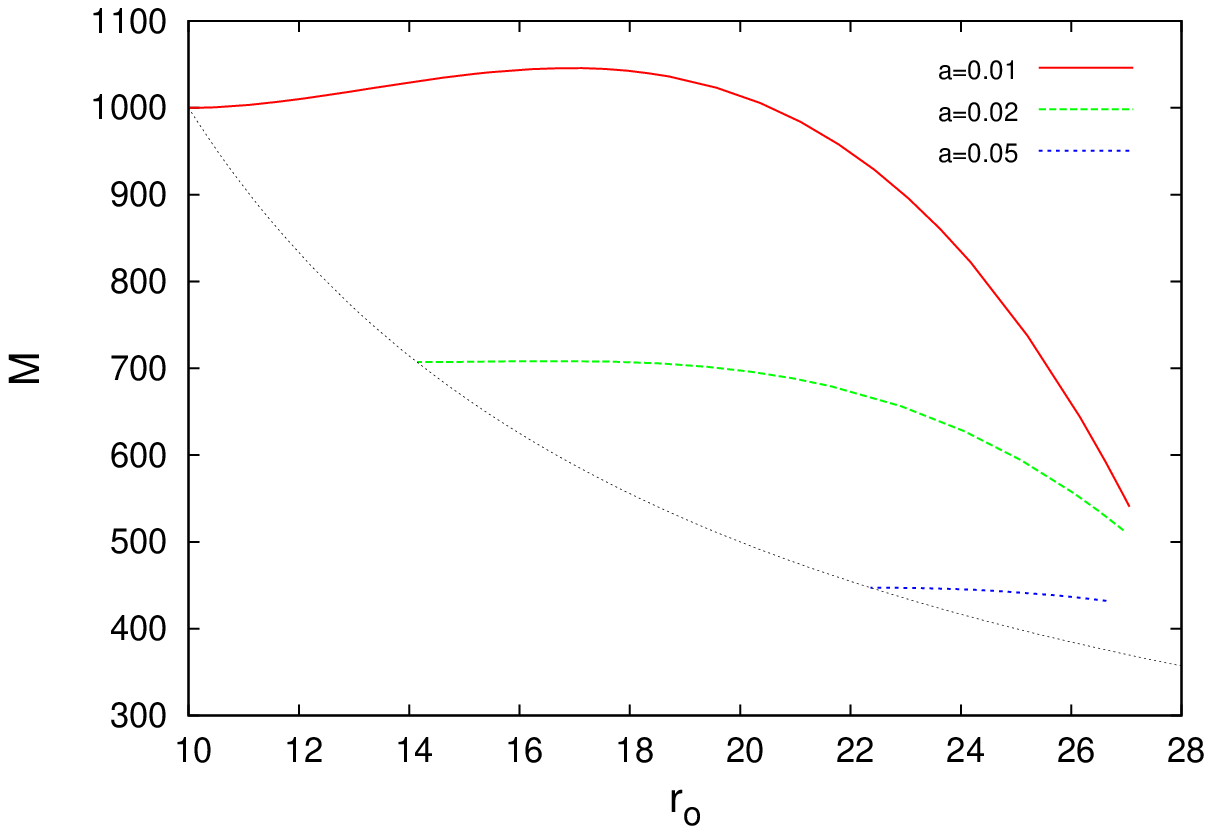}
\label{S2b}
}
}
\vspace{-0.5cm}
\mbox{\hspace{-1.5cm}
\subfigure[][]{
\includegraphics[height=.27\textheight, angle =0]{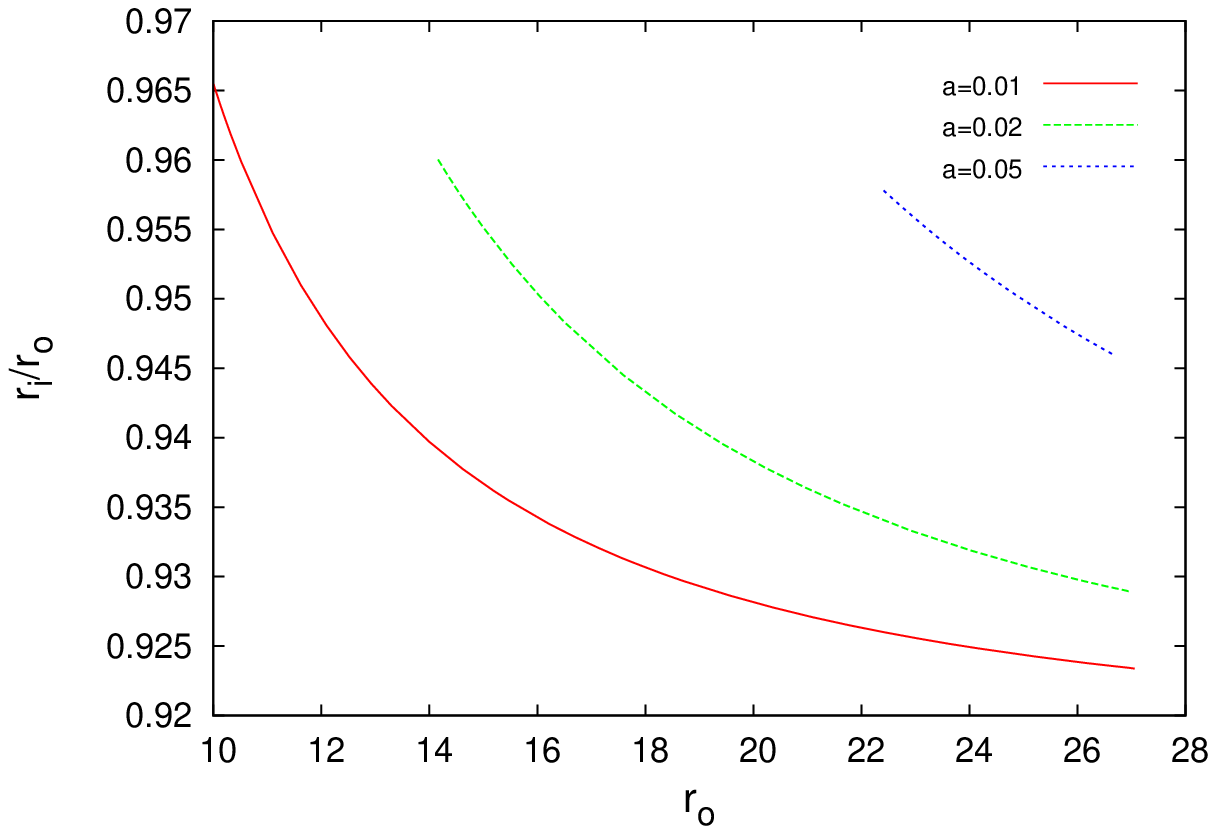}
\label{S2c}
}
\subfigure[][]{\hspace{-0.5cm}
\includegraphics[height=.27\textheight, angle =0]{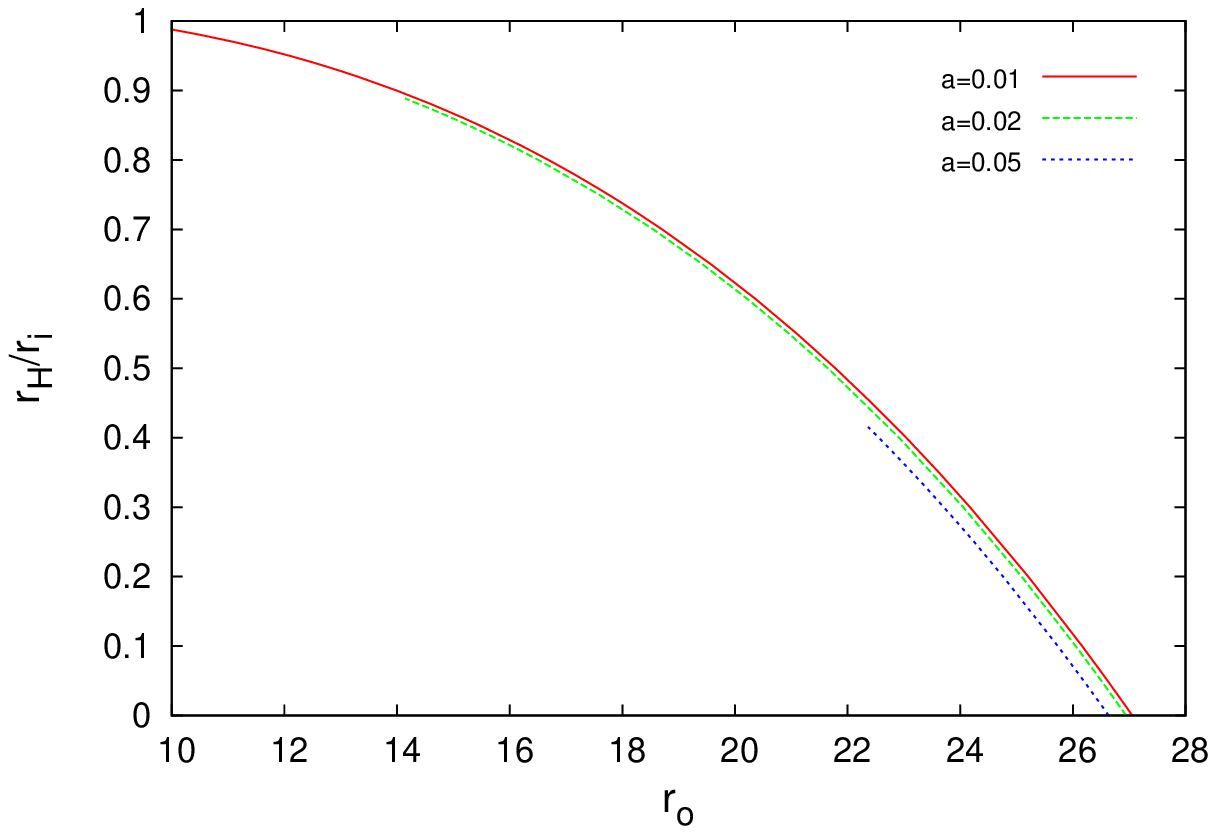}
\label{S2d}
}
}

\vspace{0.5cm}
\mbox{\hspace{-1.5cm}
\subfigure[][]{
\includegraphics[height=.27\textheight, angle =0]{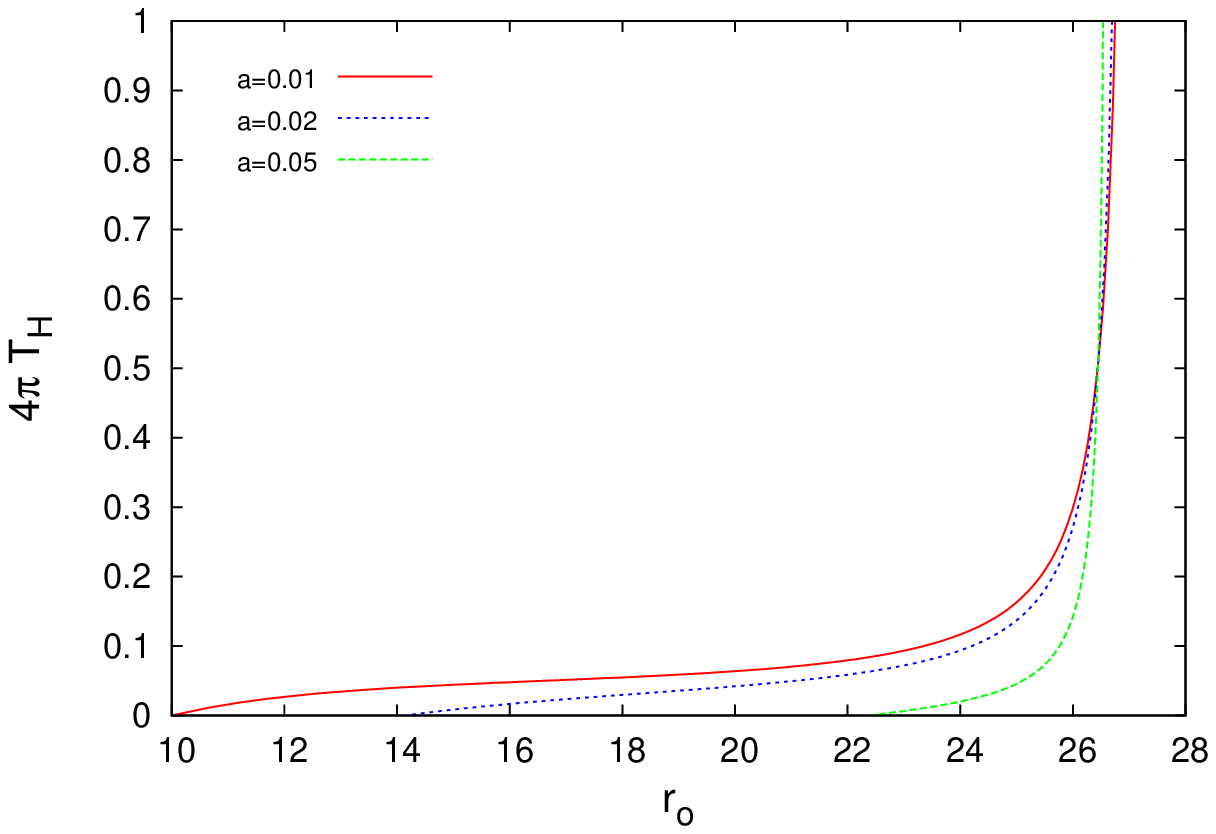}
\label{S2e}
}
\subfigure[][]{\hspace{-0.5cm}
\includegraphics[height=.27\textheight, angle =0]{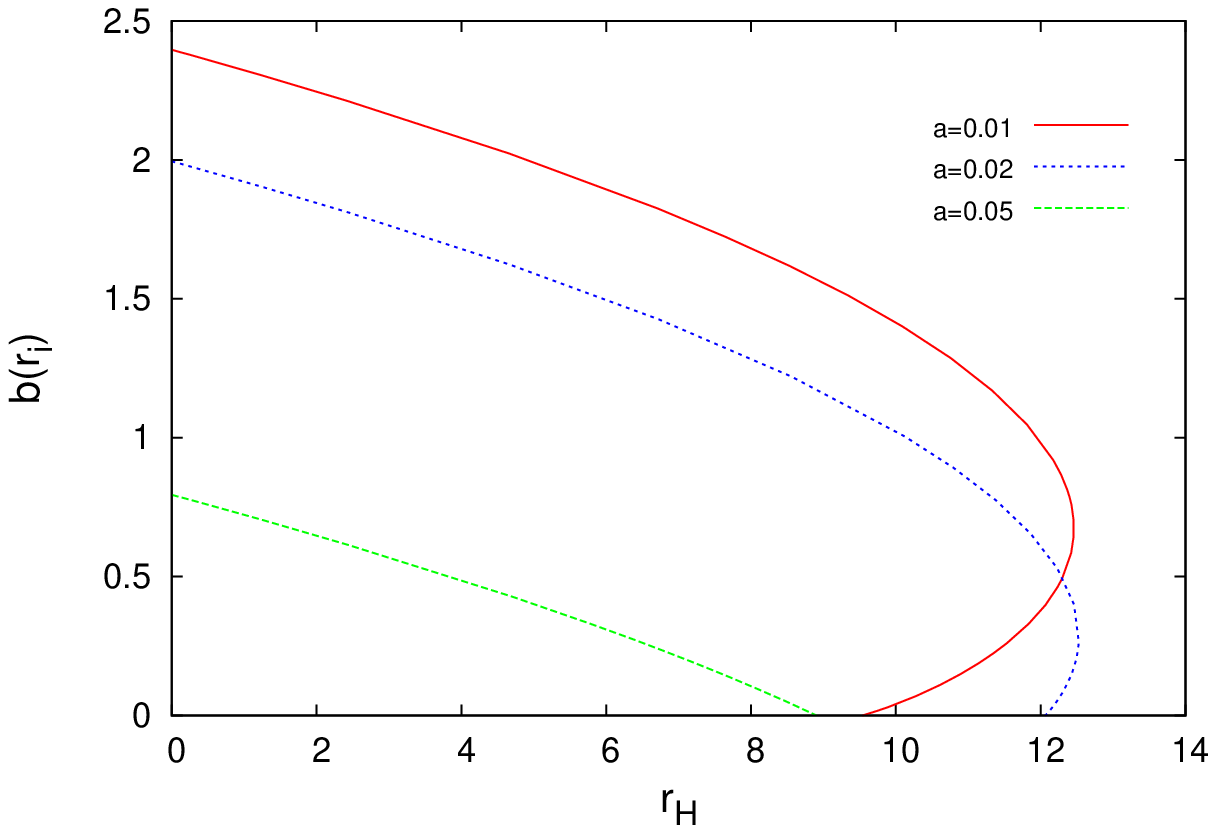}
\label{S2f}
}
}

\end{center}
\caption{Properties of boson shells with charge $Q=100$
and Schwarzschild like black holes in their interior:
(a) the event horizon radius $r_{\rm H}$ 
versus the outer shell radius $r_{\rm o}$;
(b) the mass $M$ vs.~$r_{\rm o}$,
the thin dotted curve corresponds to the
extremal limit where a throat is formed;
(c) the ratio $r_{\rm i}/r_{\rm o}$ vs.~$r_{\rm o}$;
(d) the ratio $r_{\rm H}/r_{\rm i}$ vs.~$r_{\rm o}$;
(e) the temperature $T_{\rm H}$ vs.~$r_{\rm o}$;
(f) the value $b(r_{\rm i})$ of the gauge field function $b(r)$ at the 
inner shell radius $r_{\rm i}$ vs.~$r_{\rm H}$.
The gravitational coupling $a=\alpha^2$ assumes the values
0.01, 0.02, 0.05, 0.10 and 0.15.
\label{bhS2}
}
\end{figure}

Solutions with larger charge are only possible for small
values of the gravitational coupling.
Let us for definiteness now consider solutions with charge $Q=100$,
whose properties are demonstrated in Fig.~\ref{bhS2}.
For the largest values of the gravitational coupling
that are possible for this charge, the branches of solutions
with a Schwarzschild like black hole in the interior exhibit the same 
pattern as discussed above for the case of small charge and large
gravitational coupling.
These branches exhibit a monotonic behaviour,
where the horizon radius 
and the mass increase with decreasing outer shell radius.
The branches end when a throat is formed 
at the outer shell radius,
when $r_{\rm o}= \alpha^2 M = \alpha Q$ is satisfied.
This is demonstrated in Fig.~\ref{bhS2} for $\alpha^2=0.05$.
Clearly, with this monotonic behaviour
the solutions exhibit uniqueness in this parameter range.

For smaller values of the gravitational coupling, however,
the pattern changes, allowing for non-monotonic branches
and thus, again, for nonuniqueness of the black hole solutions.
The new pattern is seen in Fig.~\ref{bhS2} for $\alpha^2=0.02$ and 0.01.
These large charge solutions have large outer and inner shell radii.
With a ratio $r_{\rm i}/r_{\rm o}$ close to 1, they seem like rather thin shells.
(The proper distance from $r_{\rm i}$ to $r_{\rm o}$ can, of course, become
arbitrarily large, as an extremal configuration is approached.)

As the horizon radius increases along these branches, the outer shell radius
decreases while the mass increases. Interestingly, 
now the scaled mass and charge become equal,
$\alpha^2 M = \alpha Q$,
before the outer shell radius has sufficiently shrunk
to satisfy the conditions for extremal Reissner-Nordstr\"om solutions,
thus $r_{\rm o} > \alpha^2 M = \alpha Q$. 
Consequently, a throat cannot yet be formed.
As the horizon is increased further,
the mass increases further,
and the exterior solution becomes the respective outer part
of a Reissner-Nordstr\"om black hole solution,
since now $\alpha^2 M > \alpha Q$.

The horizon then reaches a maximal value, 
similar to the case when spirals arise. 
Beyond this value the horizon, the outer shell radius
and the mass decrease, while the ratio $r_{\rm H}/r_{\rm i}$
continues to increase. However, since both the outer shell radius
and the mass decrease, they can and do reach
the extremal Reissner-Nordstr\"om values
$r_{\rm o} = \alpha^2 M = \alpha Q$.
Here a throat is formed at the outer shell radius
and the branches end.

Unlike the case of full spirals, however, we here have only
two solutions for a given mass and charge,
since a single bifurcation is present.
These two black hole space-times have the same set of global charges
but are otherwise distinct solutions of the Einstein-matter equations,
and thus black hole uniqueness is again seen not to hold in this model
of scalar electrodynamics coupled to gravity.
{
Note that also pure Reissner-Nordstr\"om black holes exist 
in some parameter range (i.e.~when $M > |Q|/\alpha$),
which possess the same global charges as the boson shells with
interior black hole.}

\section{Boson shells with charged interior} 

We now replace the inner empty Minkowski space of the boson shell
by a point charge or by a charged Reissner-Nordstr\"om
like black hole and consider the effect of these inner
charges $Q_{\rm i}$ respectively $Q_{\rm H}$
on the physical properties of these solutions
and on their domain of existence.

\subsection{Boson shells with point-like charges}

\begin{figure}[h]
\begin{center}
\vspace{-0.5cm}
\mbox{\hspace{-1.5cm}
\subfigure[][]{
\includegraphics[height=.27\textheight, angle =0]{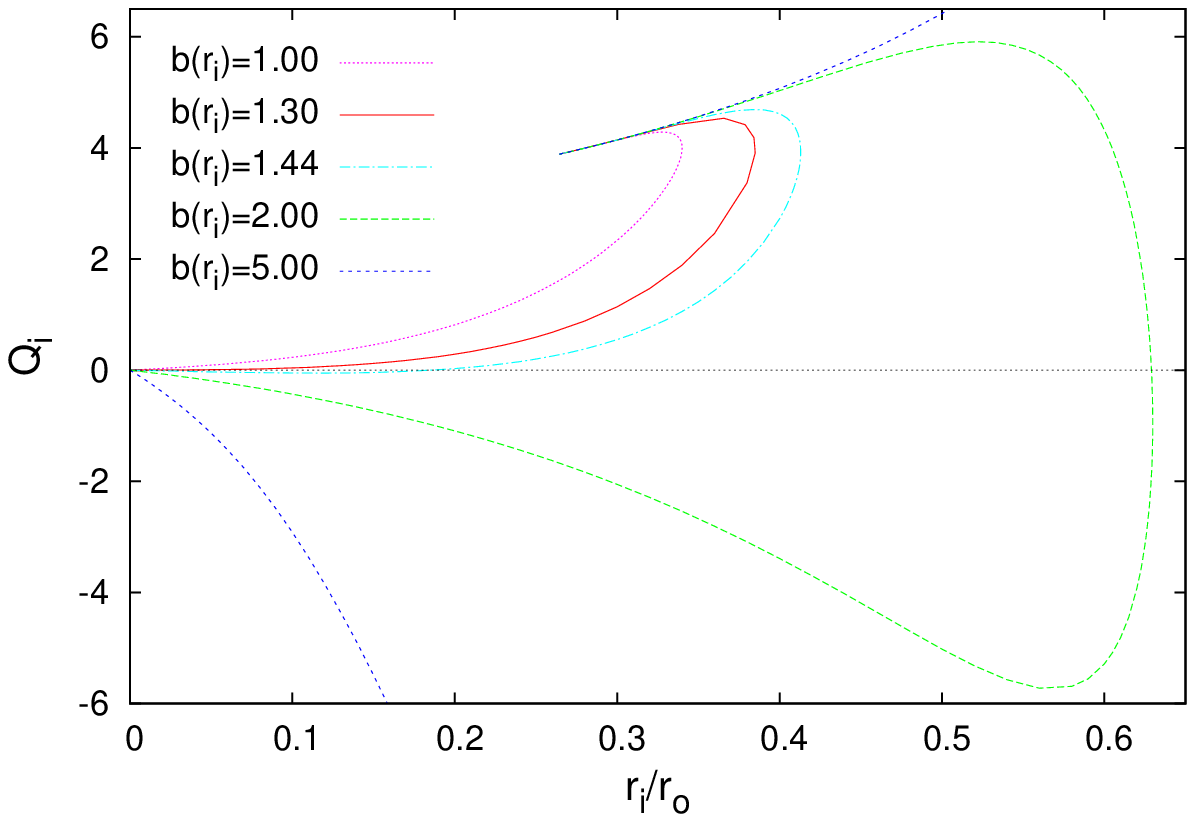}
\label{M1a}
}
\subfigure[][]{\hspace{-0.5cm}
\includegraphics[height=.27\textheight, angle =0]{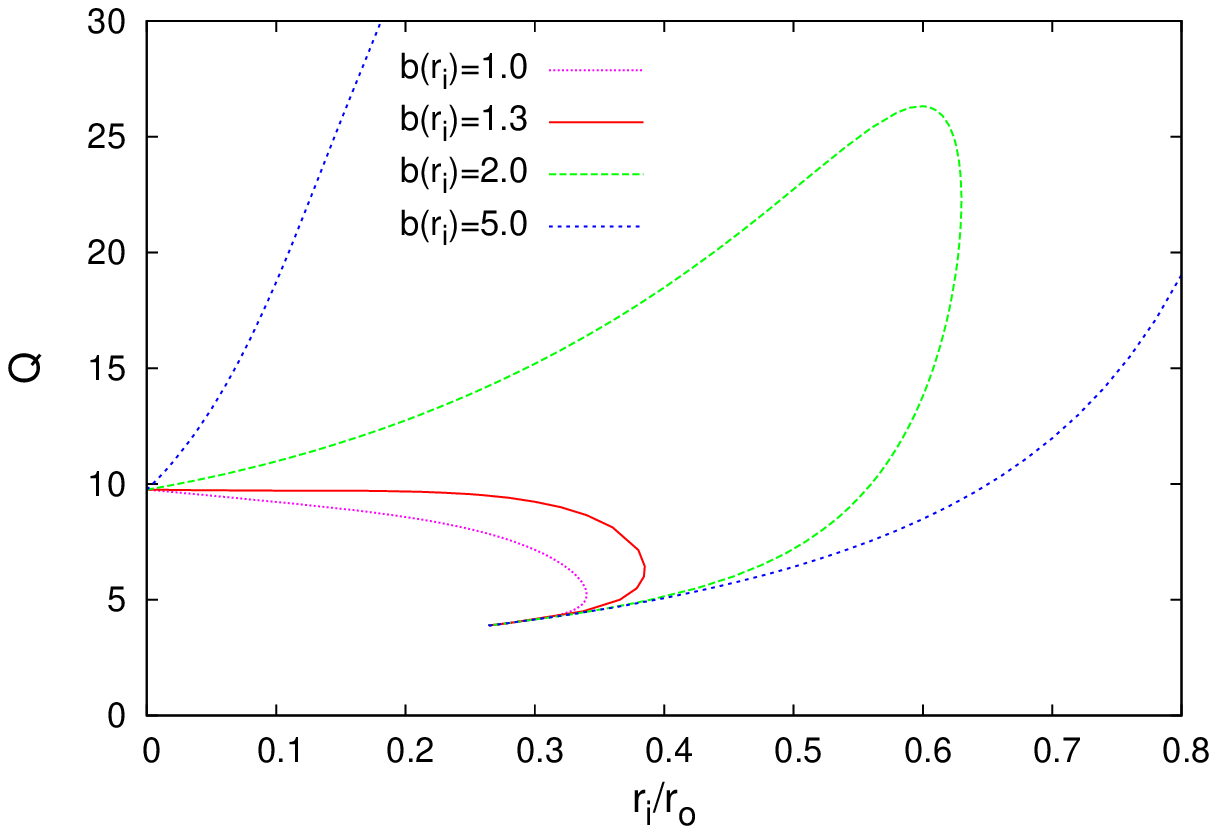}
\label{M1b}
}
}
\mbox{\hspace{-1.5cm}
\subfigure[][]{
\includegraphics[height=.27\textheight, angle =0]{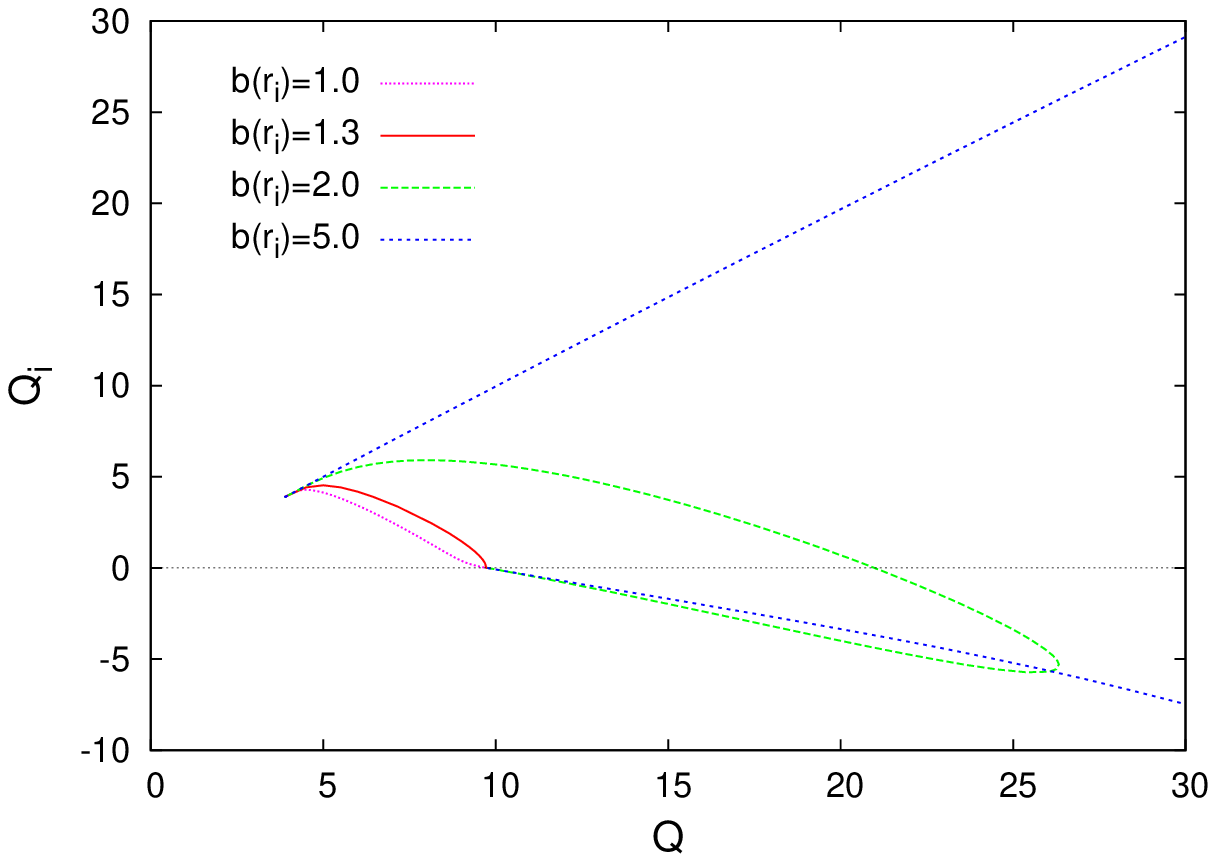}
\label{M1c}
}
\subfigure[][]{\hspace{-0.5cm}
\includegraphics[height=.27\textheight, angle =0]{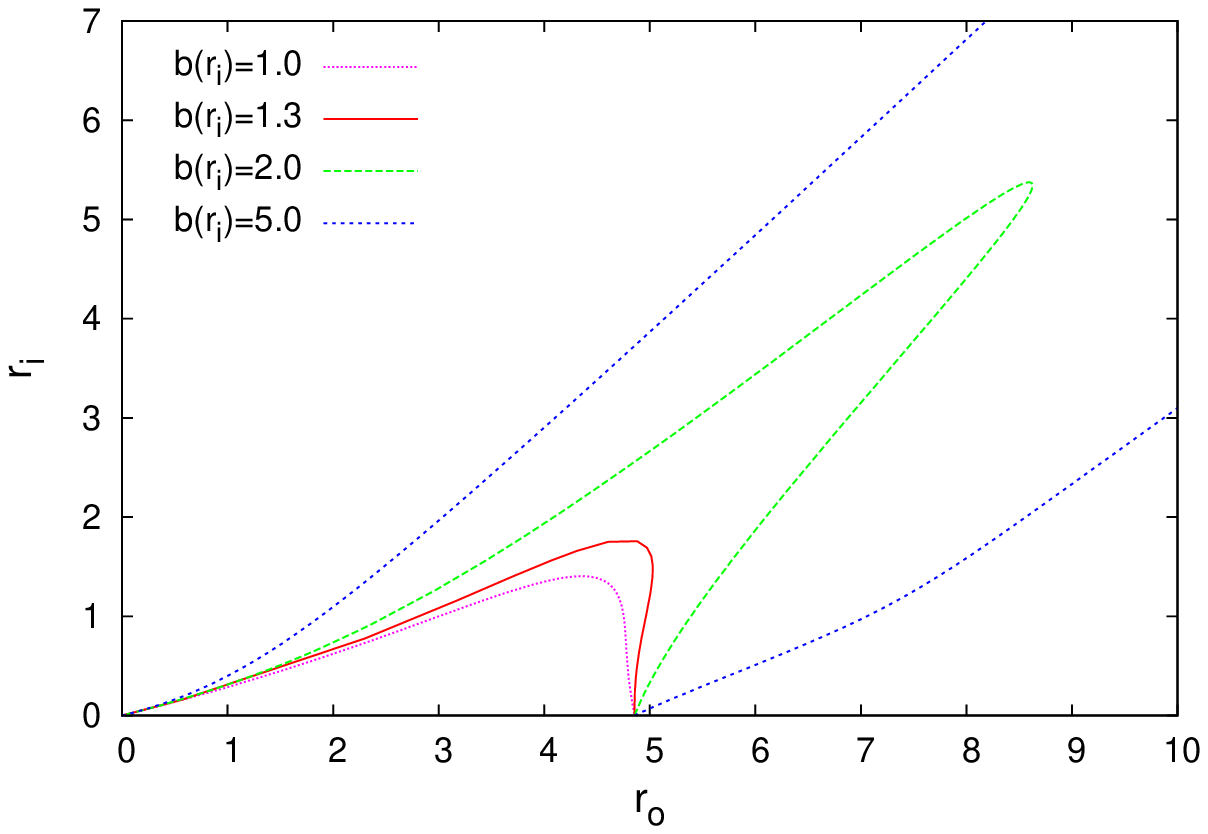}
\label{M1d}
}
}

\end{center}
\vspace*{-0.5cm}
\caption{Properties of boson shells in flat space
with a point-like charge $Q_{\rm i}$ in their interior:
(a) the charge $Q_{\rm i}$ versus the ratio of inner and outer shell radii
$r_{\rm i}/r_{\rm o}$;
(b) the charge $Q$ vs.~$r_{\rm i}/r_{\rm o}$;
(c) $Q_{\rm i}$ vs.~$Q$;
(d) $r_{\rm i}$ vs.$r_{\rm o}$.
The branches correspond to fixed values of the function $b(r)$
at the inner shell radius.
\label{sh2}
}
\end{figure}

To get a first understanding of the 
physical effects in the presence of a charge $Q_{\rm i}$
in the interior region $r< r_{\rm i}$ of the boson shells, 
we switch off gravity and consider flat space.
We thus put a point charge at the origin of the 
Minkowski space in the interior of the flat space boson shells.
As we turn on the charge $Q_{\rm i}$,
branches of solutions emerge from the boson shells with Minkowski interior.
When the charge of the shell and the charge in the interior
have the same sign, we obtain a repulsive force.
For like charges the branches can then be extended far out.
For opposite charges, in contrast,
the force is attractive.
This is expected to limit the relative magnitude of the charge,
that can be put into the interior.

The basic question that we would like to address 
is whether the negative charge in the interior region
of the shell can become sufficiently large
to cancel the positive charge carried by the boson shell itself
and thus yield a globally neutral solution,
i.e., a solution with $Q=0$.
To answer this question, let us consider branches of solutions,
where the global charge and the charge in the interior are varied.
For convenience, we choose fixed values of the gauge field
function $b(r)$ at the inner shell radius,
while we vary the charges. 
This procedure then reveals the domain of existence of this
type of solutions.

As seen in Fig.~\ref{sh2},
we cannot reach globally neutral solutions.
When the negative charge in the interior reaches 
sizeable values, the global charge does not decrease.
Instead an even larger positive charge carried by the shell is needed
to accommodate the negative charge in the interior.
Thus the global charge does not decrease but increase.
Obviously, the attraction of the unlike charges is too big
to allow for globally neutral solutions consisting of
a charged boson shell and an oppositely charged point charge.
The coupling to gravity does not change this basic fact,
as discussed below.


\subsection{Boson shells with charged black holes: $Q_{\rm H} \ne0$}

Let us now turn to gravitating boson shells,
which harbour a charged black hole in their interior, $r<r_{\rm i}$.
These branches of solutions with 
a Reissner-Nordstr\"om like black hole in their interior
emerge from the solutions with Schwarzschild like black holes,
when the black hole is endowed with a small charge $Q_{\rm H}$,
its horizon charge.
The horizon charge $Q_{\rm H}$ is then
increased or decreased from zero, while the remaining
parameters $\alpha$, $Q$ and $r_{\rm H}$ are held fixed.

Since the parameter space is rather large,
we focus in the following on three regions of the parameter space,
covering the basic features which these solutions possess.

\subsubsection{Small charge, small gravitational coupling}

\begin{figure}[p]
\begin{center}
\vspace{-1.5cm}
\mbox{\hspace{-1.5cm}
\subfigure[][]{
\includegraphics[height=.27\textheight, angle =0]{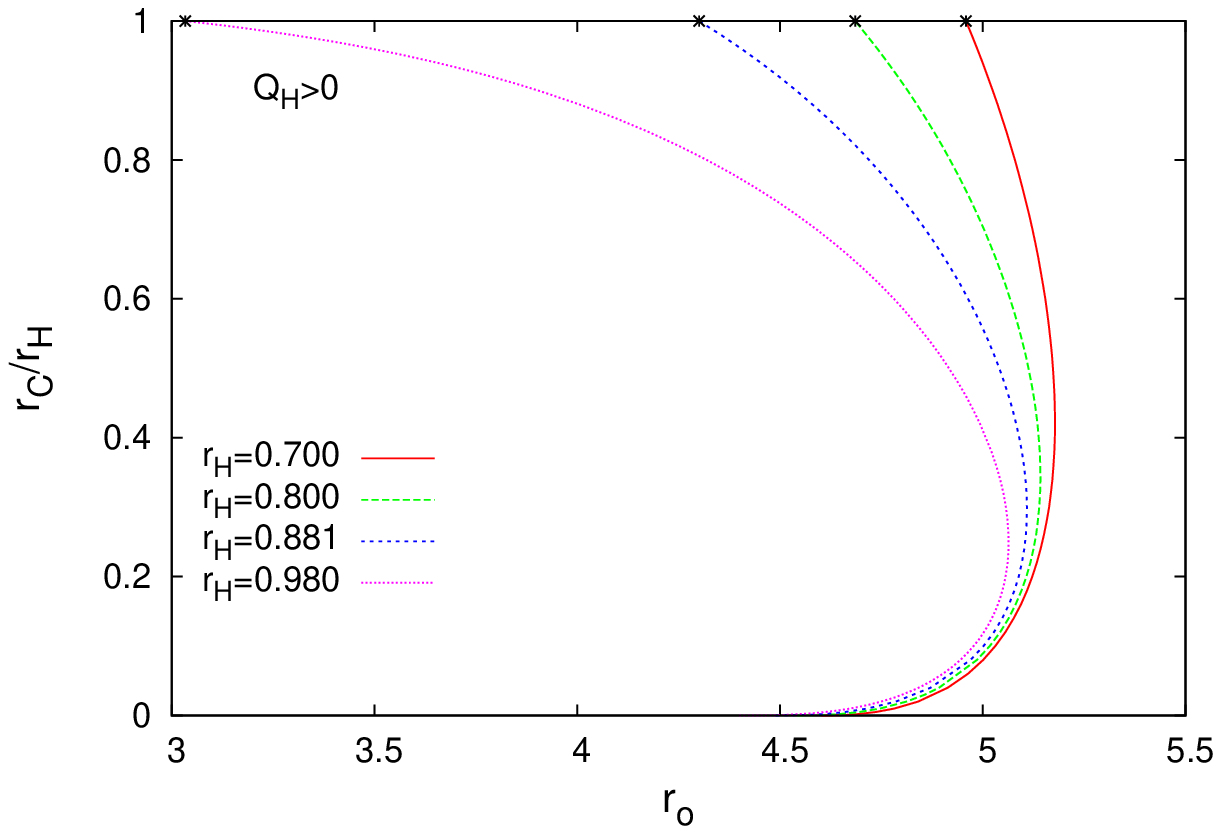}
\label{F1a}
}
\subfigure[][]{\hspace{-0.5cm}
\includegraphics[height=.27\textheight, angle =0]{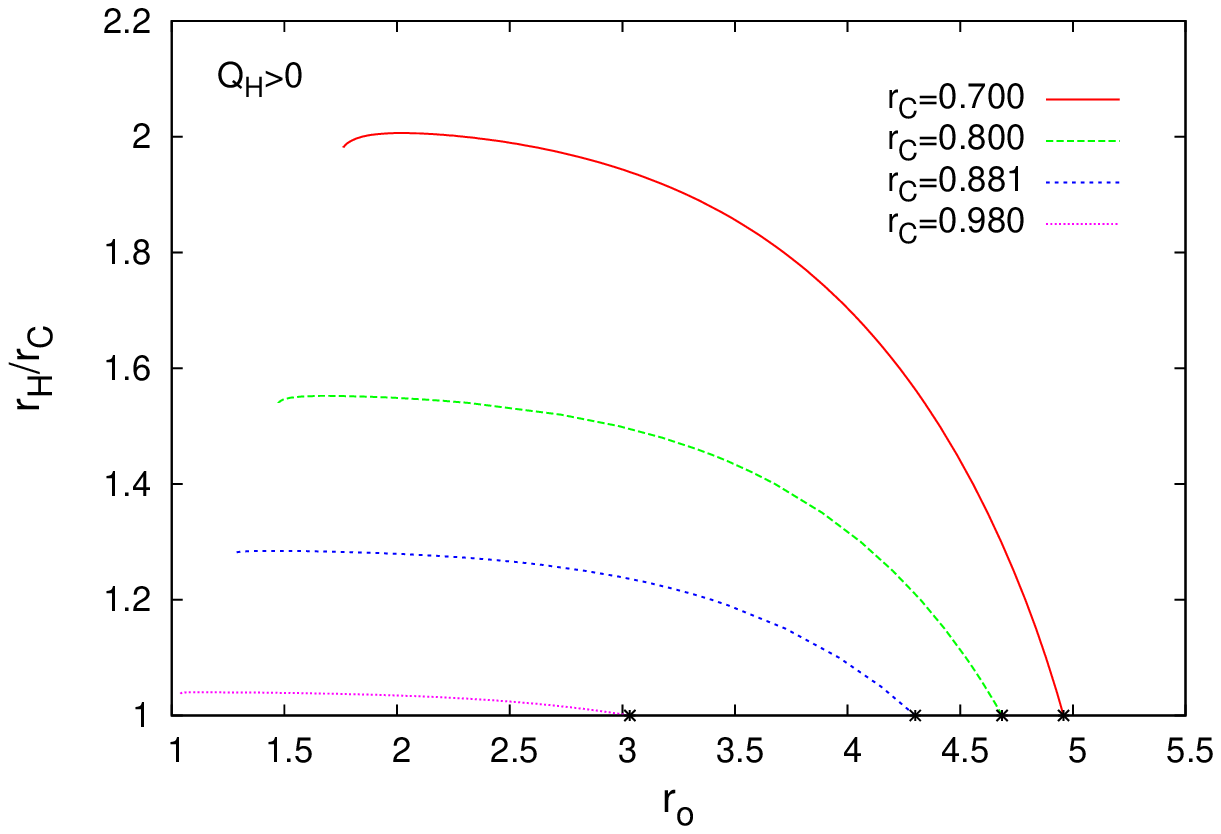}
\label{F1b}
}
}
\vspace{-0.5cm}
\mbox{\hspace{-1.5cm}
\subfigure[][]{
\includegraphics[height=.27\textheight, angle =0]{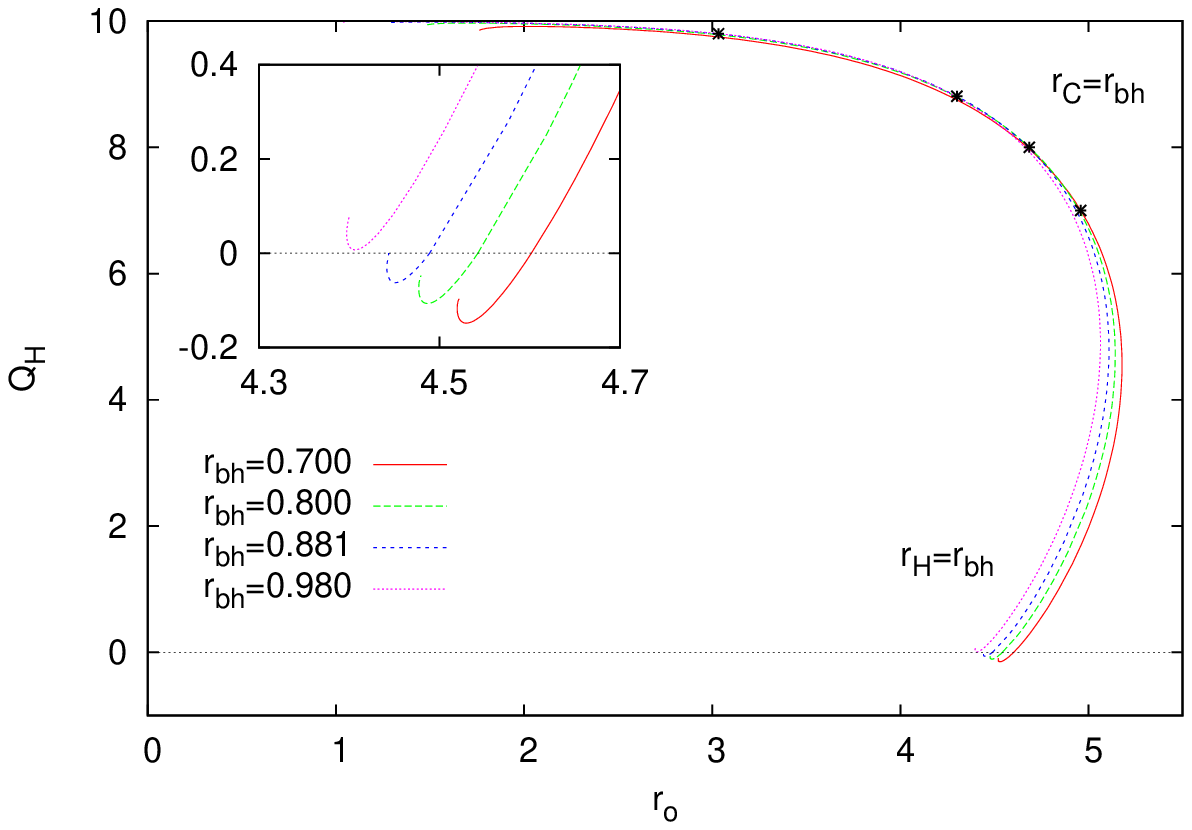}
\label{F1c}
}
\subfigure[][]{\hspace{-0.5cm}
\includegraphics[height=.27\textheight, angle =0]{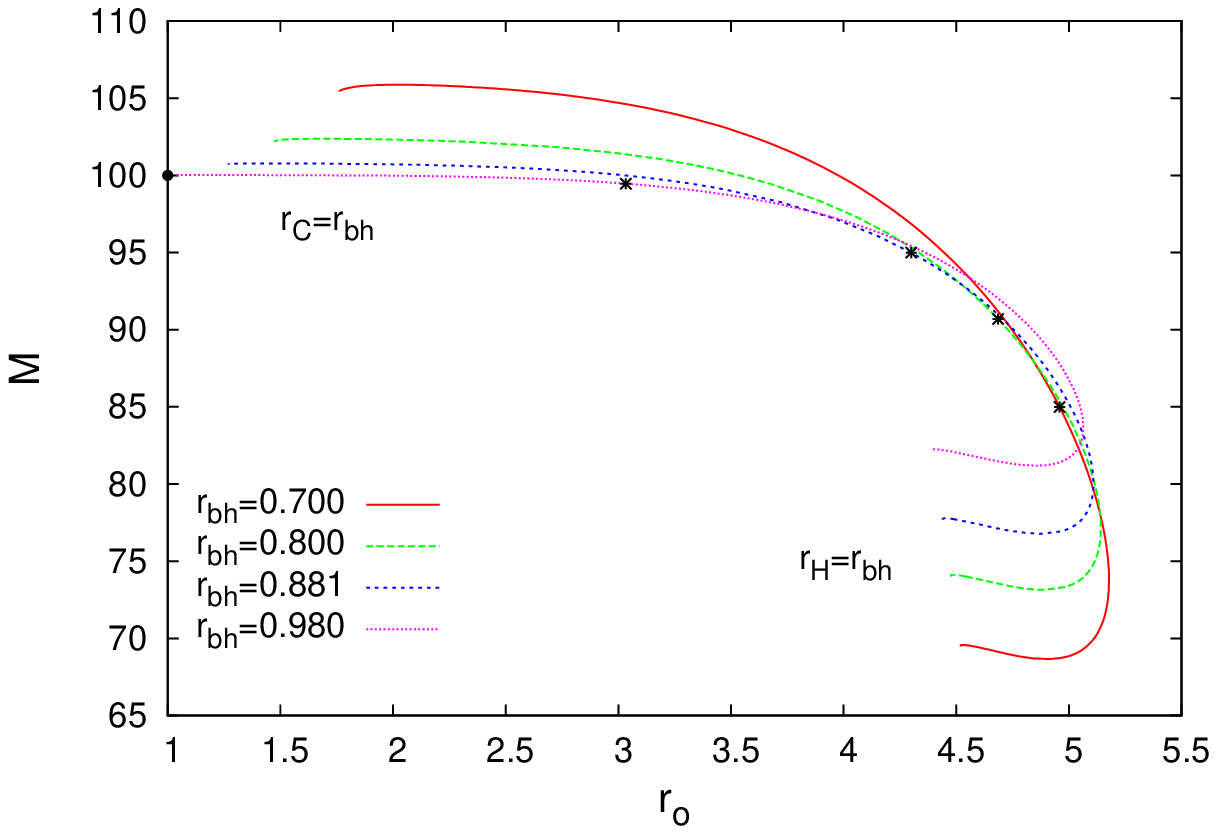}
\label{F1d}
}
}
\vspace{-0.5cm}
\mbox{\hspace{-1.5cm}
\subfigure[][]{
\includegraphics[height=.27\textheight, angle =0]{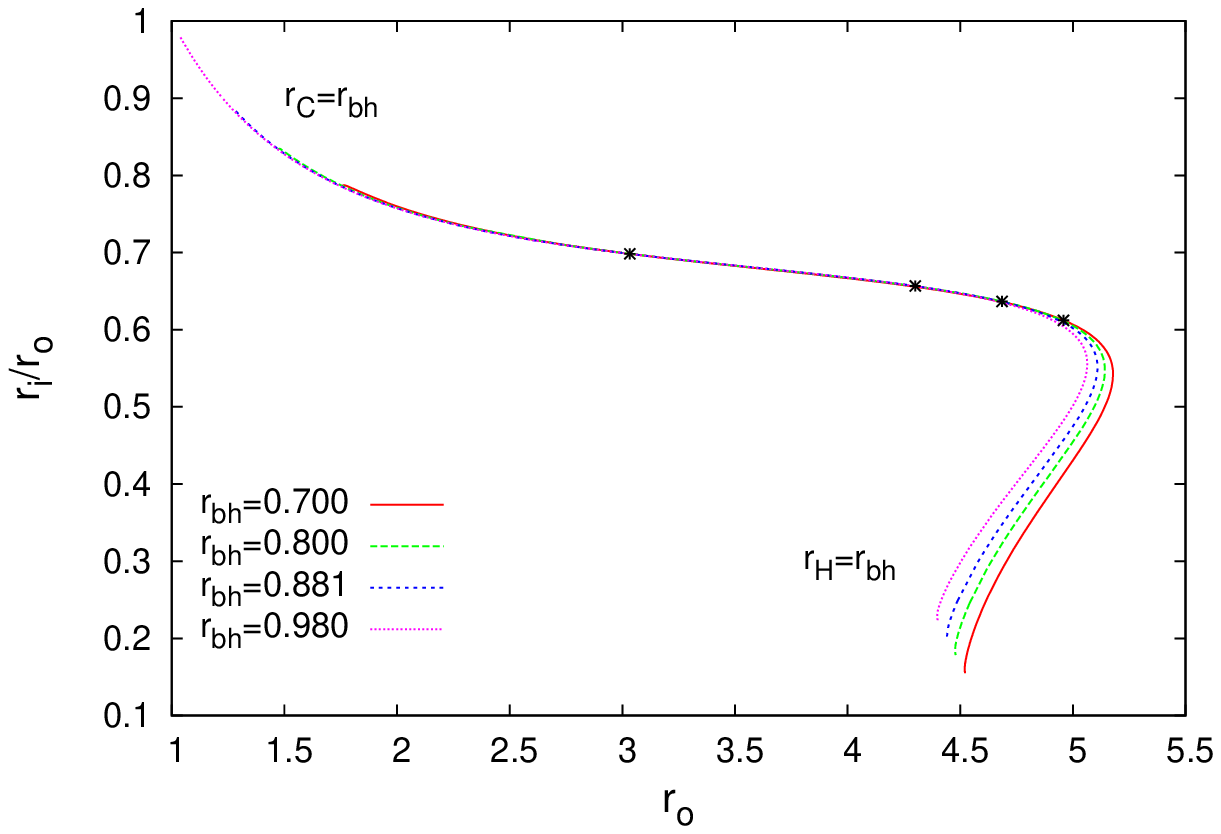}
\label{F1e}
}
\subfigure[][]{\hspace{-0.5cm}
\includegraphics[height=.27\textheight, angle =0]{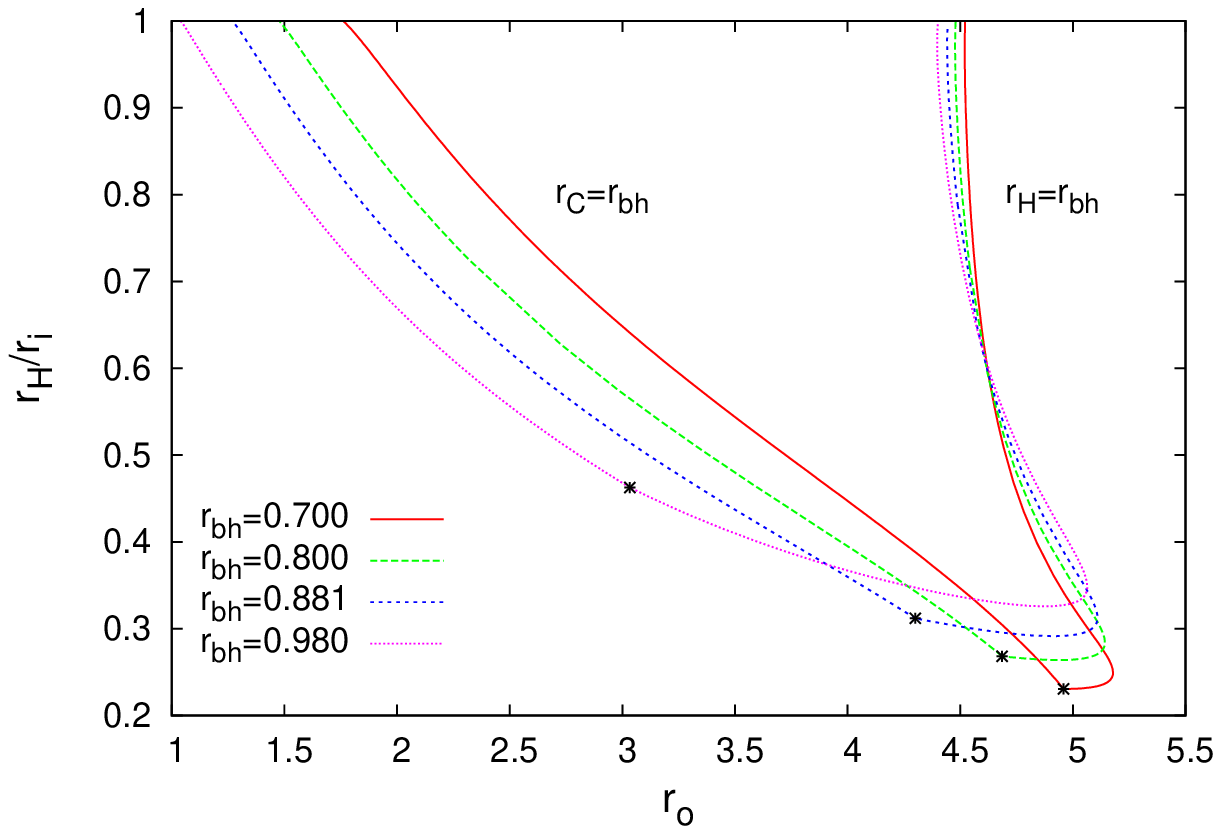}
\label{F1f}
}
}
\end{center}
\caption{Properties of boson shells with 
Reissner-Nordstr\"om like black hole solutions in their
interior shown versus the outer shell
radius $r_{\rm o}$
for charge $Q=10$ and gravitational coupling constant $a=\alpha^2=0.01$:
(a) the ratio of Cauchy horizon to event horizon $r_{\rm C}/r_{\rm H}$;
(b) the ratio of Cauchy horizon to event horizon $r_{\rm H}/r_{\rm C}$;
(c) the horizon charge $Q_{\rm H}$;
(d) the mass $M$,
the dot corresponds to the
extremal limit where a throat is formed;
(e) the ratio of inner to outer shell radius $r_{\rm i}/r_{\rm o}$;
(f) the ratio of event horizon to inner shell radius $r_{\rm H}/r_{\rm i}$.
The asterisks mark extremal black holes where $r_{\rm C}/r_{\rm H}=1$.
\label{bh1}
}
\end{figure}

\begin{figure}[t]
\begin{center}
\vspace{-1.5cm}
\mbox{\hspace{-1.5cm}
\subfigure[][]{
\includegraphics[height=.27\textheight, angle =0]{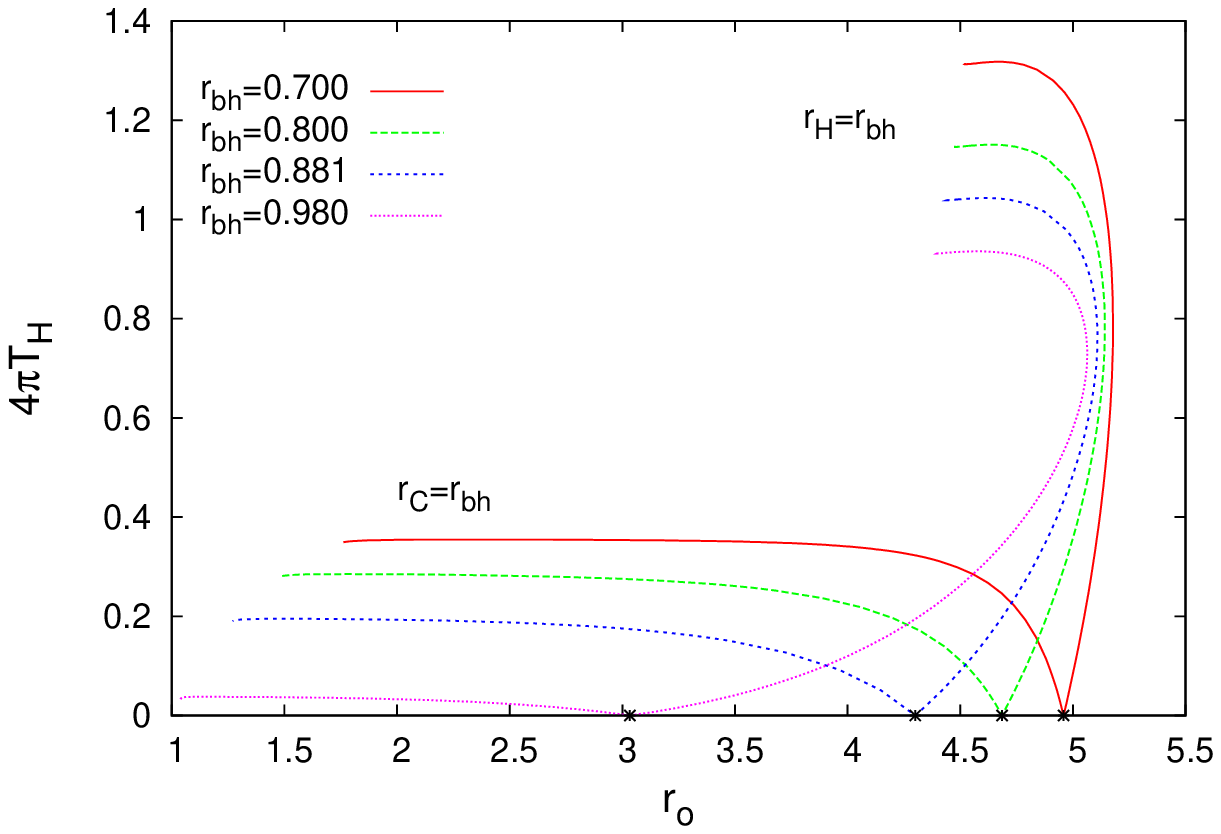}
\label{F2a}
}
\subfigure[][]{\hspace{-0.5cm}
\includegraphics[height=.27\textheight, angle =0]{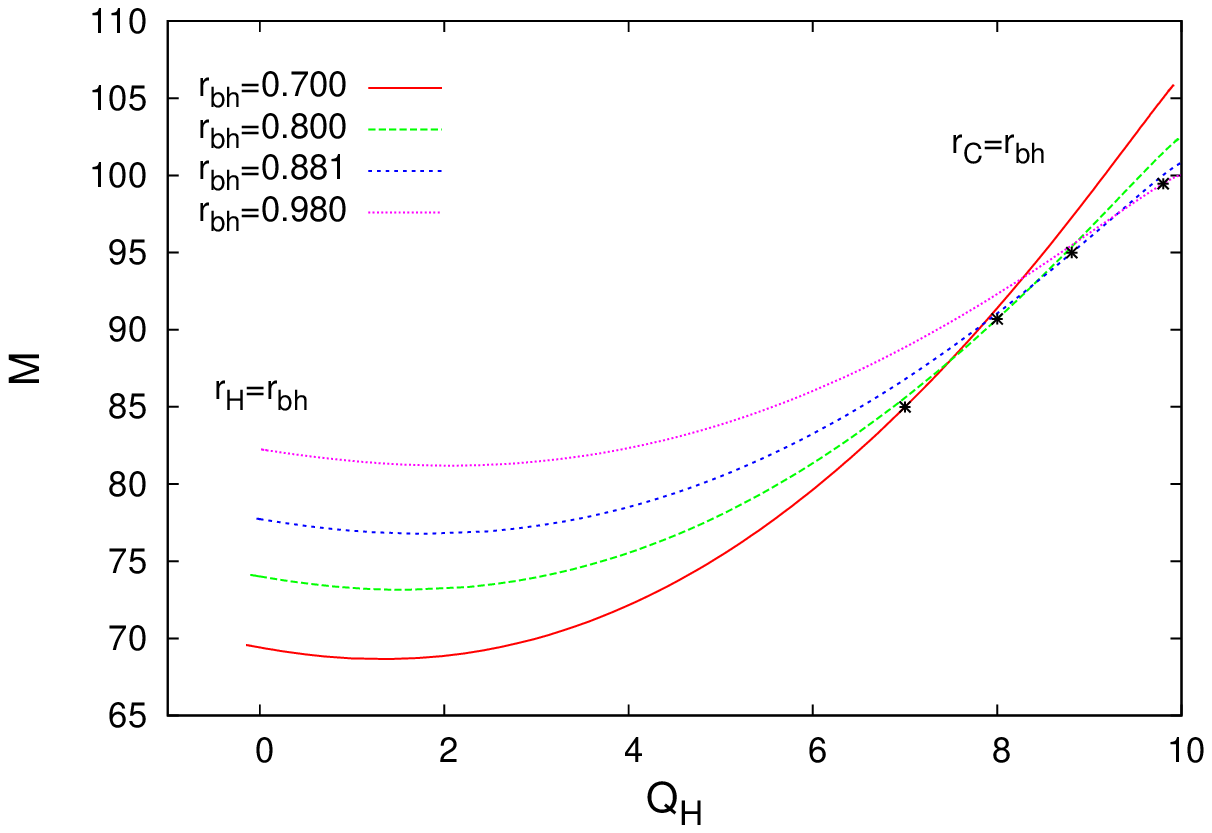}
\label{F2b}
}
}
\vspace{-0.5cm}
\mbox{\hspace{-1.5cm}
\subfigure[][]{
\includegraphics[height=.27\textheight, angle =0]{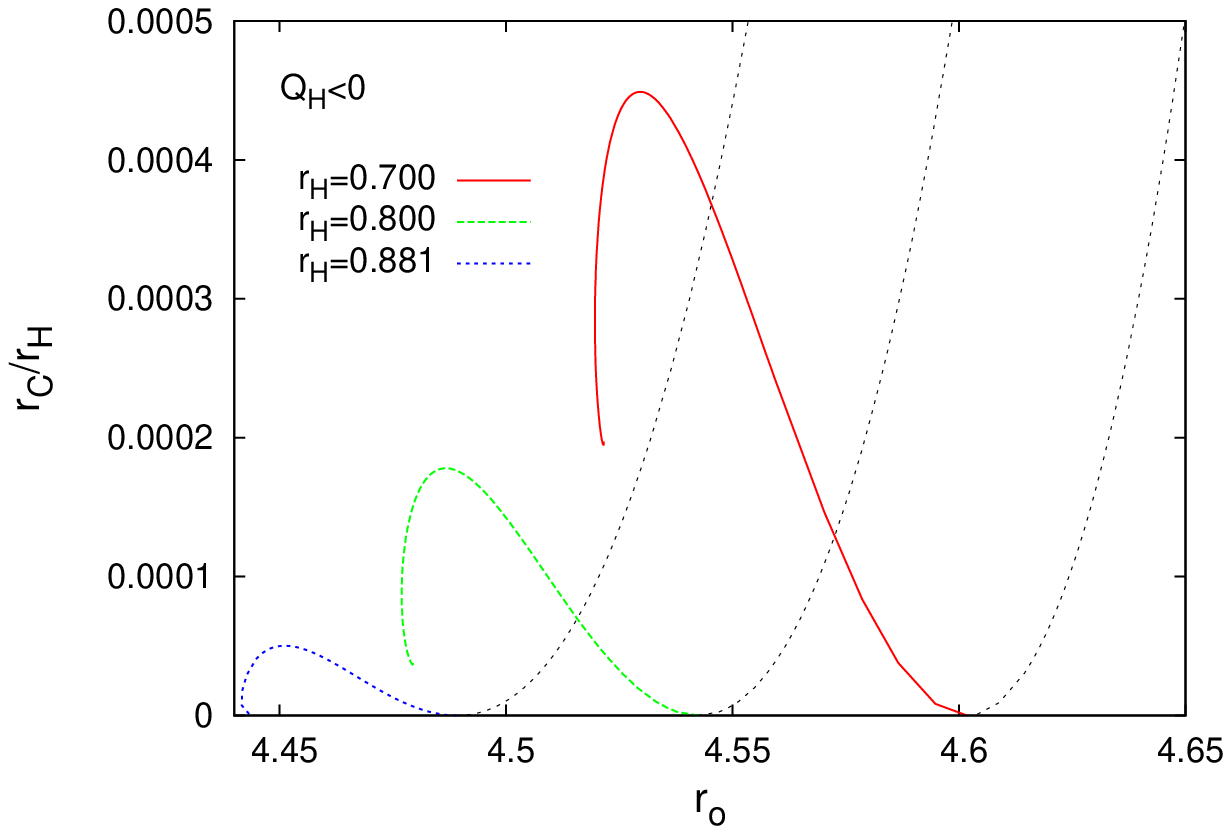}
\label{F2c}
}
\subfigure[][]{\hspace{-0.5cm}
\includegraphics[height=.27\textheight, angle =0]{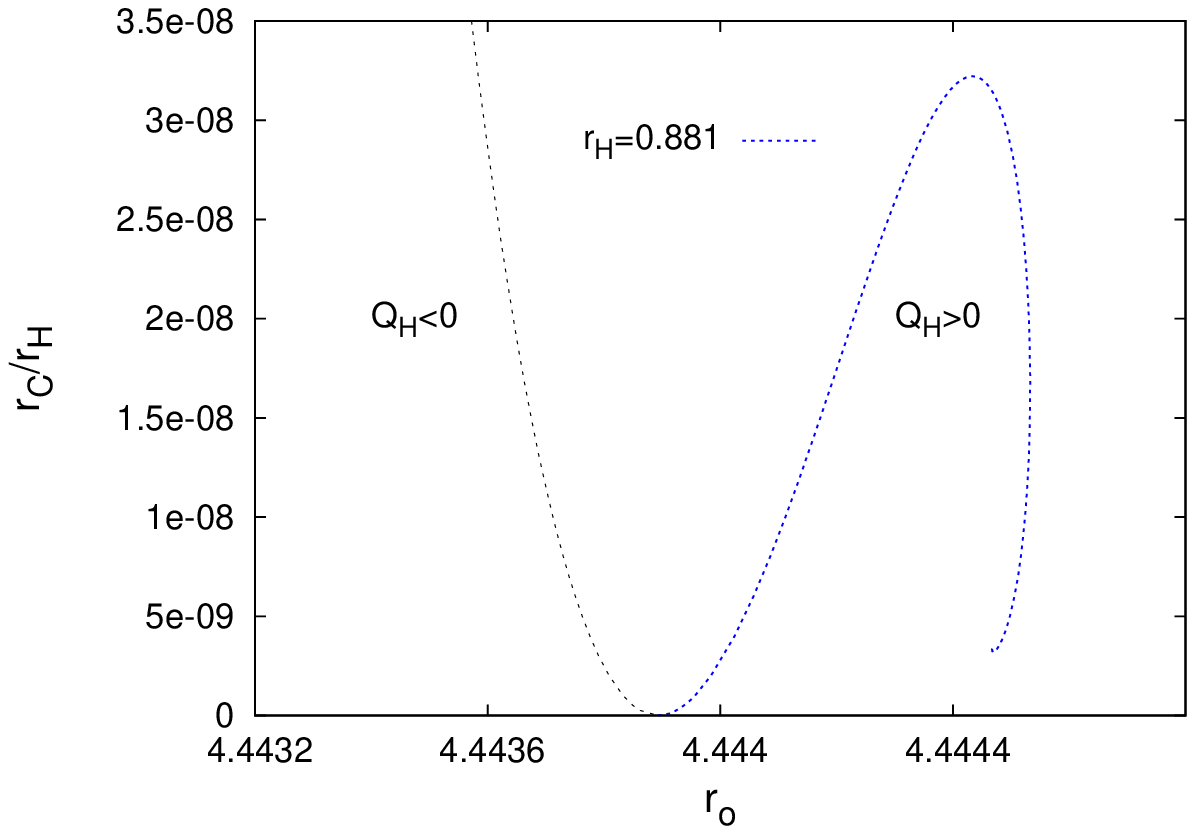}
\label{F2d}
}
}
\end{center}
\caption{Properties of boson shells with
Reissner-Nordstr\"om like black hole solutions in their
interior shown versus the outer shell
radius $r_{\rm o}$
for charge $Q=10$ and gravitational coupling constant $a=\alpha^2=0.01$:
(a) the temperature $T_{\rm H}$ versus $r_{\rm o}$;
(b) the mass $M$ versus the horizon charge $Q_{\rm H}$;
(c) zooming in on $r_{\rm C}/r_{\rm H}$ versus $r_{\rm o}$ for $Q_{\rm H}<0$;
(d) zooming in on $r_{\rm C}/r_{\rm H}$ versus $r_{\rm o}$ for $Q_{\rm H}>0$.
\label{bh2}
}
\end{figure}

Let us first consider the case of small charge {and}
small gravitational coupling,
choosing (as in the Schwarzschild case) 
the values $Q=10$ and $\alpha^2=0.01$.
We exhibit some of the properties of these solutions in Fig.~\ref{bh1}.
In Fig.~\ref{F1a} we show the ratio of the Cauchy horizon
radius $r_{\rm C}$ to the event horizon radius $r_{\rm H}$ for
branches of solutions with fixed values of the event horizon $r_{\rm H}$.
For positive horizon charge $Q_{\rm H} >0$,
these branches extend from the respective solutions
with a Schwarzschild like interior where $r_{\rm C}=0$,
up to the solutions with an extremal charged black hole in the interior.
The extremality condition can be expressed in terms
of the event horizon $r_{\rm H}$ and the Cauchy horizon $r_{\rm C}$
of the inner black hole
\begin{equation}
\frac{r_{\rm C}}{r_{\rm H}}= 1 \ .
\end{equation}
The extremal endpoints of the branches are indicated in the
figures by small asterisks.
As seen in Fig.~\ref{F1c},
along these branches the horizon charge $Q_{\rm H}$ rises monotonically
from zero to the extremal value.

In order to reach shells 
with even larger charges in their interior,
we can either consider
Reissner-Nordstr\"om like solutions without horizons,
i.e., naked singularities inside the shells.
Alternatively, we can switch the role of the two horizons,
and keep the Cauchy horizon $r_{\rm C}$ fixed, while we vary
the event horizon $r_{\rm H}$.
For the latter choice, we exhibit in Fig.~\ref{F1b}
the continuously extended branches of solutions, 
exhibiting the ratio of the event horizon radius $r_{\rm H}$ to
the Cauchy horizon radius $r_{\rm C}$.
The extremal solutions, where
the two parts of the branches match, are marked by the asterisks.

{
In the following we denote by $r_{\rm bh}$ the value of the 
radius which is kept fixed, 
i.e.~$r_{\rm bh}=r_{\rm H}$ along the lower parts, where $r_{\rm C}$ is varied,
and $r_{\rm bh}=r_{\rm C}$ along the upper parts, where $r_{\rm H}$ is varied.
}
Figs.~\ref{F1c}-\ref{F1f} then exhibit both parts of the branches,
the fixed $r_{\rm H}$ as well as the fixed $r_{\rm C}$ parts,
with both parts merging 
when the interior black hole solutions become extremal.
As seen in \ref{F1c},
the horizon charge $Q_{\rm H}$ becomes very large
along the fixed $r_{\rm C}$ parts of the branches,
and assumes values rather close to the global charge $Q$.
Thus the horizon charge becomes almost as large as possible.
Like $Q_{\rm H}$ the mass $M$ varies
smoothly along both parts of these branches,
as seen in Fig.~\ref{F1d}.

Let us now consider the endpoints of these branches of solutions.
When the Cauchy horizon is kept fixed, while the event horizon is
increased, the boson shells become smaller,
since the outer shell radius $r_{\rm o}$ decreases monotonically.
At the same time, the shell thickness $r_{\rm o}-r_{\rm i}$ decreases,
yielding an increasing ratio $r_{\rm i}/r_{\rm o}$, as seen in Fig.~\ref{F1e}.
Most revealing is, however, the increase of the
ratio $r_{\rm H}/r_{\rm i}$ of the event horizon to the inner shell radius,
shown in Fig.~\ref{F1f}.
Since this ratio goes to one, as the branches reach their endpoint,
this signals that the branches end in spirals.
A close inspection of the branches of solutions in the vicinity
of the endpoints indeed reveals the presence of spirals.

The temperature $T_{\rm H}$ of these solutions is shown in Fig.~\ref{F2a}.
The temperature vanishes when the interior black hole solutions
become extremal, otherwise the temperature is finite.
To address the nonuniqueness of the solutions, we inspect Fig.~\ref{F2b},
where the mass $M$ is shown as a function of the horizon charge $Q_{\rm H}$
for fixed global charge $Q$.
Interestingly, now a continuous set of solutions possesses
the same global charges $M$ and $Q$.
Such a set can for instance be obtained by
varying the horizon radius $r_{\rm H}$. 
Moreover, even if we were
to add the horizon charge $Q_{\rm H}$ as a further charge
to characterize the black hole space-times, uniqueness would not be recovered,
as seen from the crossings of the branches in Fig.~\ref{F2b}.
{
Note that also pure Reissner-Nordstr\"om black holes exist 
in some parameter range (i.e.~when $M > |Q|/\alpha=100$ in Fig.~\ref{F2b}),
which possess the same global charges as the boson shells with
interior black hole.}

Let us finally turn to solutions with negative charges in their interior.
For that purpose we reconsider Fig.~\ref{F2c}.
Clearly, the inner black hole solutions, that have so far been
endowed with positive charge, can also be endowed with negative charge,
leading to branches with negative horizon charge $Q_{\rm H}<0$.
However, the amount of negative charge that can be immersed in the interior
of these solutions is only rather small as seen in Fig.~\ref{F2c}.
(Because $Q_{\rm H} = - \sqrt{r_{\rm H} r_{\rm C}} /\alpha$, small $r_{\rm C}$ corresponds to small
$Q_{\rm H}$ when $r_{\rm H}$ and $\alpha$ are fixed.)

Indeed, for solutions in the considered parameter range
the horizon charge cannot be decreased far
and thus remains close to zero. 
Since for negative horizon charge $Q_{\rm H} <0$
the ratio $r_{\rm C}/r_{\rm H}$ remains small as well,
we zoom into this region in Fig.~\ref{F2c}.
Zooming in even further in Fig.~\ref{F2d} we see that
for certain parameters the horizon charge
can even change back to become positive again
before such a branch with fixed horizon radius $r_{\rm H}$ ends.

Addressing now the endpoints of these brnaches, we note that
for small values of the event horizon radius $r_{\rm H}$, 
where only one solution with 
vanishing horizon charge exists, the branches with negative horizon charge 
end in spirals, e.g., for~$r_{\rm H}=0.7$ and $0.8$.
In contrast, if there are two solutions with vanishing horizon charge for a
given event horizon radius $r_{\rm H}$, 
then the branch of solutions with negative
horizon charge connects these two solutions. 
Moreover, in this case a second branch with positive
horizon charge exists, which emerges from the second solution
with vanishing horizon charge and ends in a spiral
as seen in Fig.~\ref{F2c} and Fig.~\ref{F2d} for $r_{\rm H} = 0.881$.
This pattern continues when more than two solutions 
with vanishing horizon charge exist for a given 
horizon radius $r_{\rm H}$, i.e., in the spirals,
leading to more branches with positive respectively negative horizon charges,
connecting pairs of solutions with vanishing horizon charge.
Note finally that in Fig.~\ref{bh1} and Fig.~\ref{bh2}
the branch with $r_{\rm H}=0.98$
does not possess a limit with zero horizon charge,
and ends in a spiral with $Q_{\rm H}>0$.

\subsubsection{Small charge, larger gravitational coupling}

\begin{figure}[p]
\begin{center}
\vspace{-1.5cm}
\mbox{\hspace{-1.5cm}
\subfigure[][]{
\includegraphics[height=.27\textheight, angle =0]{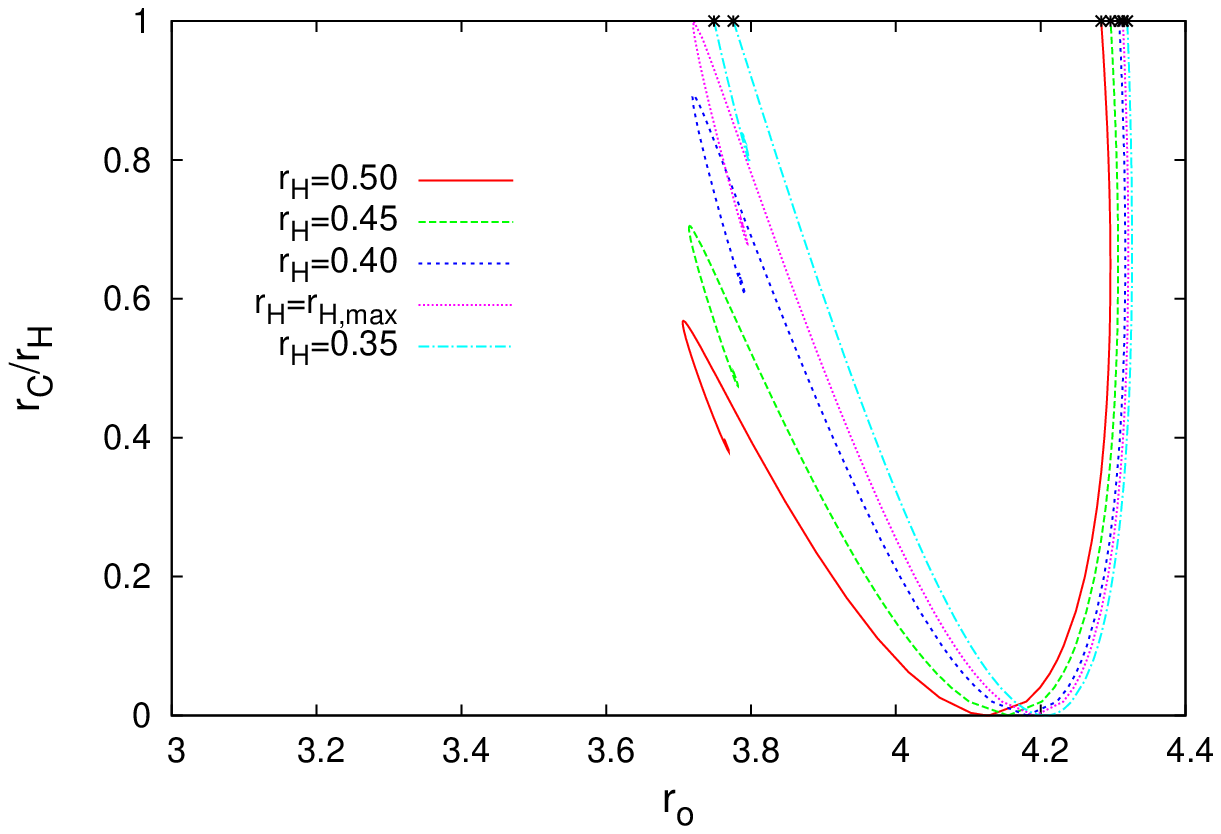}
\label{F3a}
}
\subfigure[][]{\hspace{-0.5cm}
\includegraphics[height=.27\textheight, angle =0]{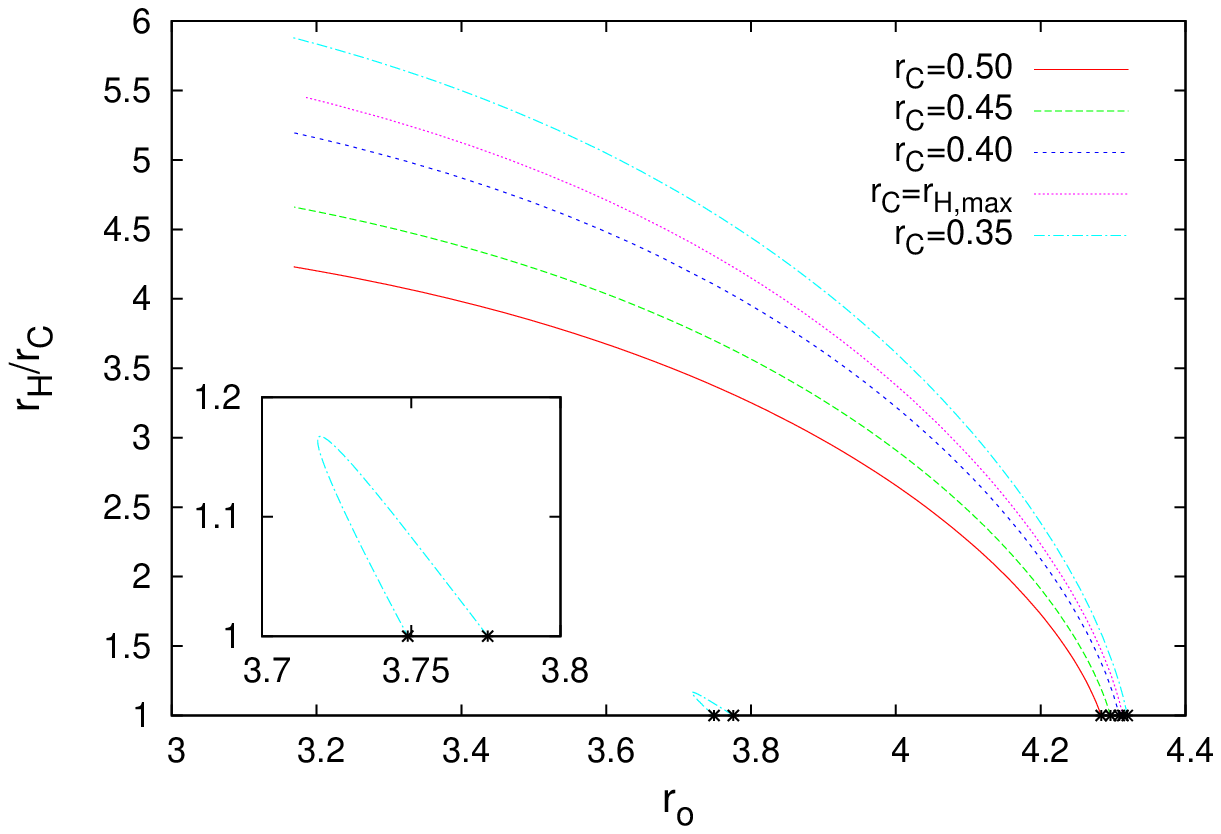}
\label{F3b}
}
}
\vspace{-0.5cm}
\mbox{\hspace{-1.5cm}
\subfigure[][]{
\includegraphics[height=.27\textheight, angle =0]{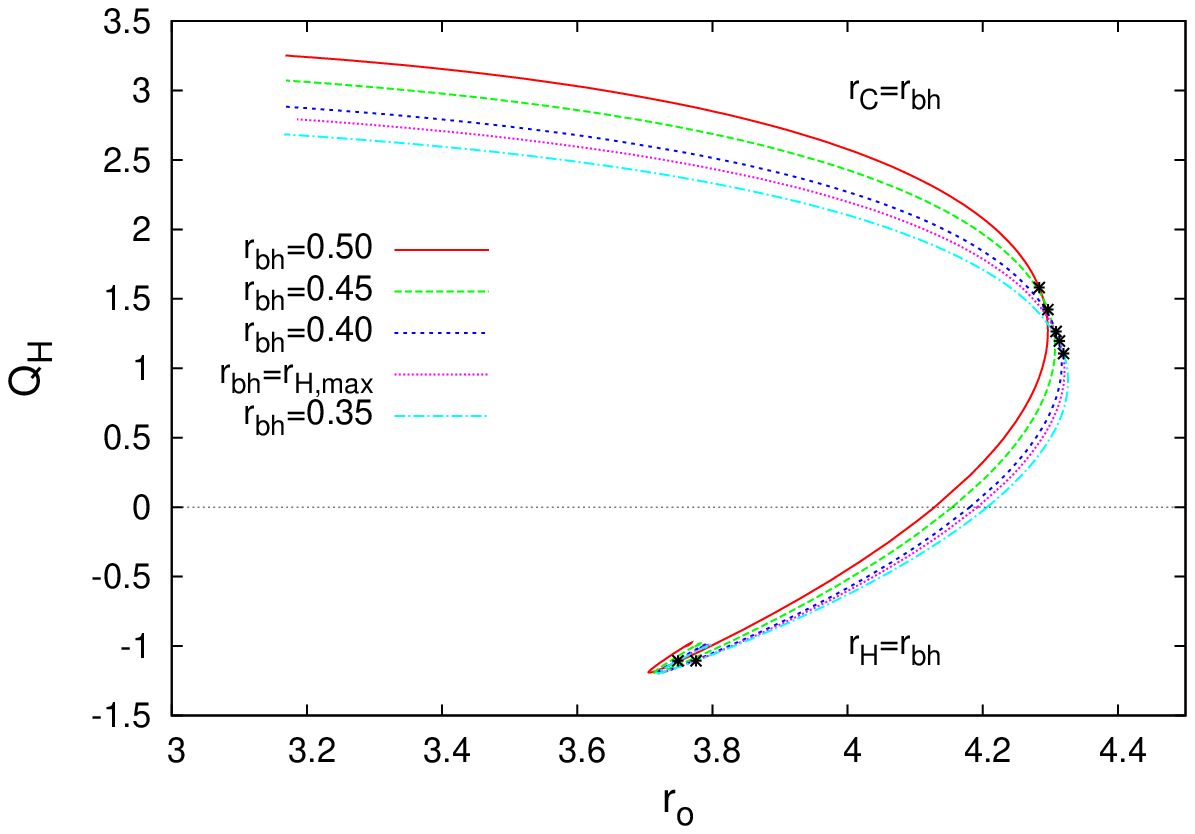}
\label{F3c}
}
\subfigure[][]{\hspace{-0.5cm}
\includegraphics[height=.27\textheight, angle =0]{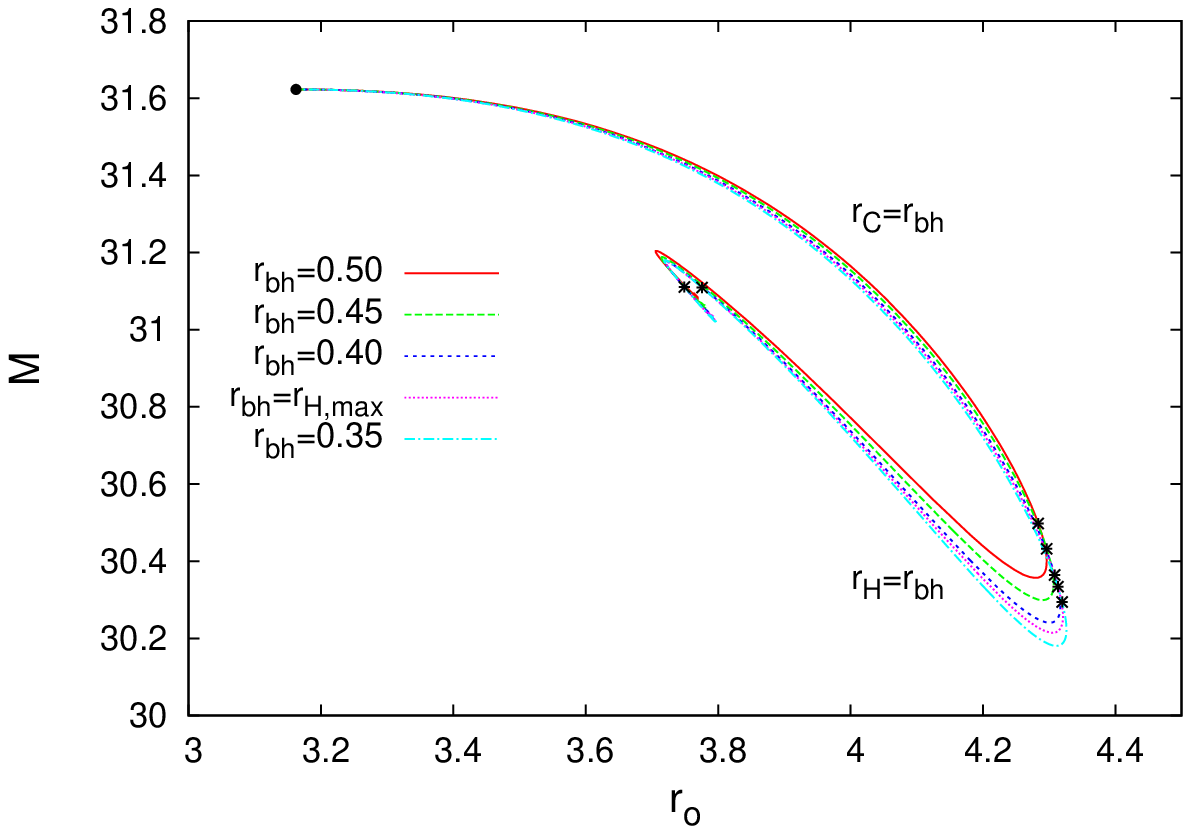}
\label{F3d}
}
}
\vspace{-0.5cm}
\mbox{\hspace{-1.5cm}
\subfigure[][]{
\includegraphics[height=.27\textheight, angle =0]{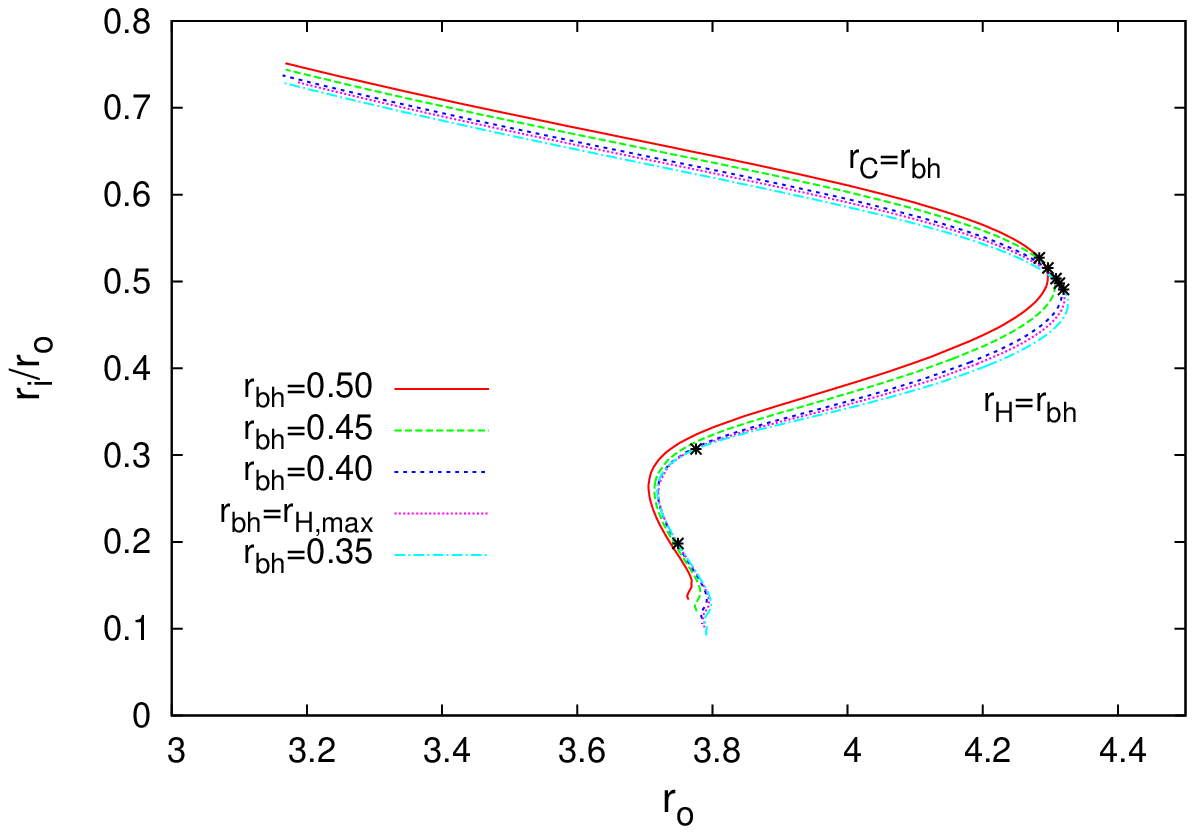}
\label{F3e}
}
\subfigure[][]{\hspace{-0.5cm}
\includegraphics[height=.27\textheight, angle =0]{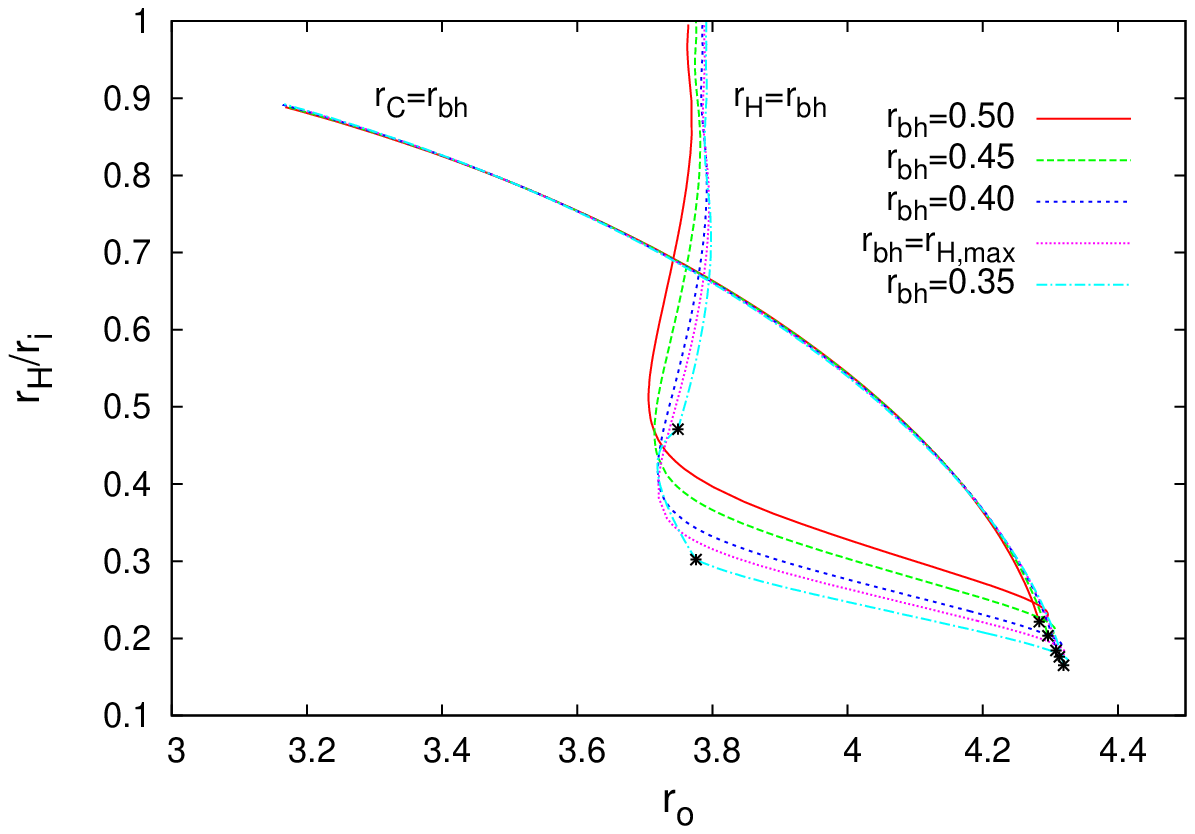}
\label{F3f}
}
}
\end{center}
\caption{Properties of boson shells with
Reissner-Nordstr\"om like black hole solutions in their
interior shown versus the outer shell
radius $r_{\rm o}$
for charge $Q=10$ and gravitational coupling constant $a=\alpha^2=0.1$:
(a) the ratio of Cauchy horizon to event horizon $r_{\rm C}/r_{\rm H}$;
(b) the ratio of Cauchy horizon to event horizon $r_{\rm H}/r_{\rm C}$;
(c) the horizon charge $Q_{\rm H}$;
(d) the mass $M$,
the dot corresponds to the
extremal limit where a throat is formed;
(e) the ratio of inner to outer shell radius $r_{\rm i}/r_{\rm o}$;
(f) the ratio of event horizon to inner shell radius $r_{\rm H}/r_{\rm i}$.
Note that $a=\alpha^2$, and the asterisks mark extremal black holes
where $r_{\rm C}/r_{\rm H}=1$.
\label{bh3}
}
\end{figure}

\begin{figure}[h]
\begin{center}
\vspace{-0.5cm}
\mbox{\hspace{-1.5cm}
\subfigure[][]{
\includegraphics[height=.27\textheight, angle =0]{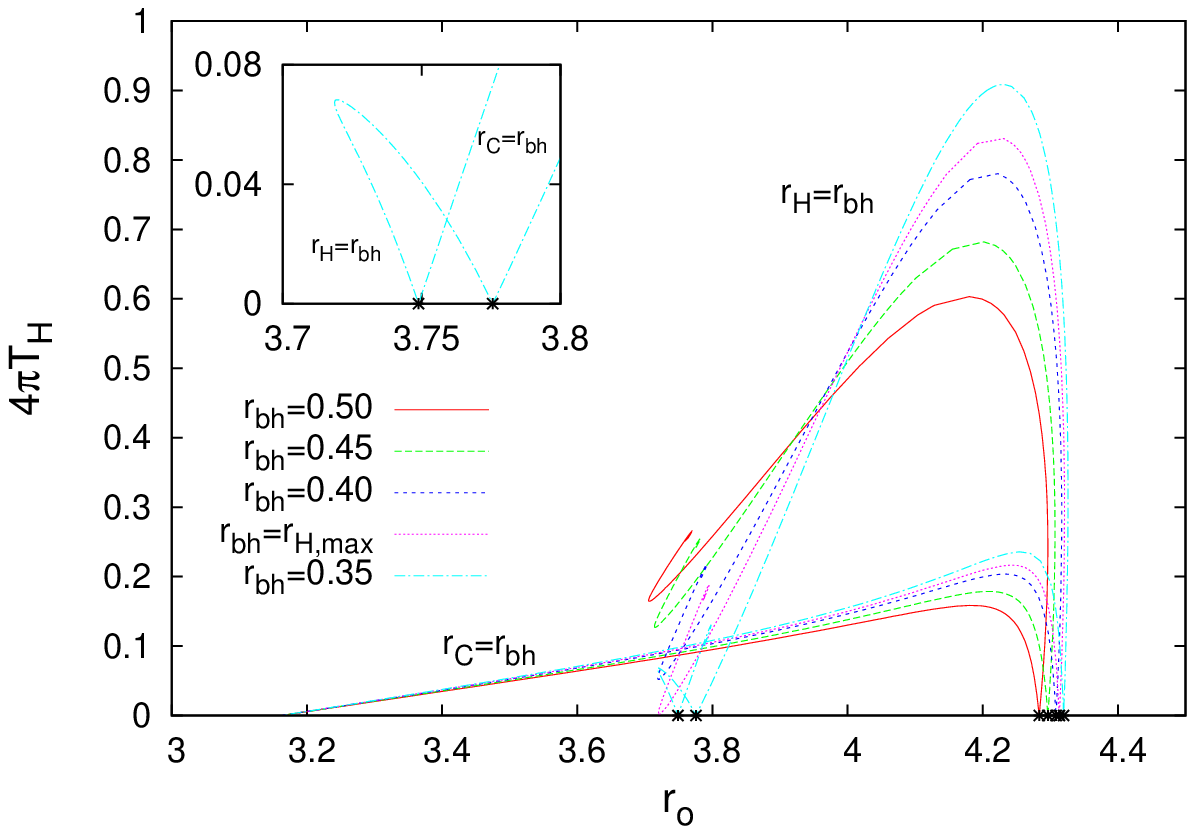}
\label{F5a}
}
\subfigure[][]{
\includegraphics[height=.27\textheight, angle =0]{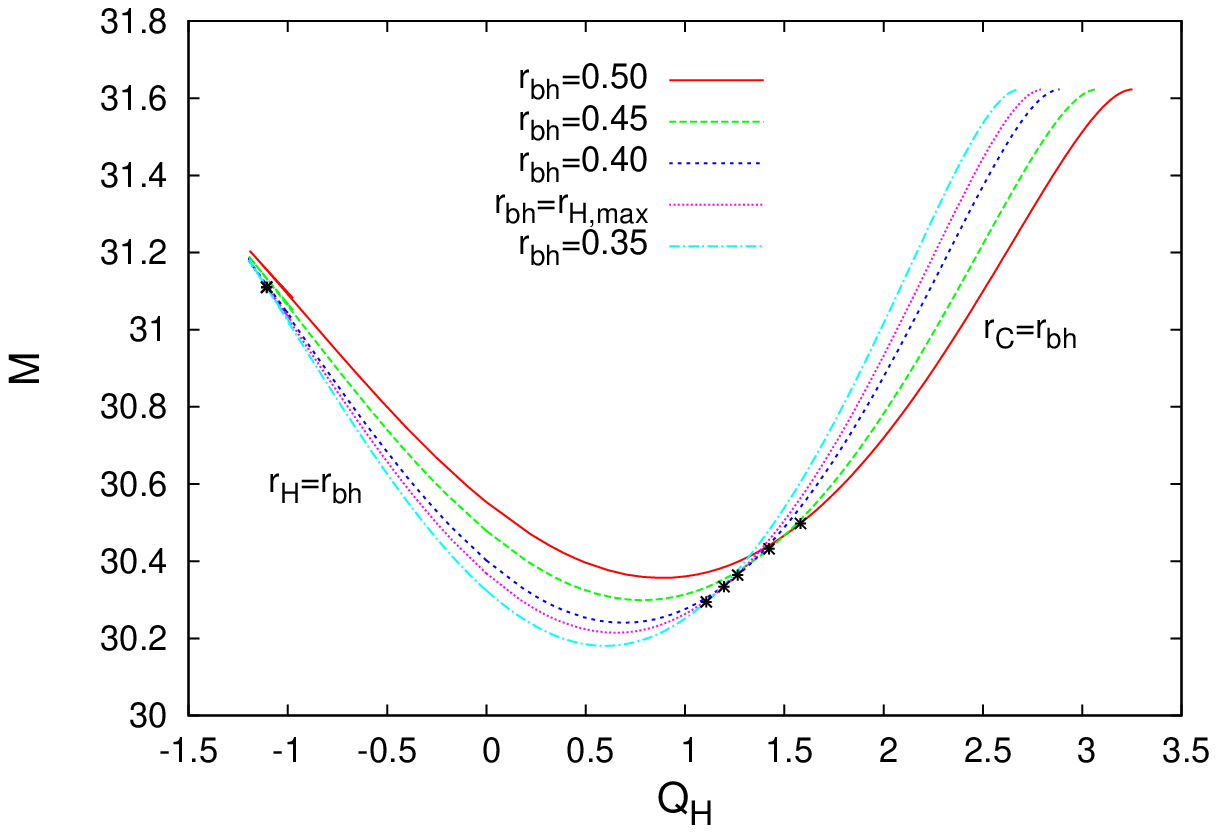}
\label{F5b}
}
}
\end{center}
\vspace*{-0.5cm}
\caption{\label{h5}
Properties of boson shells with
Reissner-Nordstr\"om like black hole solutions in their
interior shown versus the outer shell
radius $r_{\rm o}$
for charge $Q=10$ and gravitational coupling constant $a=\alpha^2=0.1$:
(a) the temperature $T_{\rm H}$ versus $r_{\rm o}$;
(b) the mass $M$ versus the horizon charge $Q_{\rm H}$.
}
\end{figure}

Next we turn to solutions for larger gravitational coupling $a=0.1$ and 
small charge $Q=10$. 
In Fig.~\ref{F3a} we show the
ratio $r_{\rm C}/r_{\rm H}$
versus the outer shell radius $r_{\rm o}$
for several fixed values of $r_{\rm H}$,
and in Fig.~\ref{F3b}
the ratio $r_{\rm H}/r_{\rm C}$ for several fixed values of $r_{\rm C}$.
As before, branches of solutions with charged black holes
emerge from the solutions with Schwarzschild like black holes.
Interestingly, as seen in Fig.~\ref{F3c},
now the negative horizon charge can assume a
considerably larger magnitude as compared to the previous case.
Thus the negative horizon charge branches extend much further.
These branches end in spirals, when the ratio of
the horizon radius to the inner shell radius $r_{\rm H}/r_{\rm i}$
tends to one, as seen in Fig.~\ref{F3f}.

The branches with positive horizon charge 
and fixed horizon radius extend up to the extremal case, where
$r_{\rm C}/r_{\rm H}=1$. 
However, for negative horizon charge the 
branches can now also reach the extremal case,
when (for the parameters employed) the event
horizon radius is below $r_{\rm H, max}=0.37823$.
Increasing the magnitude of the charge in the interior further,
can then yield boson shells with naked singularities
in their interior, carrying positive or negative charge.

Alternatively, extending these branches beyond the extremal solutions
by keeping the Cauchy horizon fixed, while varying
the event horizon, we see that the negative horizon charge
solutions only form small loops
connecting two extremal solutions,
as shown in Fig.~\ref{F3b} for $r_{\rm C}=0.35$.
In contrast, the branches with positive horizon charge black holes
extend further, yielding boson shells of considerably smaller size.

Whereas in the previous case described in section {4.2.1} 
the fixed $r_{\rm C}$ branches ended also in spirals, 
we here observe the alternative endpoint scenario, 
namely the branches end
when a throat is formed at the outer shell radius.
This is obvious from Fig.~\ref{F3b} and Fig.~\ref{F3d},
since the formation of a throat occurs when the conditions
$r_{\rm o} = \alpha^2 M = \alpha Q$ hold.
Clearly, the endpoints of all branches occur at 
an outer shell radius of $r_{\rm o} = \alpha Q$,
where the mass assumes the value $M=Q/\alpha$.

The formation of the throat at these endpoints
is also seen in the temperature
$T_{\rm H}$, exhibited in Fig.~\ref{F5a}.
Evidently, the temperature goes to zero when the interior
black hole solutions become extremal, as indicated by
the asterisks. But the temperature also goes to zero
at the end points of the branches,
where a throat is formed.

Nonuniqueness, finally, is addressed in Fig.~\ref{F5b},
where the mass is exhibited versus the horizon charge.
Again, we see a continuous nonuniqueness of the solutions
containing charged black holes,
as long as only the global charges are considered.
The crossings of the branches, on the other hand, show
that, as in the previous case, even with 
the introduction of a further charge $Q_{\rm H}$
uniqueness is not regained.

\subsubsection{Large charge}

\begin{figure}[p]
\begin{center}
\vspace{-1.5cm}
\mbox{\hspace{-1.5cm}
\subfigure[][]{
\includegraphics[height=.27\textheight, angle =0]{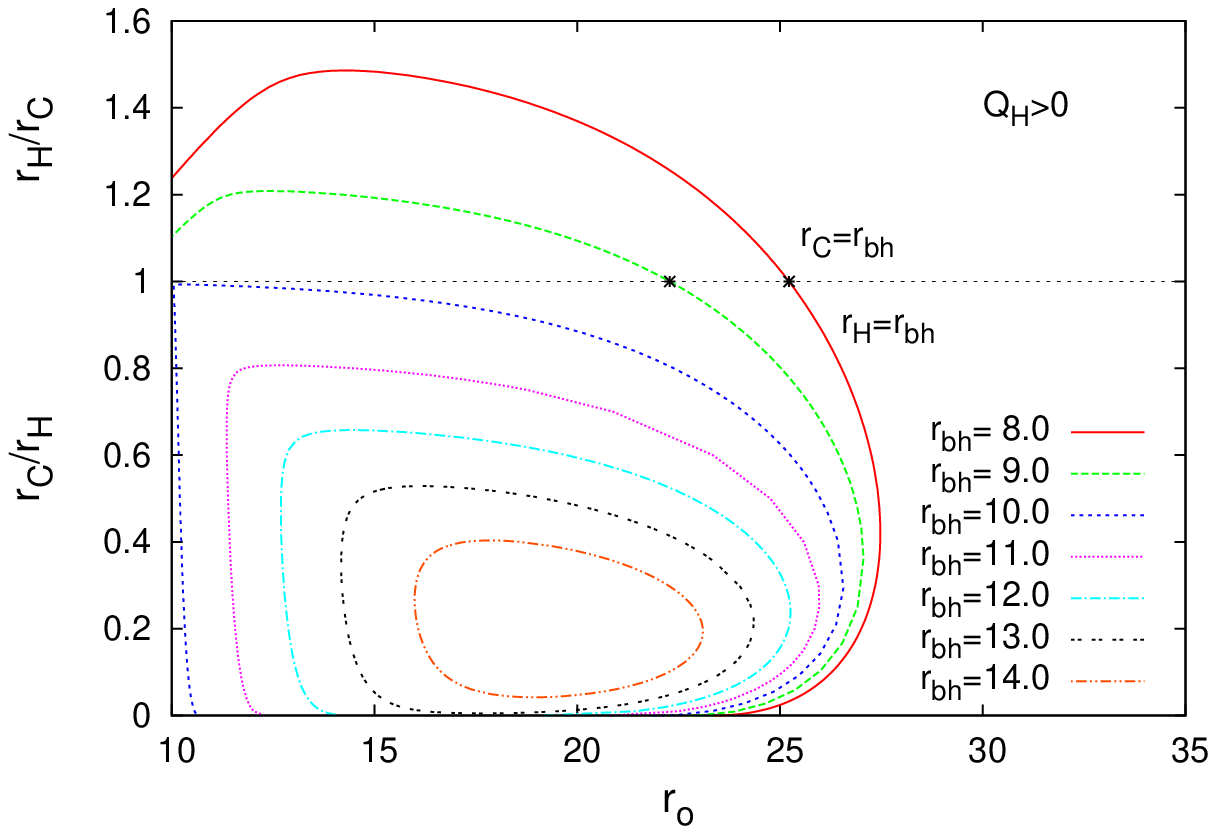}
\label{F6a}
}
\subfigure[][]{\hspace{-0.5cm}
\includegraphics[height=.27\textheight, angle =0]{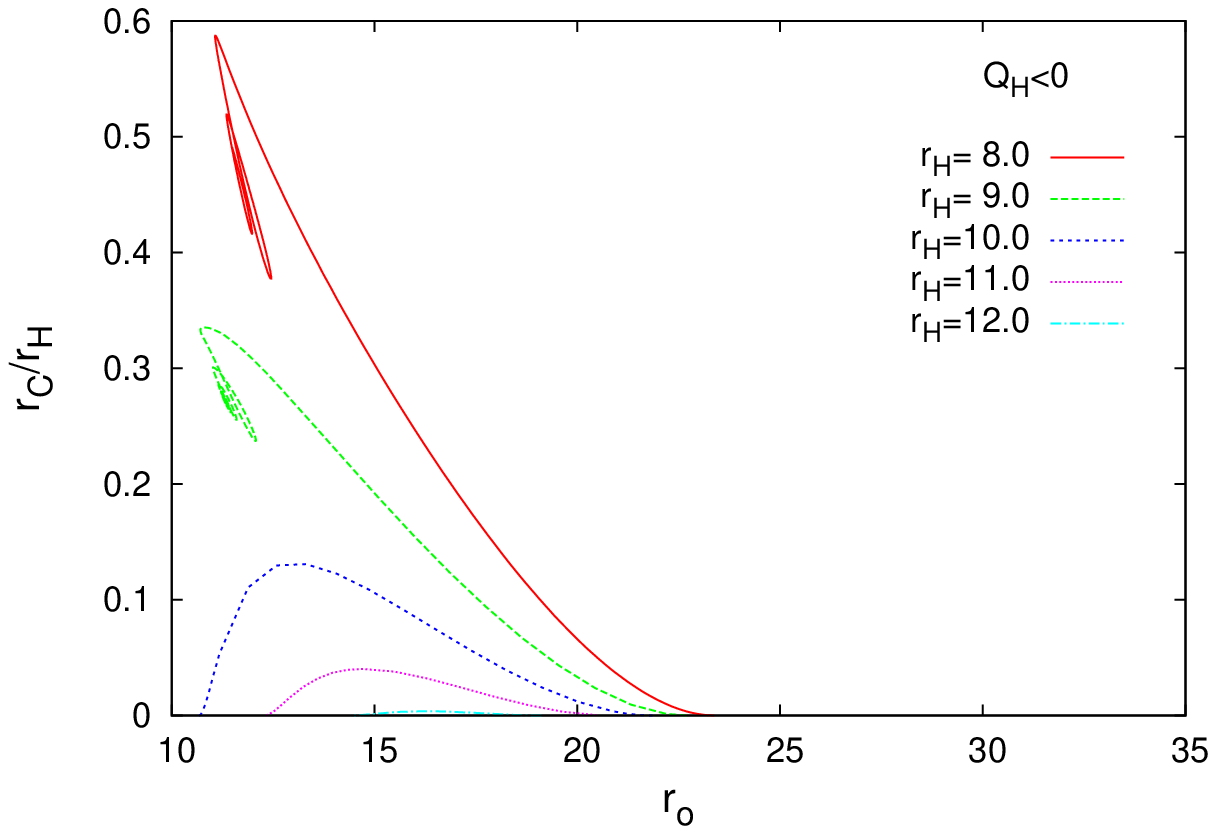}
\label{F6b}
}
}
\vspace{-0.5cm}
\mbox{\hspace{-1.5cm}
\subfigure[][]{
\includegraphics[height=.27\textheight, angle =0]{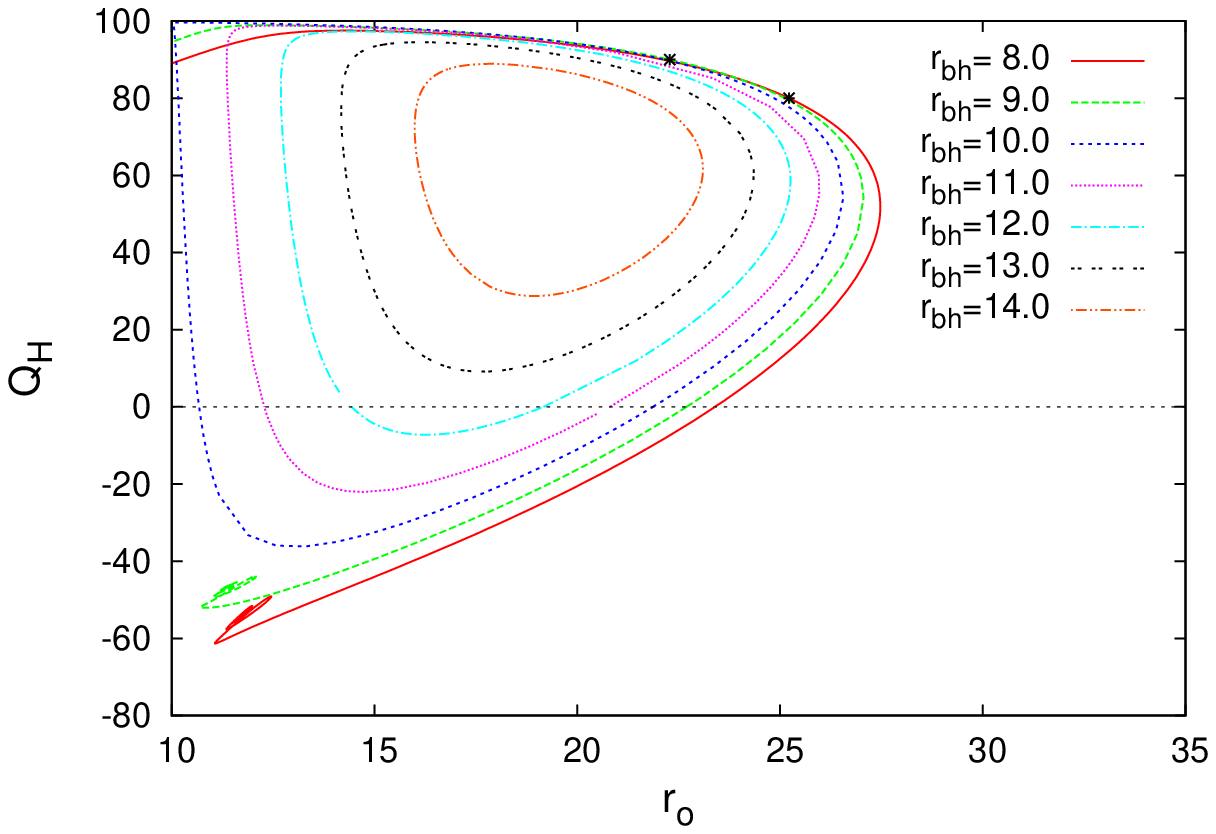}
\label{F6c}
}
\subfigure[][]{\hspace{-0.5cm}
\includegraphics[height=.27\textheight, angle =0]{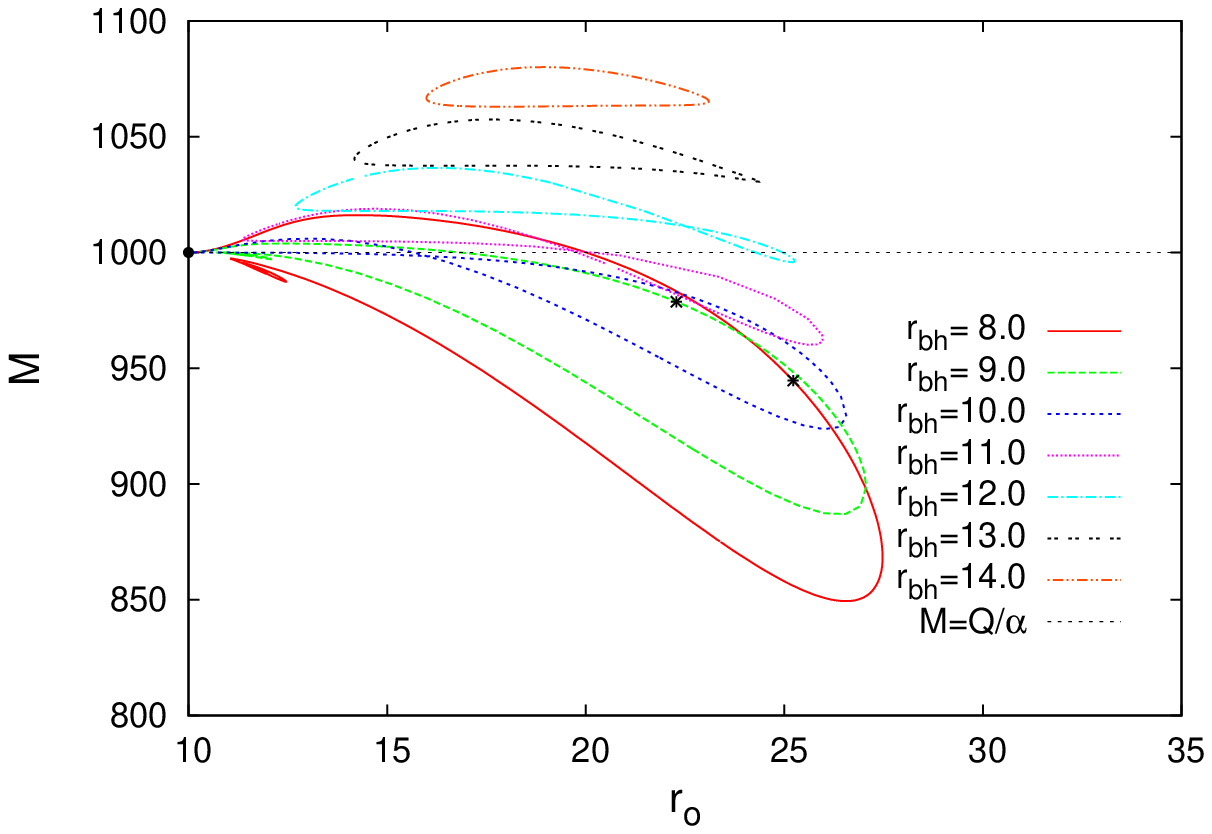}
\label{F6d}
}
}
\vspace{-0.5cm}
\mbox{\hspace{-1.5cm}
\subfigure[][]{
\includegraphics[height=.27\textheight, angle =0]{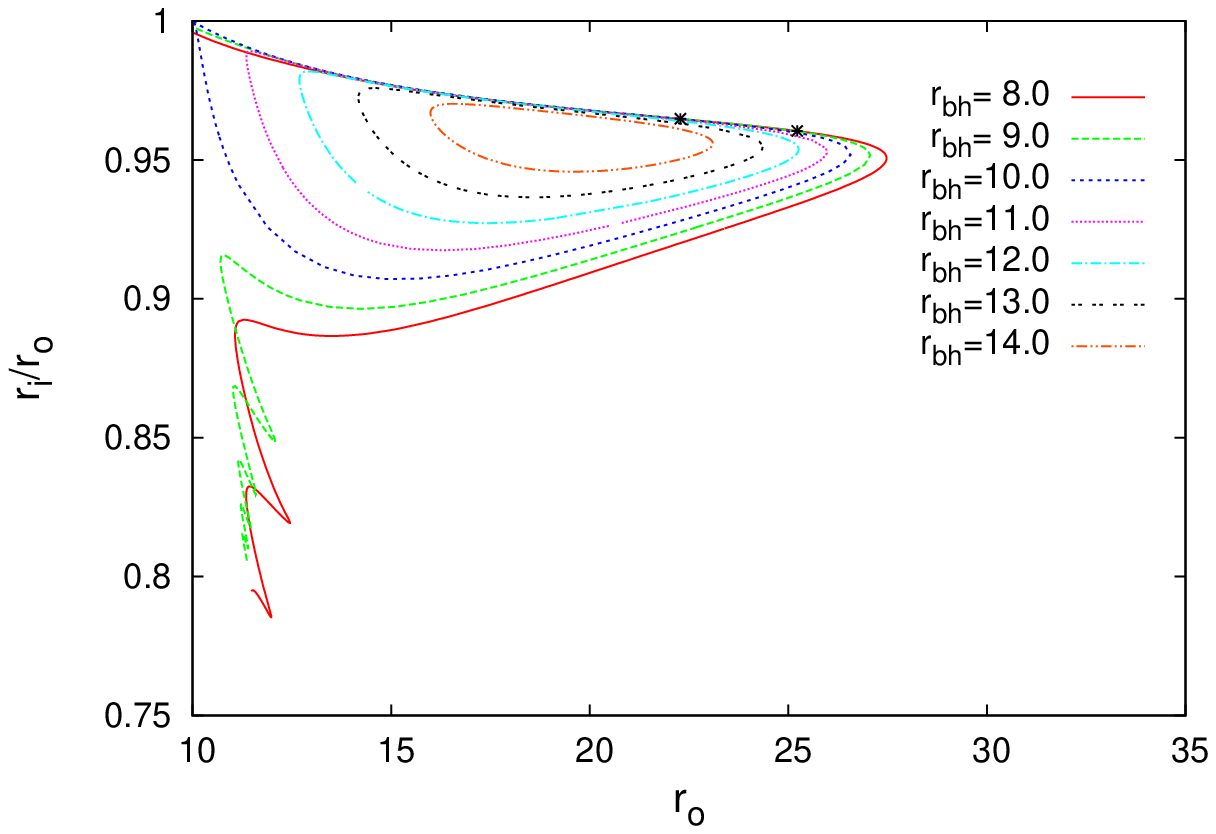}
\label{F6e}
}
\subfigure[][]{\hspace{-0.5cm}
\includegraphics[height=.27\textheight, angle =0]{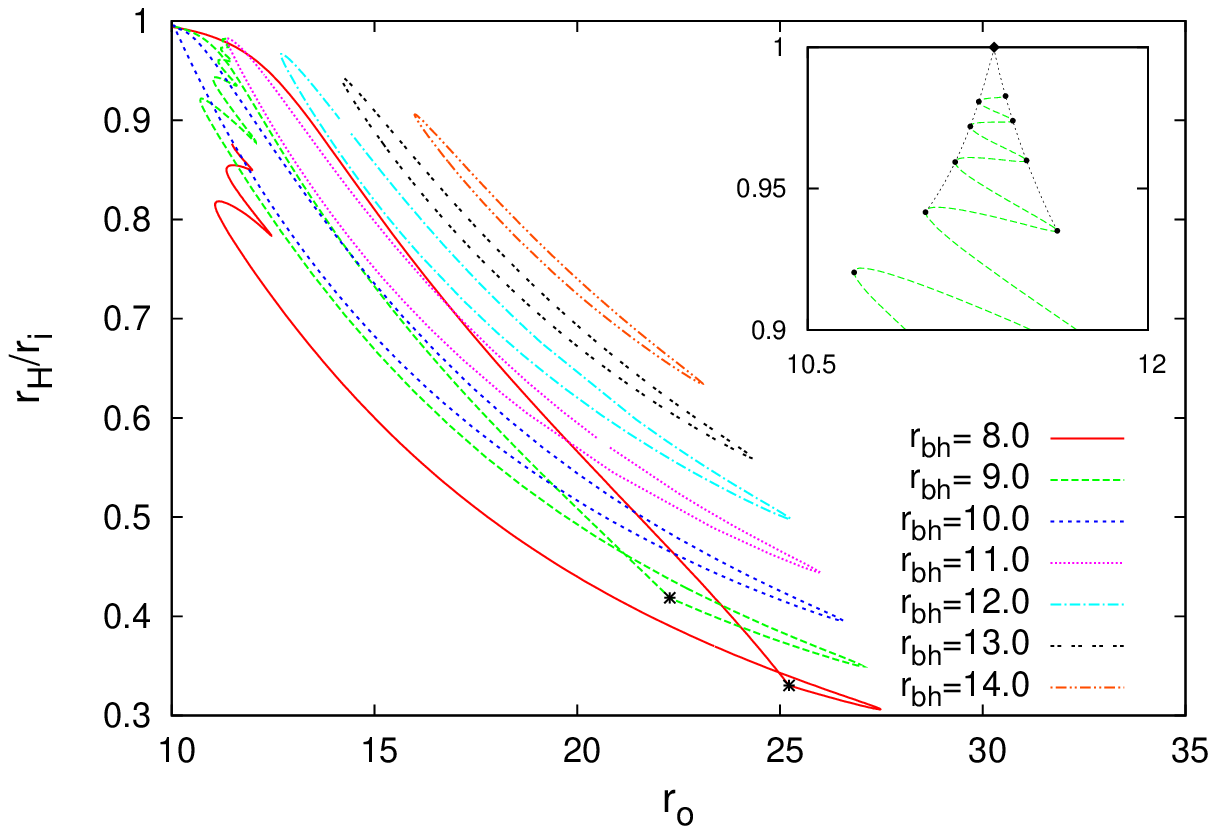}
\label{F6f}
}
}
\end{center}
\caption{Properties of boson shells with
Reissner-Nordstr\"om like black hole solutions in their
interior shown versus the outer shell
radius $r_{\rm o}$
for charge $Q=100$ and gravitational coupling constant $a=\alpha^2=0.01$:
(a) the ratio of Cauchy horizon to event horizon $r_{\rm C}/r_{\rm H}$;
(b) the ratio of Cauchy horizon to event horizon $r_{\rm H}/r_{\rm C}$;
(c) the horizon charge $Q_{\rm H}$;
(d) the mass $M$,
the dot corresponds to the
extremal limit where a throat is formed;
(e) the ratio of inner to outer shell radius $r_{\rm i}/r_{\rm o}$;
(f) the ratio of event horizon to inner shell radius $r_{\rm H}/r_{\rm i}$.
{ The inlet shows  $r_{\rm H}/r_{\rm i}$ together with the envelope (dotted lines)
and the extrapolated endpoint (diamond) for $r_{\rm H}=r_{\rm bh}=9.0$.}
Note that $a=\alpha^2$, and the asterisks mark extremal black holes
where $r_{\rm C}/r_{\rm H}=1$.
\label{bh6}
}
\end{figure}

\begin{figure}[t]
\begin{center}
\vspace{-1.5cm}
\mbox{\hspace{-1.5cm}
\subfigure[][]{
\includegraphics[height=.27\textheight, angle =0]{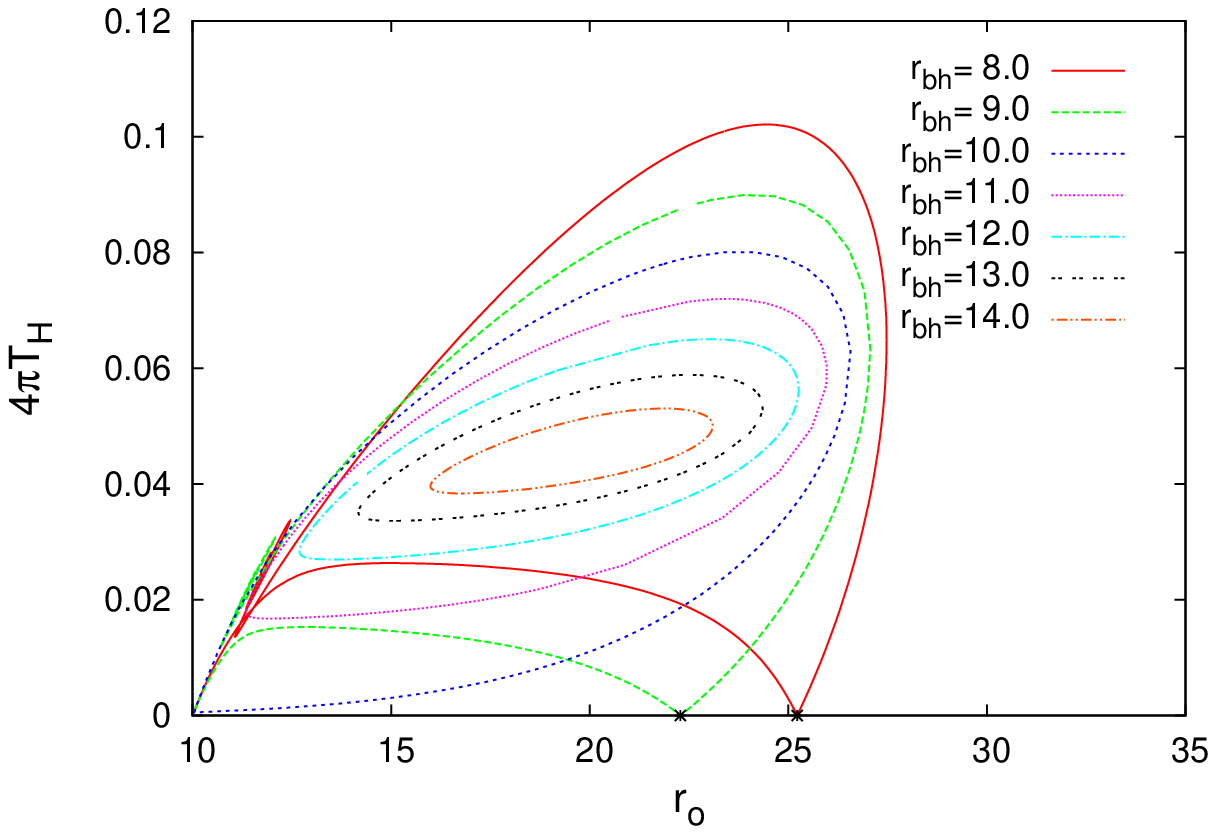}
\label{F7a}
}
\subfigure[][]{\hspace{-0.5cm}
\includegraphics[height=.27\textheight, angle =0]{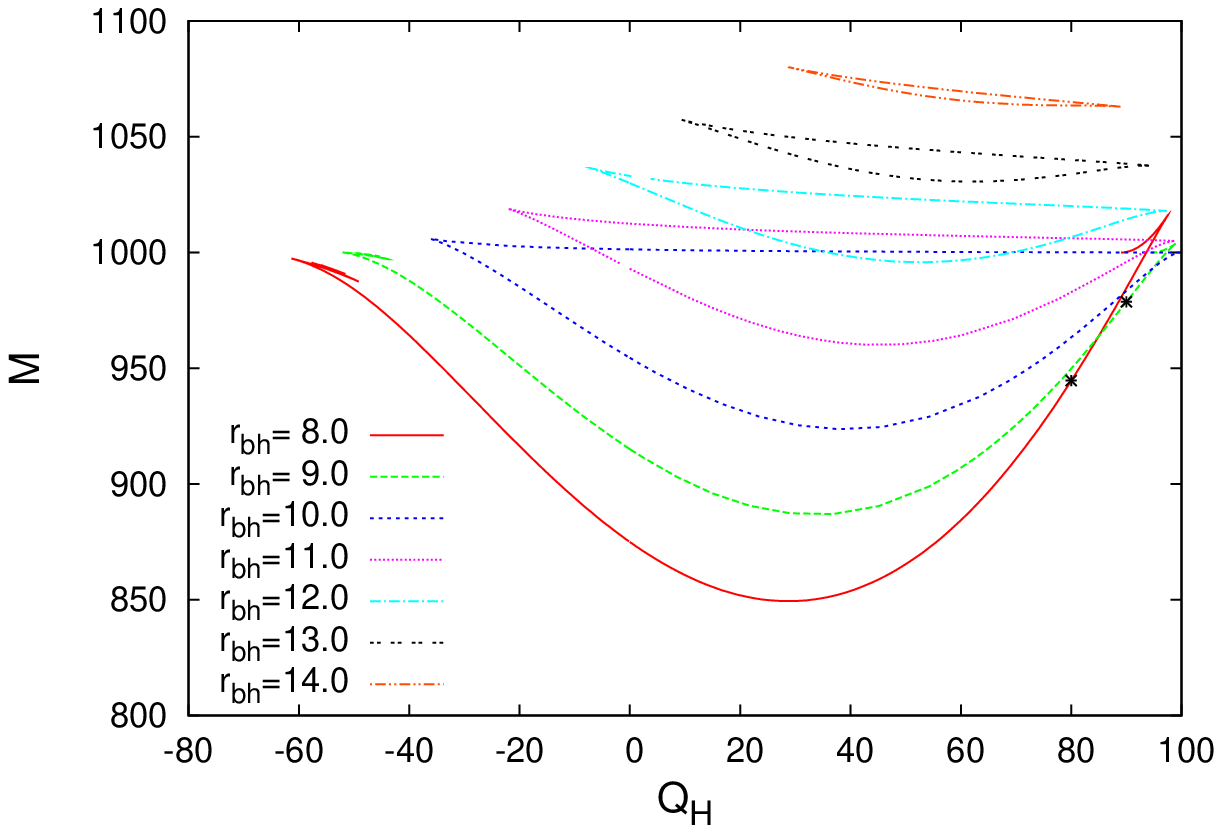}
\label{F7b}
}
}
\end{center}
\vspace*{-0.5cm}
\caption{Properties of boson shells with
Reissner-Nordstr\"om like black hole solutions in their
interior shown versus the outer shell
radius $r_{\rm o}$
for charge $Q=100$ and gravitational coupling constant $a=\alpha^2=0.01$:
(a) the temperature $T_{\rm H}$ versus $r_{\rm o}$;
(b) the mass $M$ versus the horizon charge $Q_{\rm H}$.
}
\label{bh7}
\end{figure}

Next we turn to black hole solutions with large charge $Q=100$
and small gravitational coupling $a=0.01$.
As in the above cases
branches of solutions with 
positively and negatively charged interior black holes
emerge from the solutions with Schwarzschild like interior.
The properties of these solutions are exhibited in Fig.~\ref{bh6}
and Fig.~\ref{bh7}.

Considering first solutions with a positively charged interior
black hole, we expect the same type of behaviour as above.
With increasing horizon charge, the branches of
fixed horizon size end when the inner black hole solution becomes extremal.
From there these branches can be continued by retaining the
Cauchy horizon fixed while varying the event horizon.
We observe this expected behaviour indeed,
but only for the smaller values of the fixed horizon radius.
These branches end, when a throat is formed
at the outer shell radius, i.e., when
$r_{\rm o}=\alpha^2 M = \alpha Q$.

However, for larger values of the fixed horizon radius,
no longer an extremal black hole is encountered.
Instead, the branches reach a maximal value
of $r_{\rm C}/r_{\rm H}$ and then bend backwards
towards smaller values of the horizon charge
and larger values of the shell size.
The value $r_{\rm H}=10$ shown in the figures,
is slightly above the critical value $r_{\rm H,cr}$,
where this new type of behaviour sets in.
These branches then reach a second solution
with a neutral interior, different from the
starting solution.

When the value of the fixed horizon radius is increased further,
beyond a critical value there no longer exists
a solution with a neutral interior black hole,
and the branch of solutions with charged interior black holes
closes on itself.
Above a maximal value of the horizon radius
also solutions with charged
interior black holes no longer exist.

Thus the behaviour of these branches of solutions depends on 
how many solutions with a neutral black hole in the interior
exist for a given global charge and fixed horizon radius.
The critical radius $r_{\rm H,cr}$ is precisely given
by the transition point from one to two neutral interior solutions,
readily identifiable in Fig.~\ref{bhS1}.
Here, as the throat is formed at the outer shell radius,
the inner black hole solution becomes extremal,
thus $r_{\rm C}/r_{\rm H}=1$, but the horizon is still slightly smaller
than the inner shell radius, and the
inner shell radius is slightly smaller than the
outer shell radius, $r_{\rm H}<r_{\rm i}<r_{\rm o}$.
This space-time therefore satisfies the conditions
for an extremal solution at two places, at $r_{\rm H}$ and at $r_{\rm o}$.

Let us now extend the above analysis to solutions
with negative horizon charges.
The ratio $r_{\rm C}/r_{\rm H}$ for this type of solutions
is considered explicitly in Fig.~\ref{F6b},
while the other figures contain solutions with both
positive and negative horizon charges.
When there is only one solution with a neutral black hole in the interior, 
i.e., when $r_{\rm H}< r_{\rm H, cr}$, 
then the branches with 
negative horizon charge show a spiral-like behaviour.
On the other hand, 
when there are two solutions with a neutral black hole in the interior
i.e., when $r_{\rm H}> r_{\rm H,\rm cr}$,
then the branches with negative (or positive) horizon charges connect both 
of these solutions with vanishing horizon charge. 
{ 
The spiral-like behaviour indicates that the ratio of the event horizon
radius to the inner radius $r_{\rm H}/r_{\rm i}$ tends to one.
This is demonstrated in detail for $r_{\rm H}=9.0$ in the inlet 
of Fig.~\ref{F6f}, where we show the first ten branches.
To extrapolate the endpoint (diamond) we made a fourth order approximation 
using the data at the minima and maxima of $r_{\rm o}$.}

In Fig.~\ref{F7a} we exhibit the temperature of these solutions.
The temperature tends to zero when the inner black hole
becomes extremal, 
which is indicated by the asterisks in the figure.
The temperature also tends to zero, when a throat forms
at the outer horizon radius. For the given values of the charge
and gravitational coupling constant, this happens
when the outer shell radius reaches $r_{\rm o}=10$.
In the figure we see the throat formation
for $r_{\rm C}=8$ and 9.
The continuous nonuniqueness of the solutions
containing charged black holes
is demonstrated in Fig.~\ref{F7b}.
{
Note that also pure Reissner-Nordstr\"om black holes exist 
in some parameter range (i.e.~when $M > |Q|/\alpha=1000$ in Fig.~\ref{F7b}),
which possess the same global charges as the boson shells with
interior black hole.}

\section{Energy conditions}

We now consider the energy conditions for these
boson shell solutions with black holes in their interior.
To that end,
we define $X^\mu(x)$ as a unit timelike vector field, $X^\mu X_\mu < 0$,
and $T_{\mu\nu}$ is the stress energy tensor.
Then the weak and strong energy conditions read
\begin{center}
\begin{tabular}{lc}
weak energy condition   & $T_{\mu\nu}X^\mu X^\nu \geq 0 $ \\
strong energy condition & ($T_{\mu\nu}-\frac{1}{2}T g_{\mu\nu})X^\mu X^\nu \geq 0 $ .
\end{tabular}
\end{center}
The dominant energy condition requires that for 
all future directed timelike $X^\mu(x)$, 
the vector $-T^\mu_\nu  X^\nu(x)$ is future directed timelike or null.

If $X^\mu$ represents the four velocity of an observer
then the weak energy condition states that the energy density as measured 
by any observer cannot be negative, whereas the dominant energy condition 
states that the speed of the energy flow as measured by any observer 
cannot exceed the speed of light.
The strong energy condition guarantees that for 
a hypersurface orthogonal congruence the 
change of the expansion with respect to proper time
is negative.

To analyze the weak, strong and dominant energy conditions for 
the boson shell solutions we follow Wald \cite{wald}.
For a diagonal stress-energy tensor
\begin{equation}
T_\mu^\nu = {\rm diag}(-\rho, P_r, P_\theta, P_\vphi)
\label{Tdiag}
\end{equation}
the weak energy condition can be expressed as
\begin{equation}
\rho \geq 0 \ \ \ {\rm and} \ \ \ 
\rho+P_i \geq 0 \ , \ \ \ i=r,\theta,\vphi \ ,  
\label{WEC}
\end{equation}
the strong energy condition requires
\begin{equation}
\rho+\sum_i P_i \geq 0 \ \ \ {\rm and} \ \ \ 
\rho+P_i \geq 0 \ , \ \ \ i=r,\theta,\vphi \ ,  
\label{SEC}
\end{equation}
and for the dominant energy condition 
\begin{equation}
\rho \geq |P_i| \ , \ \ \ i=r,\theta,\vphi \ ,  
\label{DEC}
\end{equation}
has to hold.

In the interior ($r\leq r_{\rm i}$) and exterior ($r\geq r_{\rm o}$)
of the shell all energy conditions are satisfied, since here 
the space-time is of Reissner-Nordstr\"om type. 
In order to analyze a possible violation of the energy conditions
it is sufficient to restrict to the shell $r_{\rm i} < r < r_{\rm o}$.
For the boson shell solutions the energy density and the pressures
are given by
\begin{eqnarray}
\rho & = & -T_0^0 = \frac{1}{2}\left(N h'^2+2|h|\right)
                   +\frac{1}{2 A^2 N}\left(N b'^2+b^2 h^2\right) \ , 
\label{ECrho}\\
P_r & = & T_r^r = \frac{1}{2}\left(N h'^2-2|h|\right)
                   +\frac{1}{2 A^2 N}\left(-N b'^2+b^2 h^2\right) \ , 
\label{ECPr}\\
P_\theta & = & T_\theta^\theta = -\frac{1}{2}\left(N h'^2+2|h|\right)
                   +\frac{1}{2 A^2 N}\left(N b'^2+b^2 h^2\right) \ , 
\label{ECPt}\\
P_\vphi & = & T_\vphi^\vphi = T_\theta^\theta \ . 
\label{ECPp}
\end{eqnarray}

From Eqs.~(\ref{ECrho})-(\ref{ECPp})
it can immediately be seen that the weak energy condition is 
satisfied. 
For the dominant energy condition we find
\begin{eqnarray}
\left(\frac{N h'^2}{2}+\frac{b^2 h^2}{2 A^2 N}\right)
      +\left(h+\frac{b'^2}{2 A^2}\right)
& \geq &
\left|
\left(\frac{N h'^2}{2}+\frac{b^2 h^2}{2 A^2 N}\right)
      -\left(h+\frac{b'^2}{2 A^2}\right)
\right|	
\nonumber \\
\frac{1}{2}\left(N h'^2+2 h\right)
                   +\frac{1}{2 A^2 N}\left(N b'^2+b^2 h^2\right)	   
& \geq &
\left|
-\frac{1}{2}\left(N h'^2+2 h\right)
                   +\frac{1}{2 A^2 N}\left(N b'^2+b^2 h^2\right)
\right|	
\nonumber
\end{eqnarray}
Both inequalities are of the form $\xi_1^2+\xi_2^2 \geq |\xi_1^2-\xi_2^2|$,
which is always satisfied. 
Hence the dominant energy condition is also fulfilled.
For the strong energy condition we note that 
\begin{equation}
\rho+\sum_i P_i = -2 h +\frac{1}{A^2 N}\left( N b'^2 + 2 h^2 b^2)\right)
\label{SEcond}
\end{equation}
may become negative for some solutions.
For the Schwarzschild case, e.g. $b'=0$ and $h^2 << h$ near $r=r_{\rm i}$. 
Thus for these solutions the strong energy condition is violated. 
However, for the Reissner-Nordstr\"om case  
$b'\neq 0$ at $r=r_{\rm i}$. 
Thus the strong energy condition may be satisfied 
if ${b'}^2$ at $r=r_{\rm i}$ is large enough.

\begin{figure}[t]
\begin{center}
\vspace{-1.5cm}
\mbox{\hspace{-1.5cm}
\subfigure[][]{
\includegraphics[height=.27\textheight, angle =0]{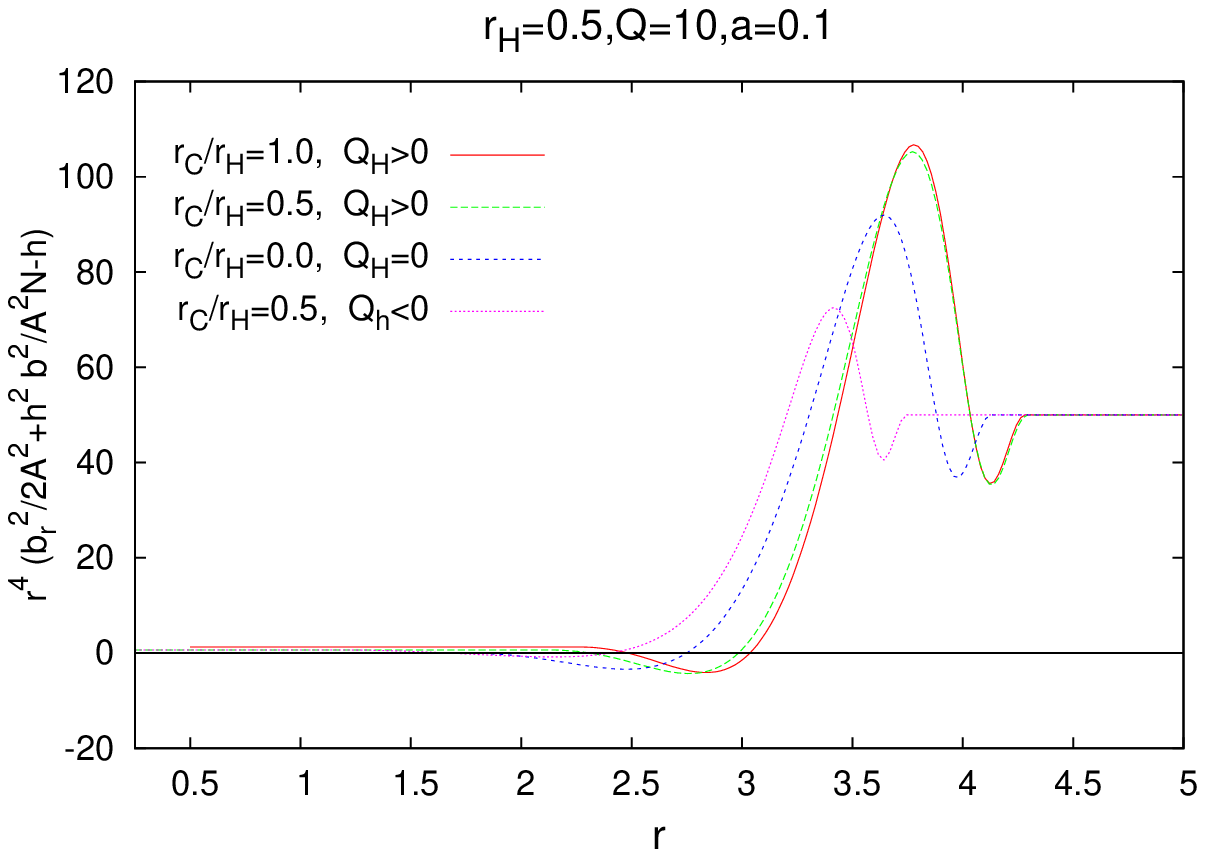}
\label{F9a}
}
\subfigure[][]{\hspace{-0.5cm}
\includegraphics[height=.27\textheight, angle =0]{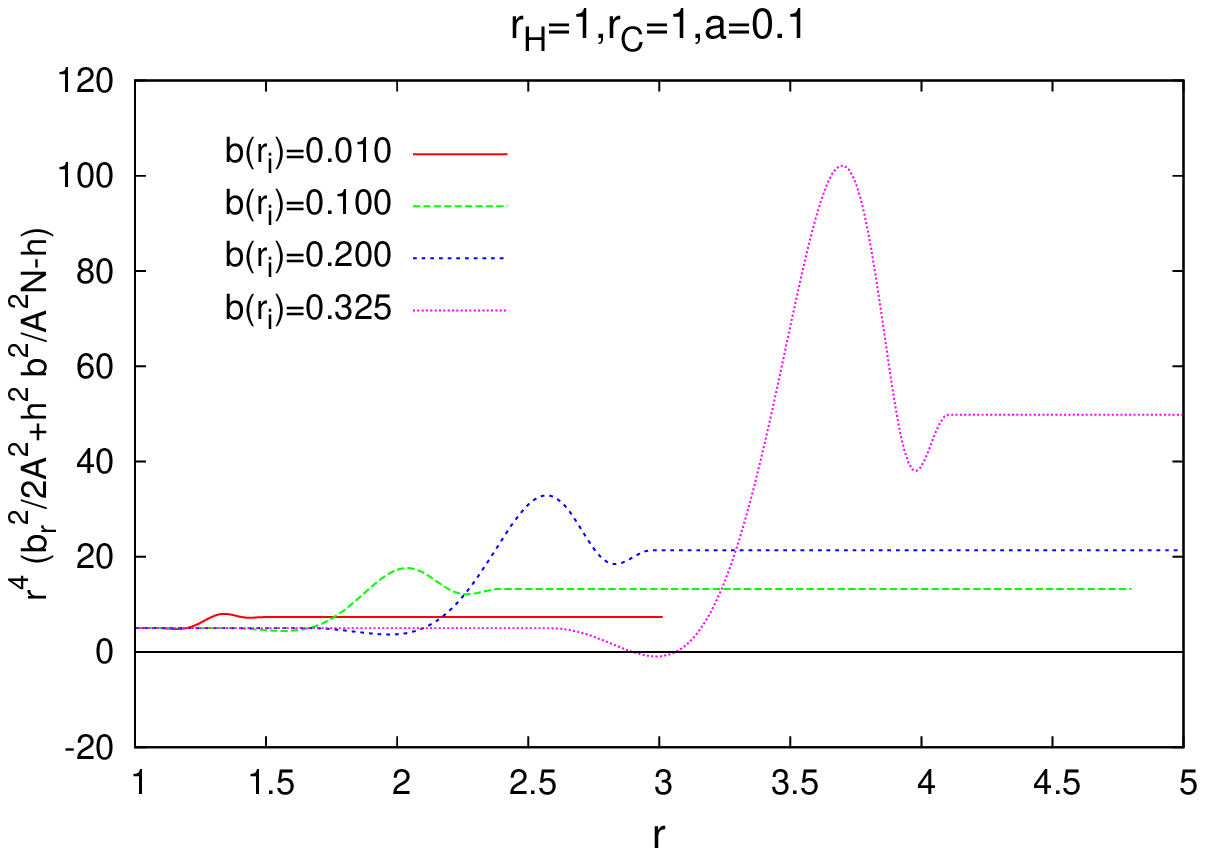}
\label{F9b}
}
}
\end{center}
\caption{Violation resp.~nonviolation of the strong energy condition
($b_r=b'$).
\label{bh9}
}
\end{figure}

We show in Fig.~\ref{F9a}
examples of solutions for which the strong energy condition is 
violated in some bounded region of space-time. However, there are
also solutions for which the strong energy condition is 
everywhere satisfied, as seen in Fig.~\ref{F9b}.
(For convenience we show the quantity 
$r^4 \left( b'^2/2 A^2+ h^2 b^2/2A^2 N -h)\right)$).

\section{Conclusions}

We have considered boson shells in scalar electrodynamics
with a $V$-shaped scalar potential,
implying that the scalar field is finite only in a compact 
shell-like region, $r_{\rm i} < r < r_{\rm o}$.
Coupling the scalar field to gravity
leads to self-gravitating boson shells
with an empty Minkowski-like interior region $r<r_{\rm i}$.
However, the interior region need not be empty.
In flat space, the interior region can contain a point charge.
In curved space, the interior region can either contain 
a black hole or a naked singularity.

While the flat space solutions can grow without bound,
the domain of existence of the self-gravitating solutions  
is limited. 
The reason is, that gravity does not allow for regular solutions,
which have a large mass concentrated in a (too) small region.
Indeed,
when the solutions approach the boundary of the domain of existence,
a throat develops at the outer radius of the shells,
making the exterior region of the space-time equal
to the exterior region of an extremal Reissner-Nordstr\"om
black hole.

When the boson shells carry like charges in their interior,
these give rise to electromagnetic repulsion of the surrounding boson shell.
When the charges located in the interior of the shell are of opposite
sign, however, further attraction arises that supplements the
gravitational attraction.
When the resulting attraction becomes too large,
shell-like solutions are no longer possible.
In particular,
the attraction resulting from opposite charges is too big,
to allow for globally neutral shell-like solutions.

Analyzing the physical properties of these types of solutions
we have found, that 
space-times with a black hole inside the boson shell can possess
the same values of the global charge $Q$ and the same
values of the global mass $M$,
while they differ in other physical properties.
Consequently, uniqueness does not hold for these black hole space-times.
Instead they possess
scalar hair in the form of charged compact shells.
When the inner black holes are neutral, there can only be
denumerably many such solutions with the same global charges.
These are present for sets of solutions
with a spiral-like behaviour.
In contrast, when the inner black holes carry charge
a continuous nonuniqueness of the solutions arises.
The black hole uniqueness theorems 
\cite{Bekenstein:1971hc,Bekenstein:1995un,Mayo:1996mv}
are evaded for these solutions because of the special type of potential,
making the scalar field vanish identically
outside the boson shell.
{ 
Note that in the proof of non-existence of black holes with 
scalar hair \cite{Mayo:1996mv} the asymptotic form of the scalar field 
is essential. For our solutions however, the scalar field
vanishes at some finite radius $r_{\rm o}$ and stays identically zero
for $r>r_{\rm o}$. Therefore the proof of \cite{Mayo:1996mv} does not
apply to the compact boson shells harbouring black holes.}

The next step will be to consider rotation \cite{Arodz:2009ye}.
While rotating boson stars are well-known
\cite{Mielke:2000mh,Schunck:2003kk,Yoshida:1997qf,Kleihaus:2005me,Kleihaus:2007vk},
the inclusion of rotation for self-gravitating boson shells
will present an interesting generalization.
However, the implementation and subsequent
numerical construction of such rotating boson shells
which may also harbour black holes
still represents a challenge.

{
So far we have considered only solutions which have an empty
Minkowski like interior, or possess
point-like charges,
Schwarzschild like black holes
or Reissner-Nordstr\"om like solutions in the interior of the boson shells.
It would be interesting to
consider also compact boson stars inside the boson shells.
This would lead to space-times with an onion like shell structure:
a compact boson star surrounded by 
a Reissner-Nordstr\"om like solution surrounded 
by a boson shell surrounded by 
a Reissner-Nordstr\"om like solution.
In principle this kind of solutions can be extended further to 
include also multiple boson shells.
}

\vspace{0.5cm}
{\bf Acknowledgement}

\noindent
BK gratefully acknowledges support by the DFG,
ML by the DLR.

\section{Appendix}

\begin{figure}[h]
\begin{center}
\mbox{\hspace{-1.5cm}
\subfigure[][]{
\includegraphics[height=.27\textheight, angle =0]{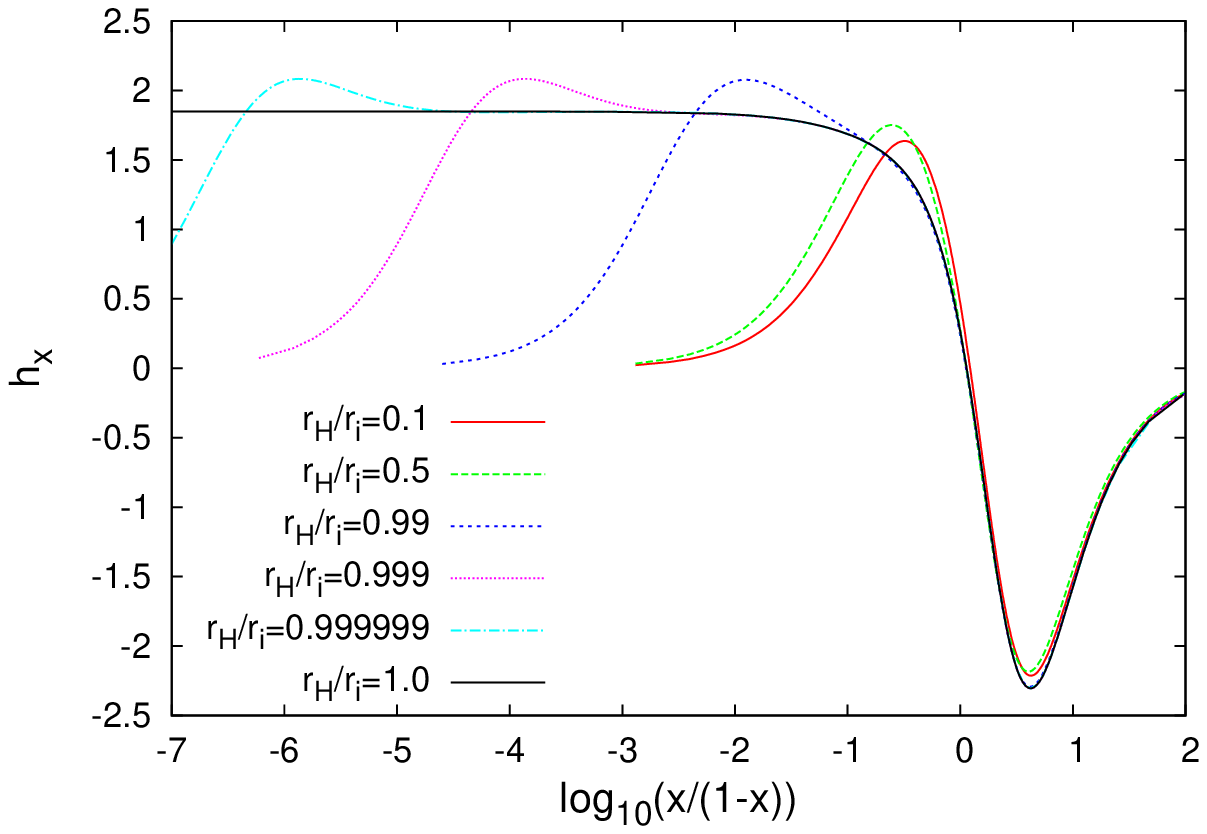}
\label{xNf1a}
}
\subfigure[][]{
\includegraphics[height=.27\textheight, angle =0]{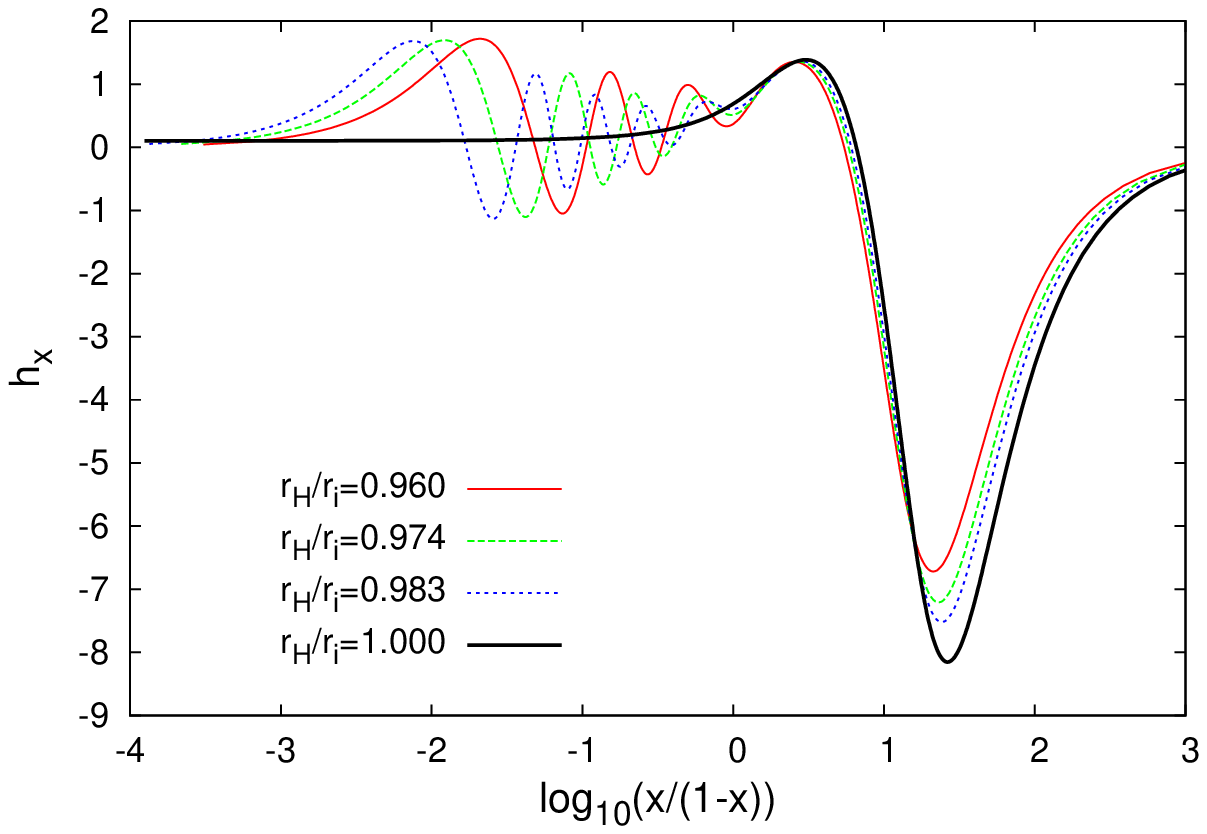}
\label{xNf2a}
}
}
\end{center}
\vspace{-0.5cm}
\caption{
(a) Function $h_x= dh/dx$ versus 
$\log_{10}(x/(1-x))$ with radial coordinate $x=(r-r_{\rm i})/(r_{\rm o}-r_{\rm i})$
for boson shells with Schwarzschild like black hole solutions
in their interior for charge $Q=10$ and gravitational coupling 
$a=\alpha^2 = 0.01$
and several values of $r_{\rm H}/r_{\rm i}$.
(b) The same as (a) for boson shells with negatively charged 
Reissner-Nordstr\"om like black hole 
solutions in their interior for charge $Q=100$, gravitational coupling 
$a=\alpha^2 = 0.01$ and event horizon radius $r_{\rm H}=9.0$.  
\label{sNh1}
}
\end{figure}

We here elucidate the limiting behaviour of the solutions,
when the branches of boson shells harbouring black holes end in spirals.
A particular case for the occurrence of
a spiral pattern was shown in Fig.~\ref{S1a},
where sets of boson shell solutions with Schwarzschild like
black holes in their interior were exhibited for 
small gravitational coupling.
As discussed above, such a spiral pattern arises in
the limit when the event horizon and the 
inner boundary of the shell approach each other
and finally coincide, i.e., when $r_{\rm H}/r_{\rm i} \to 1$.

In order to demonstrate how the solutions behave in this limit, 
we choose two typical examples for gravitational coupling constant $a=0.01$.
The first one has charge $Q=10$ and a Schwarzschild-like
black hole in the interior; the second one has charge $Q=100$ and 
a  negatively charged Reissner-Nordstr\"om like black hole in the interior. 
In both cases we consider a sequence of solutions with increasing values of 
the ratio $r_{\rm H}/r_{\rm i}$.

It is convenient to employ the radial coordinate $x=(r-r_{\rm i})/(r_{\rm o}-r_{\rm i})$. 
In Fig.~\ref{xNf1a} we show the function $h_x= dh/dx$ versus
$\log_{10}(x/(1-x))$. 
In the first example 
we observe that $h_x$ assumes its maximum at 
decreasing values of $x$ as the ratio $r_{\rm H}/r_{\rm i}$ tends to one.
Away from the maximum, the function $h_x$ approaches a limiting 
function labeled `$r_{\rm H}/r_{\rm i}=1$' in Fig.~\ref{xNf1a}.
In the second example we observe that the number of extrema of $h_x$ 
increases with increasing $r_{\rm H}/r_{\rm i}$. But similar to the 
previous example it seems that the locations of the extrema tend to zero 
(except for the two outer most extrema) as $r_{\rm H}/r_{\rm i}$
tends to one. This is demonstrated in Fig.~\ref{xNf2a}.
Note that in this example
the largest ratio $r_{\rm H}/r_{\rm i}=0.983$ is still not 
very close to the limiting value of one.

The limiting solutions are obtained with boundary conditions different
from the ones required for the boson shell solutions.
Thus we no longer impose $h'(r_{\rm i})=0$, but require regularity at the horizon, 
i.e., $|h''(r_{\rm H})|< \infty$, together with $N(r_{\rm H})=0$. 
We also require $h((r_{\rm H})=0$ and ${\rm sign}(h)=1$ at the horizon.
{
On the other hand we need ${\rm sign}(h)=0$ at the outer radius $r_{\rm o}$,
for compactness of the solution. Therefore, although the limiting solutions are
solutions of the boundary value problem, the 
potential of the scalar field is not consistently defined.}
Note that the condition ${\rm sign}(h)=1$ at the horizon
also is not consistent with ${\rm sign}(0)=0$, 
which we used for the boson shell solutions. 
Thus the limiting solution
does not belong to the set of boson shell solutions.
In fact, we observe from Fig.~\ref{sNh1} 
that the boson shell solutions converge
to the limiting solutions only pointwise, i.e.,
at all $r$ except $r_{\rm i}$.

\end{document}